\documentclass[12pt,preprint]{aastex}
\slugcomment{To be submitted to {\it The Astrophysical Journal}}
\shortauthors{Kirkpatrick}
\shorttitle{Lithium Test}

\begin{document}

\title{A Sample of Very Young Field L Dwarfs and Implications for the Brown Dwarf ``Lithium Test'' at Early Ages\altaffilmark{1}}

\author{J.\ Davy Kirkpatrick\altaffilmark{2},
Kelle L.\ Cruz\altaffilmark{3},
Travis S.\ Barman\altaffilmark{4},
Adam J.\ Burgasser\altaffilmark{5},
Dagny L.\ Looper\altaffilmark{6},
C.\ G.\ Tinney\altaffilmark{7},
Christopher R.\ Gelino\altaffilmark{8}, 
Patrick J.\ Lowrance\altaffilmark{8},
James Liebert\altaffilmark{9},
John M.\ Carpenter\altaffilmark{3},
Lynne A.\ Hillenbrand\altaffilmark{3},
John R.\ Stauffer\altaffilmark{10}}

\altaffiltext{1}{Most of the spectroscopic data presented herein were obtained at the W.M. Keck Observatory, which is operated as a scientific partnership among the California Institute of Technology, the University of California and the National Aeronautics and Space Administration. The Observatory was made possible by the generous financial support of the W.M. Keck Foundation. Other spectroscopic data were collected at the Subaru Telescope, the twin telescopes of the Gemini Observatory, the Magellan-Clay Telescope, the Kitt Peak National Observatory Mayall Telescope, and the Cerro Tololo Interamerican Observatory Blanco Telescope.}

\altaffiltext{2}{Infrared Processing and Analysis Center, MS 100-22, California 
    Institute of Technology, Pasadena, CA 91125; davy@ipac.caltech.edu}
\altaffiltext{3}{Department of Astronomy, MS 105-24, California Institute of Technology, Pasadena, CA 91125}
\altaffiltext{4}{Lowell Observatory, Planetary Research Centre, 1400 West Mars Hill Road, Flagstaff, AZ 86001}
\altaffiltext{5}{Massachusetts Institute of Technology, 77 Massachusetts Avenue, Building 37, Cambridge, MA 02139}
\altaffiltext{6}{Institute for Astronomy, University of Hawai'i, 2680 Woodlawn Drive, Honolulu, HI 96822}
\altaffiltext{7}{Department of Astrophysics, School of Physics, University of New South Wales, NSW 2052, Australia}
\altaffiltext{8}{Spitzer Science Center, MS 220-6, California Institute of Technology, Pasadena, CA 91125}
\altaffiltext{9}{Steward Observatory, 933 North Cherry Avenue, University of Arizona, Tucson, AZ 85721}
\altaffiltext{10}{Spitzer Science Center, MS 314-6, California Institute of Technology, Pasadena, CA 91125}

\begin{abstract}

Using a large sample of optical spectra of late-type dwarfs, we identify a subset of late-M through L field dwarfs that, because of the presence of low-gravity features in their spectra, are believed to be unusually young. From a combined sample of 303 field L dwarfs, we find observationally that 7.6$\pm$1.6\% are younger than 100 Myr. This percentage is in agreement with theoretical predictions once observing biases are taken into account. We find that these young L dwarfs tend to fall in the southern hemisphere (Dec$<$0$^\circ$) and may be previously unrecognized, low-mass members of nearby, young associations like Tucana-Horologium, TW Hydrae, $\beta$ Pictoris, and AB Doradus. We use a homogeneously observed sample of roughly one hundred and fifty 6300-10000 \AA\ spectra of L and T dwarfs taken with the Low-Resolution Imaging Spectrometer at the W.\ M.\ Keck Observatory to examine the strength of the 6708-\AA\ \ion{Li}{1} line as a function of spectral type and further corroborate the trends noted by \cite{kirkpatrick2000}. We use our low-gravity spectra to investigate the strength of the \ion{Li}{1} line as a function of age. The data weakly suggest that for early- to mid-L dwarfs the line strength reaches a maximum for a few$\times$100 Myr, whereas for much older (few Gyr) and much younger ($<$100 Myr) L dwarfs the line is weaker or undetectable. We show that a weakening of lithium at lower gravities is predicted by model atmosphere calculations, an effect partially corroborated by existing observational data. Larger samples containing L dwarfs of well determined ages are needed to further test this empirically. If verified, this result would reinforce the caveat first cited in \cite{kirkpatrick2006} that the lithium test should be used with caution when attempting to confirm the substellar nature of the youngest brown dwarfs.

\end{abstract}


\section{Introduction}

Determining the difference between an object stably burning hydrogen (a main sequence 
star) and an object of lower mass incapable of such stable fusion (a brown dwarf) is
not always straightforward. Although objects with main sequence spectral types of 
O, B, A, F, G, or K are always stars and objects with a spectral type of 
T are always brown dwarfs, dwarfs of type M or L comprise a 
mixture of stellar and substellar objects.

By definition, distinguishing a brown dwarf from a star requires knowledge 
that the object is not burning hydrogen in its core. There is no direct way to 
probe the interiors of these objects, so other techniques are required.
The most common indirect method used is 
the ``lithium test'' (\citealt{rebolo1992}). This test
recognizes the fact that the temperature ($T > 3{\times}10^6$K; \citealt{burrows1997, 
nelson1993}) needed to sustain core hydrogen fusion in 
the lowest mass stars is 
only slightly higher than that needed to burn lithium ($T \approx 
2{\times}10^6$K; \citealt{pozio1991}). If lithium does not burn, that means hydrogen 
is also not being fused. Fortuitously lithium, once destroyed, is not easily manufactured 
again in stellar interiors, so stars and brown dwarfs will never have more 
than their natal abundance of this element. In principle, the presence or absence of lithium 
in the spectrum can thus provide knowledge of inner temperatures.

In practice there are several drawbacks when applying the lithium test to 
M and L dwarfs: 

(1) The ground-state resonance line of \ion{Li}{1} is located at
6708 \AA, a portion of the spectrum where the flux is quite low for late-M and L dwarfs.
The low flux levels means that large telescopes 
are needed for observation of the line. Even then, very few late-type 
dwarfs are bright enough that they can be acquired at high resolution
so to obtain a sizable sample of spectra, moderate resolution 
($\Delta{\lambda}\approx 5-10$ \AA) is required. Fortunately, this is sufficient for  
measuring \ion{Li}{1} equivalent widths of a few \AA\ or greater provided that the
signal-to-noise ratio is high. 

(2) As stated above, there is a ${\sim}1{\times}10^6K$ difference between the temperatures
for lithium ignition and normal hydrogen ignition. In other words, some brown dwarfs are capable
of burning lithium and for those the lithium test will fail. These higher mass brown dwarfs 
are unrecognizable as substellar based on this test alone.

(3) A sufficient amount of time needs to elapse for the natal supply of lithium to
be extinguished. The 
\ion{Li}{1} line may therefore still be present not because the object is substellar,
but because it is a star too young to have fused its lithium store. In a
very young, low-mass M star, lithium may still be present in the spectrum 
either 
%(a)
because the core has not yet reached lithium-burning
temperatures or
%(b) because the object has not yet cooled to the temperatures at
%which it is fully convective, and is thus unable to recirculate the surface lithium to the core
%where it can be destroyed, or (c) 
because there has been insufficient time for convective
recirculation to have destroyed all the lithium in the star and in particular all of the 
lithium in the photosphere (but this recirculation is
expected to be rapid compared to the other two timescales considered 
here; \citealt{basri1998}). Despite these
potential pitfalls, \cite{basri1998} has demonstrated
that any object with lithium absorption and Teff $<$ 2670K is unambiguously a brown dwarf. Thus if we
confine our focus to objects with temperatures cooler than this,
the lithium test can be applied regardless of these considerations. Using the
relation of optical dwarf spectral type to temperature from \cite{kirkpatrick2008} --
$T_{eff}(K) = 3759 - 135x$, where, e.g, x=0 for M0, 9 for M9, 10 for L0, and 18 for L8 --
we find that an effective temperature of 2670K corresponds to a dwarf of optical type M8. For the
remainder of this paper, we will therefore restrict our discussion of the lithium test
to dwarfs with optical types of M8 and later.

(4) It is 
expected that the 6708-\AA\ line will disappear at very cool temperatures as
lithium forms molecular species. \cite{lodders1999} has shown that below $T_{eff} \approx 1500$K 
Li-bearing molecules like 
LiCl and LiOH begin to form. For brown dwarfs of solar age, this temperature corresponds to
optical spectral types of late-L through mid-T (see Figure 7 of \citealt{kirkpatrick2005}).
Stars of solar abundance do not exist at temperatures this low (see Figure 10
of \citealt{kirkpatrick2005}), so we can safely assume that objects in this range of spectral
types are
all brown dwarfs. For these the lithium test is not needed. This theoretical prediction of
lithium disappearance at these types should nonetheless be tested with empirical data.

(5) Young ($<$100 Myr) brown dwarfs have radii larger than older stars and brown dwarfs of the same
spectral type because they are still contracting to their final structural configuration
(\citealt{burrows1997}). These young brown dwarfs are also lower in mass than stars and older brown
dwarfs of the same spectral type. Both the larger radius and the lower mass mean that these
objects will have lower surface gravities, which should be detectable even with low-resolution
spectroscopy. (See \S4.1.) Lower gravity will weaken the \ion{Li}{1} line due to the lower
atmospheric pressure, making it much harder 
to detect. This has been observed by \cite{kirkpatrick2006} in the low-gravity, early-L dwarf
2MASS J01415823$-$4633574, which shows no \ion{Li}{1} line down to an equivalent width
of 1 \AA. Weak \ion{Li}{1} is seen despite the fact that this object is believed to be a young 
(1-50 Myr), low-mass (6-25 M$_{Jupiter}$) brown dwarf and despite the fact that other brown dwarfs
of similar spectral type are known that show prominent \ion{Li}{1} lines -- such as
the L0 dwarf 2MASS J11544223$-$3400390 with a \ion{Li}{1} equivalent width of
3 \AA.

We can test points (4) and (5) using a unique set of spectroscopic data.
Over the past several years we have amassed a library of optical spectra of 
L and T dwarfs primarily using the W.\ M.\ Keck Observatory. Data from
\cite{kirkpatrick1999, kirkpatrick2000, burgasser2000, kirkpatrick2001, wilson2001, 
burgasser2003, burgasser2003b, thorstensen2003, mcgovern2004, kirkpatrick2006} 
combined with new data presented in this paper now gives us a library of $\sim$150 
spectra spanning the wavelength regime 6300-10000 \AA\ and covering spectral types 
L0 through T8. With such a large set of homogeneous spectra, we can 
better characterize the behavior of the \ion{Li}{1} line as a function of spectral type.
Moreover, this large set of spectra enables us to identify peculiar objects,
particularly those with unusual spectra indicative of low gravity. As stated above, low gravity in
L and T dwarfs is an indicator of youth, so these spectra also enable us to
characterize the \ion{Li}{1} strength as a function of age.

In this paper we present the newest spectra in our optical spectral library (\S2),
analyze objects with detected H$\alpha$ (\S3), and identify 
objects with low-gravity (young) signatures (\S4). We perform
further analysis of the young objects (\S5) and
build statistics on the behavior of \ion{Li}{1} line
strengths as a function of both spectral type and age (\S6). Conclusions are given in \S7.

\section{Object Selection, Observation, and Classification}

\subsection{The Two Samples}

Over the past several years we have conducted Keck Low Resolution 
Imaging Spectrometer (LRIS, \citealt{oke1995}) observations of two different samples. The first is a 
color-selected sample of southern hemisphere L dwarf candidates chosen from 2MASS. 
The second sample is a collection of late-L dwarfs, early-T dwarfs, and a few other 
interesting cases drawn from the literature. We present details on each sample below.

\subsubsection{The Southern L Dwarf Sample from 2MASS}

In July 2000, one of us (CGT) wished to begin a parallax program for 
ultra-cool dwarfs using facilities at the Anglo Australian Observatory (AAO).
At this time, very few L dwarfs were recognized south of Dec=$0^{\circ}$, so we performed a photometric search for L dwarf candidates using data from the Two Micron All Sky Survey (2MASS, \citealt{skrutskie2006}). This search was performed in two steps. First, the 2MASS Second Incremental Data Release (IDR2) Point Source Catalog\footnote{Available at \url{http://irsa.ipac.caltech.edu}.} was searched for objects meeting the following criteria: $-50^{\circ} <$ Dec $< +5^{\circ}$, $8.0 < K_s < 13.6$, $J-K_s > 1.30$, $R-K_s > 6.5$ (where optical magnitudes are available from the USNO-A), and $|b| > 25^{\circ}$. In order to weed out fainter objects at early-L types, a further cut was employed to drop objects with $K_s > 13.0$ if $J-K_s < 1.6$. To supplement this data set, we also performed on the 2MASS Point Source Working Database a search for southern hemisphere data taken after 20 Feb 1999 (the cutoff date for the 2MASS IDR2) but before July 2000. Selection criteria on this set stipulated that the candidates have $-50^{\circ} <$ Dec $< +5^{\circ}$, $|b| > 25^{\circ}$, $J-K_s > 1.30$, $R-K_s > 5.5$ (where optical magnitudes are available from the USNO-A), and $9.5 < K_s < 13.6$. Visual inspection of the DSS, XDSS, and 2MASS images was used to drop objects contaminated by bright stars or galaxies, lying in nebular regions such as the Lynds 134 cloud complexes, or having bright R-band counterparts on the optical plates. The latter objects are not always eliminated using the $R-K_s$ selection criterion because the USNO-A Catalog has as one of its constraints that any real object must be detected on both the blue and red plates; as some late-type giants will be R-band only sources on the optical sky plates, these will not be included in the USNO-A (\citealt{monet2003}).

Our final candidate list of 68 objects is given in Table~\ref{l_candidates}. Objects 
verified spectroscopically to be late-M or L dwarfs are given in the upper portion of the table; those objects known not to be late-M or L dwarfs are given in the lower portion. 2MASS source designations are listed in column 1, and photometric data from the final 2MASS All-Sky Release Point Source Catalog (\citealt{cutri2003}), which postdates our selection, are given in columns 2-5. Using this later catalog as the data source, some of the objects we originally selected may no longer fall within our constraints (such as the $J-K_s$ color of 1.29 for 2MASS J04433761+0002051). Columns 6-7 give the optical spectral type and reference for the classification. For objects discovered previously by other surveys, alternate source designations are given in column 9.

As for the AAO L dwarf parallax program itself, observations of our confirmed candidates continue. Results from the program will be presented in a forthcoming paper.

\subsubsection{Interesting Objects Pulled from the Literature}

The goal of the second sample was to obtain optical spectra of late-M, L, and T dwarfs that either had no previous spectroscopic follow-up in the optical or that would benefit from an optical spectrum with higher signal-to-noise or covering a new epoch. These interesting objects\footnote{In the text we abbreviate sources names with a prefix such as 2MASS, DENIS, or SDSS and a suffix of the form Jhhmm$\pm$ddmm, where hhmm is the truncated J2000 Right Ascension in hours and minutes and $\pm$ddmm is the truncated J2000 Declination in degrees and minutes. Full designations are given in the tables.} are listed in approximate order of RA below:

\noindent {\it 2MASS J0415$-$0935} is the latest optical T dwarf standard from \cite{burgasser2003}. 

\noindent {\it 2MASS J05185$-$2828} is a discovery from the ongoing 2MASS L dwarf search of \cite{cruz2003} and because of a very unusual near-infrared spectrum was hypothesized to be a late-L + T dwarf double (\citealt{cruz2004}). Its binary nature has been confirmed via HST imaging (\citealt{burgasserhst2006}). 

\noindent {\it SDSS J0830+4828, SDSS J0837$-$0000, SDSS J0857+5708, SDSS J1021$-$0304, and SDSS J1254$-$0122} are all classified in the near-infrared as late-L or early-T dwarfs (\citealt{leggett2000,geballe2002}), and we obtained LRIS spectra of these to explore the L/T transition at far optical wavelengths. 

\noindent {\it Gl 337CD} adds to this sample near the L/T transition and is relatively bright ($K_s = 14.0$). Its optical type is L8 (\citealt{wilson2001}) whereas its near-infrared type is T0
(\citealt{burgasserunified2006}).

\noindent {\it 2MASS J1209$-$1004} adds another data point to the L/T transition. This object is the T3 near-infrared standard on the \cite{burgasserunified2006} scheme. 

\noindent {\it 2MASS J1315$-$2649} is an L dwarf with abnormally strong and variable H$\alpha$ emission (\citealt{hall2002a,gizis2002,hall2002b,riaz2007}) and we chose to observe it again to check its H$\alpha$ strength. 

\noindent {\it 2MASS J2244+2043} is an extremely red ($J-K_s = 2.45{\pm}0.16$) late-L dwarf uncovered during a hunt for red QSOs in 2MASS. We chose to observe this object to see if the optical spectrum revealed any clues regarding its unusually red near-infrared photometry.

\subsection{Spectroscopic Observations}

Optical spectra were obtained with LRIS on the 10m Keck-I Observatory atop Mauna Kea, Hawaii. A 400 lines/mm grating
blazed at 8500 \AA\ was used with a 1$\arcsec$ slit and 2048$\times$2048
CCD to produce 10-{\AA}-resolution spectra covering the range
6300 -- 10100 \AA.  The OG570 order-blocking filter was used
to eliminate second-order light.  The data were reduced and calibrated 
using standard IRAF routines. Flat-field exposures of the interior of the 
telescope dome were used to normalize the response of the detector. 

Individual stellar spectra were extracted
using the ``apextract'' routine in IRAF\footnote{IRAF is distributed by the National Optical Astronomy Observatories, which are operated by the Association of Universities for Research in Astronomy, Inc., under cooperative agreement with the National Science Foundation.}, allowing for the slight curvature of
a point-source spectrum viewed through the LRIS optics and using a template
where necessary. Wavelength calibration was achieved using neon+argon
arc lamp exposures taken after each program object. Finally,
the spectra were flux-calibrated using observations of 
standards from \citet{hamuy1994}.
Most of the data have not been corrected for telluric absorption, so the 
atmospheric
O$_2$ bands at 6867-7000, 7594-7685 \AA\ and H$_2$O bands at 7186-7273,
8161-8282, $\sim$8950-9300, $\sim$9300-9650 \AA\ are still present in
the spectra. In the sections that follow, a few cases are noted where a correction for telluric absorption was
applied by using the spectrum of a nearby field late-F/early-G star taken just before or after the spectrum of the program object. 

Table~\ref{nights} lists the UT dates of observation, program principal investigator, other observers assisting, and sky conditions for the sixteen different nights on which data were taken or attempted. The 3.5-year time span was a consequence of the fact that these L dwarf targets were used as filler targets during other programs and were thus observed only in twilight, during gaps in the main program, or during periods of cloud cover. As listed in the table, three nights -- 2000 Aug 23, 2001 Feb 20, and 2001 Nov 13 -- had poor transparency that sometimes led, as in the spectrum of the L dwarf companion Gl 618.1B, to poor signal-to-noise. Two nights were completely lost to adverse weather conditions, but on the eleven other nights conditions were clear with good seeing for generally the entire night. The list of the southern L dwarf candidates observed with Keck-LRIS during these runs is given in Table~\ref{new_spec_obs}. The discovery name is given in column 1, with the observing date and integration listed in columns 2-3. As for the sample from the literature, the list of targeted M and L dwarfs is given in Table~\ref{new_spec_ml}, and the T dwarf target list is given in Table~\ref{new_spec_t}. The first four columns in both tables give the object name, the discovery reference, the UT observation date of our Keck-LRIS spectrum, and the integration time. 

Finder charts are provided in Figure~\ref{finders} for the twelve new late-M through late-L dwarfs identified in the southern sample. Charts are made from the J-band component of the 2MASS All-Sky Release Survey Images as served through the on-request mosaicking service available at the NASA/IPAC Infrared Science Archive\footnote{Available at \url{http://irsa.ipac.caltech.edu/}}. 

\subsection{Spectroscopic Classification}

Sixty eight candidates met the criteria of the southern sample. Seventeen of these, as listed in the lower section of Table~\ref{l_candidates}, were found not to be late-type dwarfs as judged by previously published spectra or our own LRIS observations. These consisted of 10 carbon stars, 4 M giants, 2 QSOs, and one reddened, early-type star. Of the remaining fifty-one objects, listed in the upper section of Table~\ref{l_candidates}, we observed twenty-eight and took published optical classifications for the remaining twenty-three. The twenty-eight for which we obtained LRIS spectra are listed in Table~\ref{new_spec_obs} and are plotted on linear and logarithmic flux scales in Figure~\ref{allspec_lin} and Figure~\ref{allspec_log}. Spectra of objects from our literature sample, given in Table~\ref{new_spec_ml} and Table~\ref{new_spec_t}, are plotted on both linear and logarithmic scales in Figure~\ref{new_LT_lin} and Figure~\ref{new_LT_log}.

Classifications were assigned as follows. For late-M dwarfs, classifications were measured by eye using LRIS spectra of late-M dwarf secondary standards from \cite{kirkpatrick1999}. For L dwarfs, the prescription for classification presented in \cite{kirkpatrick1999} and \cite{kirkpatrick2000} was followed, using the indices\footnote{A few typesetting errors appearing in Table 7 of \cite{kirkpatrick1999} should be noted. First, the numerator of the Na-b ratio should cover the 10-\AA\ interval 8153.3-8163.3 \AA\ (the same as the numerator for Na-a) instead of the 30-\AA\ interval listed. Second, the numerator for Cs-b should be ``Av.\ of 8918.5-8928.5 and 8958.5-8968.5'' and the denominator should be ``8938.5-8948.5''. Third, the denominator of the TiO-b ratio should cover the 15-\AA\ interval 8455.0-8470.0 rather than the  35-\AA\ interval given. Only this third index, TiO-b, is actually used in the spectral typing recipe.} defined in those papers. Measurements of each of the indices are given in columns 4-9 of Table~\ref{new_spec_obs} and columns 5-10 of Table~\ref{new_spec_ml}, and the final types resulting from these indices are listed in column 10 of Table~\ref{new_spec_obs} and column 11 of Table~\ref{new_spec_ml}. After index-based types were assigned, our LRIS spectra were compared by eye to LRIS spectra of the L dwarf standards to look for anomalies. Spectral types of objects having feature strengths uncharacteristic of field objects of like type were further given a ``pec'' (peculiar) suffix. These objects are discussed further in \S4.

For the T dwarfs in Table~\ref{new_spec_t}, optical classifications were assigned using the index-based recipe outlined in \cite{burgasser2003}. Measurements for the optical T dwarf indices are given in columns 5-9 of Table~\ref{new_spec_t} and the resulting spectral types given in column 10. As with the L dwarfs these T dwarfs were compared by eye to LRIS spectra of the optical T dwarf standards to look for peculiarities and to check the final index-based types.

Because many of the objects in Table~\ref{new_spec_obs} are either new discoveries or now have optically derived spectral types for the first time, we provide distance estimates. As shown in Figure 9 of \cite{kirkpatrick2005}, optical L dwarf spectral types correlate better with absolute J-band magnitude than do near-infrared L dwarf spectral types, so optical types can be used to derive more accurate spectrophotometric distances.  Our distance estimates, listed in column 11 of Table~\ref{new_spec_obs}, are derived using the 2MASS-measured J-band magnitudes of the objects, our measured optical spectral types, and the relation between absolute J-band magnitude and optical spectral type from \cite{looper2008} for L dwarfs or the mean absolute J-band magnitudes computed by \cite{kirkpatrick1994} for M dwarfs. 

Using our LRIS spectra, we have used the {\it splot} package in IRAF to compute the pseudo-equivalent widths\footnote{These pseudo-equivalent widths, measured relative to the local (not true) continuum, will be referred to simply as ``equivalent widths'' throughout the remainder of the paper.} (or limits) for H$\alpha$ emission at 6563 \AA\ and \ion{Li}{1} absorption at 6708 \AA. In the case of limits, we measure the equivalent widths of positive and negative noise spikes in the same spectral region to ascertain the width of the largest real feature that could be masked at that signal-to-noise level. These are given in columns 12-13 of Table~\ref{new_spec_obs} and Table~\ref{new_spec_ml} and columns 11-12 of Table~\ref{new_spec_t}. (Note that we have expressed the equivalent width for both as positive.) We discuss those objects with measured H$\alpha$ emission in \S3; objects with measured lithium absorption are discussed later in \S6.

\section{Objects with H$\alpha$ Emission}

Unlike for warmer dwarfs, H$\alpha$ is not a reliable indicator of youth in late-M and later dwarfs. Most late-M dwarfs show H$\alpha$ emission as illustrated elsewhere (see Figure 6 of \citealt{gizis2000}, Figure 1 of \citealt{west2004}, and Figure 3 of \citealt{west2008}). Indeed, four of our five late-M dwarfs from Table~\ref{new_spec_obs} exhibit H$\alpha$ emission, as shown in the spectral zoom-ins of Figure~\ref{halpha}. 

Figure~\ref{halpha} also shows the L and T dwarfs from our sample that have H$\alpha$ emission. (See Table~\ref{new_spec_ml} and Table~\ref{new_spec_t}.) Three early-L dwarfs, three mid- to late-L dwarfs, and one early-T dwarf show the feature. The emission lines in 2MASS J0141$-$4633 and 2MASS J1315$-$2649 vary but are persistent. In the case of 2MASS J0141$-$4633 this persistence has been measured over only a single night (\citealt{kirkpatrick2006}). For 2MASS J1315$-$2649 our new spectrum shows an H$\alpha$ equivalent width of 160 \AA. Comparing this to the twelve separate epochs published earlier (\citealt{hall2002a,gizis2002,hall2002b,fuhrmeister2005,barrado2006}) we find that our observation places it squarely in its flaring state; the H$\alpha$ equivalent width is $\sim$20 \AA\ when the object is in quiescence (\citealt{riaz2007}). In the case of 2MASS J2057$-$0252, there are two epochs of observation separated by nearly a year; our 2000 Aug 23 measurement of 11 \AA\ equivalent width is consistent within the errors with the 2001 Jul 15 measurement of 8.44 \AA\ from \cite{schmidt2007}, who have analyzed the emission properties of the \cite{cruz2003} sample. Another object in common with the \cite{cruz2003} sample, 2MASS J0144$-$0716, was caught in a flare during our observation, as chronicled in \cite{liebert2003}. Finally, H$\alpha$ emission in the two latest objects, the L7.5 dwarf SDSS J0423$-$0414 and the T2 dwarf SDSS J1254$-$0122, has been described elsewhere (\citealt{burgasser2003}, \citealt{burgasser2005c}).

\section{Spectroscopic Effects of Low Gravity (Youth)}

\subsection{Introduction}

Many published works on late-M dwarfs have discussed the utility of gravity-sensitive spectral features in discerning lower mass brown dwarfs from higher mass stars of the same temperature or spectral type. Since the early days of astronomical spectroscopy, the strengths of CaH, the \ion{Ca}{2} triplet, and the neutral alkali lines have been identified as gravity sensitive and used to distinguish M giants from M dwarfs.  In an M giant, the much more distended photosphere has much lower gravity and pressure than the photosphere of an M dwarf of equivalent temperature. A similar effect makes a young, low-mass brown dwarf distinguishable from an older, higher-mass brown dwarf (or star) of similar temperature. Specifically, two effects are at work. First, a brown dwarf younger than $\sim$100 Myr will still be contracting to its final radius  (see Fig.\ 10 of \citealt{burrows1997}) and will have a more distended atmosphere than an older brown dwarf of the same spectral type. Second, a younger brown dwarf must necessarily have a lower mass than an older brown dwarf of the same temperature. Relative to the older object, the younger one is larger and less massive and thus has a relatively lower gravity and pressure.

One of the first papers to demonstrate these effects in low-mass stars and brown dwarfs, \cite{steele1995}, used a weakening of the \ion{Na}{1} lines in Pleiades late-M brown dwarf candidates to argue that these objects were true members of the cluster. \cite{martin1996} also noted that Pleiades brown dwarf candidates have slightly stronger VO bands than those of similarly classified field dwarfs. \cite{luhman1997} found that an M8.5 brown dwarf candidate in the $\rho$ Ophiuchi complex exhibited calcium hydride and \ion{K}{1} strengths intermediate between an M8.5 V and an M8.5 III, whereas the \ion{Na}{1} strengths were much more like the M8.5 III. \cite{luhman1998} discussed how the optical spectrum of a brown dwarf candidate in the Taurus-Auriga complex had \ion{Na}{1} and \ion{K}{1} strengths and TiO/VO ratios more similar to those of a giant while the CaH strength was intermediate between an M dwarf and an M giant. Later studies have continued to use these gravity-sensitive diagnostics to argue for youth in other late-type dwarfs.

It is important to note here that the Pleiades cluster holds a special place in the study of low-gravity effects in brown dwarf spectra because of its nearness to the Sun and the extent to which its low-mass members have been scrutinized. As discussed above, not only has it been shown that substellar objects in the Pleiades show such effects empirically, but also (as we discuss below) the Pleiades cluster falls near the aforementioned upper age boundary of $\sim$100 Myr where theory predicts these effects will be just discernible. The Pleiades is one of the few clusters where it is possible to derive an age estimate from all three of the fundamental age indicators: the pre-main sequence turn-on, the upper main sequence turn-off, and the lithium depletion boundary (Bildsten et al. 1997). We discuss each in turn.

1) The pre-main sequence turn-on age for the Pleiades is the most difficult of these to measure because the Pleiades locus is only slightly different from the zero-age main sequence even for low-mass stars and because theoretical models do not match real cluster isochrones well at V-I $>$ 2 (\citealt{stauffer2007}), making it difficult to derive accurate pre- main sequence ages for clusters with ages $\geq$ 100 Myr.  A reasonable pre- main sequence age estimate for the Pleiades is 100 Myr (\citealt{stauffer2007}), but with considerable uncertainty.  

2) Modern upper main sequence age estimates range from 77 Myr (\citealt{mermilliod1981}) for a minimal convective core overshoot parameter up to about 150 Myr (\citealt{mazzei1989}) for a relatively large convective core overshoot parameter.  The most frequently quoted upper-main sequence age is 100 Myr (\citealt{meynet1993}).  

3) It has been claimed that an age derived from the lithium depletion boundary should be more accurate than any other cluster age estimate (\citealt{bildsten1997}) because there are relatively few adjustable parameters. \cite{stauffer1998} derived a lithium depletion boundary age for the Pleiades of 125 Myr, with an estimated uncertainty of about 10 Myr. \cite{burke2004} reanalyzed the \cite{stauffer1998} data, and suggested a lithium depletion boundary age for the Pleiades of 126-148 Myr, dependent on how the bolometric correction is handled.

The following evidence from other clusters should be considered before adopting an age for the Pleiades. The $\alpha$ Persei cluster is another where it is possible to obtain age estimates from all three methods.  If the eponymous star $\alpha$ Per, an F supergiant, is indeed a member of the cluster and is coeval with the other stars, its location in color-magnitude diagrams places an upper limit on the age of the cluster of $<$80 Myr (\citealt{ventura1998}), as compared to the lithium depletion boundary age estimate of $\sim$90 Myr (\citealt{stauffer1999}). A similar tendency for the lithium depletion boundary age to be significantly older than age estimates from other methods appears to be true for other open clusters (\citealt{jeffries2001}). For that reason, we choose to treat the lithium depletion boundary age for the Pleiades as an upper limit, and to adopt a Pleiades age of 100-125 Myr for this paper.

Moving beyond the Pleiades and $\alpha$ Persei to slightly older ages, \cite{bannister2007} suggest that the early-L dwarf 2MASP J0345432+254023 may be a member of the Ursa Major/Sirius Moving Group, which has an age of 400$\pm$100 Myr. This L0 dwarf does not look unusual relative to the majority of other early-L dwarfs, a reassuring fact because it was chosen as the L0 optical spectroscopic standard by \cite{kirkpatrick1999}. \cite{jameson2008a} further identify the objects 2MASSW J0030438+313932 (optical L2), 2MASSI J1204303+321259 (optical L0), and 2MASS J15500845+1455180 (optical L2:) as being potential members of the Ursa Major Group, and none of these show spectral peculiarities (\citealt{kirkpatrick1999,cruz2003,cruz2007}). 

In summary, objects at ages $\sim$4$\times$ that of the Pleiades have spectra in which low gravity effects are no longer discernible at the $\sim$10 \AA\ resolution of our classification spectra; conversely, objects of Pleiades age show spectral hallmarks of low gravity. Hence, we adopt $\sim$100 Myr as the oldest age at which the signatures of low gravity can reliably detected in our spectra.

\subsection{Analysis of Low-Gravity (Young) Spectra}

The spectra of a number of field ultra-cool (spectral types $\ge$M7) dwarfs from the literature have spectral morphologies that can be attibuted to low gravity. We identify another four low-gravity objects using new spectra presented in this paper. Table~\ref{young_dwarfs} gives this combined list of twenty low-gravity field dwarfs. For objects having optical spectral types $>$M8 we compare in the following subsections our spectra to fiducial spectra in order to estimate ages for the peculiar objects using {\it empirical} evidence only.  

In these subsections, we provide qualitative analyses to aid in the recognition of low-gravity signatures. We feel that such ``by-eye'' assessments are, in fact, superior to classification via spectral indices because they tap the ability of the brain to consider the {\it totality} of information available in the spectra. This parallels the philosophy of William Morgan, one of the pioneers of the MKK classification system (\citealt{morgan1943}), whose reasoning was described by \cite{garrison1995} as follows:
         ``Morgan used the techniques of visual pattern recognition in a morphological
          approach to classification. In today's climate of the `deification of 
          quantification', it is sometimes difficult for people to see the power of
          such an approach, yet the human brain has evolved to be ideally suited to 
          such a methodology. Morgan was fond of using the analogy of the brain's
          ability to recognize familiar human faces. With pattern recognition, the
          result is immediate; nothing is measured, but all of the pattern information
          is compared with experience.''

This having been stated, in future papers we will outline quantitative spectroscopic classifications of our young objects. In Cruz et al.~(in prep.) we will investigate spectral indices for the optical spectra of young L dwarfs, and in Kirkpatrick et al.~(in prep.) we will establish spectral indices and a grid of spectral standards of known age to serve as comparisons to young, late-M dwarfs in both the optical and near-infrared regimes. A preview of the latter paper is shown in Figure~\ref{lateM_gravity_sequence}, which demonstrates that optical spectra at our resolution can be distinguished at the level of $\sim$1 dex in log(age).

For some of our low-gravity objects we have obtained spectra at telescopes other than Keck or have obtained Keck data for comparison objects not included in the two samples discussed earlier. Details on these additional observations are given in Table~\ref{gravity_obs}. We also include in the following discussion two low-gravity L dwarfs, G 196-3B and Gl 417B, that are companions to main sequence stars and thus have independent measures of age via their association with a well studied primary star. These are discussed below in order of spectral type. To guide the eye in Figure~\ref{lowg_lateM} through Figure~\ref{lowg_L6.5}, old field objects ($\sim$1 Gyr) are plotted in black, objects believed to have ages near that of the Pleiades ($\sim$100 Myr) are shown in red, and those believed to be significantly younger than the Pleiades ($\sim$10 Myr) are shown in magenta. Giant spectra, illustrating even lower gravities, are shown in blue.

\subsubsection{2MASS J0608$-$2753 (M8.5 pec) and SDSS J0443+0002 (M9 pec)}

2MASS J0608$-$2753 was discovered by \cite{cruz2003}, who noted that its spectrum exhibited the hallmarks of low gravity. SDSS J0443+0002 was discovered by \cite{hawley2002} but the spectrum in that paper was sufficiently noisy that the peculiarities went unnoticed. This object was recovered and reobserved as part of our southern L dwarf program.
Figure~\ref{lowg_lateM} shows our new spectra of these objects. They are compared to spectra of a normal M9 field dwarf and the late-M dwarf Teide 1. The latter object, discovered by \cite{rebolo1995}, is a member of the Pleiades and is thus assumed to have the age of $\sim$100 Myr typically assigned to the cluster. Also noticeable in the spectra of 2MASS J0608$-$2753 and SDSS J0443+0002 is the \ion{Na}{1} doublet, which is weaker than in both the normal field M9 and Teide 1, and the VO bands, which are stronger than these same two comparison spectra. The bands of CaH near 7050 \AA\ and FeH at 9896 \AA\ are somewhat weaker than in these same two comparison objects. (These same trends hold when comparing to the later Pleiad, Roque 4, in the left panel of Figure~\ref{lowg_L0}.) This evidence suggests that both 2MASS J0608$-$2753 and SDSS J0443+0002 have ages $<$100 Myr because they have gravities comparable to or lower than that of the Pleiad. 

\subsubsection{2MASS J1022+0200, DENIS J0357$-$4417, 2MASS J0241$-$0326, 2MASS J0141$-$4633, and 2MASS J2213$-$2136 (L0 pec)}

2MASS J1022+0200 was discovered by \cite{reid2008} and that spectrum is shown in the left panel of Figure~\ref{lowg_L0}. DENIS J0357$-$4417 was first published by \cite{bouy2003} who guessed a spectral type of ``$\sim$L3'' based on its measured I-J color. Our spectrum is the first reported of this object, which comes from our southern L dwarf sample, and it shows DENIS J0357$-$4417\footnote{This object is referred to as DENIS-P J035726.9$-$441730 in Table 1 of \cite{bouy2003} but as DENIS-P J035729.6$-$441730 throughout the text of that same paper. We assume the text is in error and retain the first designation because the entry in the DENIS 3rd Release clearly refers to this position, and it also coincides with the position of our independent discovery of this object in 2MASS (see Table~\ref{l_candidates}).} to have a considerably earlier type.
Our spectrum is shown in the left panel of Figure~\ref{lowg_L0}.

\cite{bouy2003} find DENIS J0357$-$4417 to be a binary with separation of 98$\pm$2.8 milliarcseconds and magnitude differences in the HST/WFPC2 filters of ${\Delta}m_{F675W}$ = 1.23$\pm$0.11 mag and ${\Delta}m_{F814W}$ = 1.50$\pm$0.11 mag. The latter difference means that at I-band the primary contributes four times more flux to the composite spectrum than the secondary does. This difference in I-band flux suggests that the L0 composite spectrum is comprised of spectra differing by roughly three spectral subclasses (e.g., an L0 and an L3 or an M9.5 and an L2.5). We have tested this by adding a standard L0 spectrum to that of a standard L3 spectrum scaled down so that it has only one-fourth the flux of the L0 in the I-band and find that the composite spectrum is only slightly different from the L0 itself. Specifically, the TiO bands at 7053 and 8432 \AA\ are somewhat shallower in the composite than in the L0 itself, but the strength of the alkali lines, the VO bands, and the hydrides are essentially unchanged. In other words, the composite spectrum is so little different than the single L0 dwarf that its spectrum would not be labelled unusual. Therefore, we can assume that the oddities seen in the composite spectrum of DENIS J0357$-$4417 are indicative of oddities in the primary spectrum itself and are not induced by the composite nature of the spectrum.

Also shown in the left panel of Figure~\ref{lowg_L0} are spectra of a normal M9 dwarf (at top), a normal L0 dwarf (at bottom), and the late-M Pleiad known as Roque 4 (discovered by \citealt{zapatero1997}). For both 2MASS J0608$-$2753 and DENIS J0357$-$4417 the \ion{K}{1} and \ion{Na}{1} lines along with the CaH bands are weaker than in a field early-L dwarf, and the VO bands are stronger than those seen in normal field late-M or early-L dwarfs.  A similar spectral morphology to the Pleiad Roque 4 suggests an age estimate for 2MASS J1022+0200 and DENIS J0357$-$4417 of very roughly 100 Myr.

2MASS J0241$-$0326 was discovered by \cite{cruz2007} and is a near duplicate of the canonical low-gravity L0 dwarf 2MASS J0141$-$4633. 2MASS J0141$-$4633 itself was discovered by \cite{kirkpatrick2006}, who studied the optical and near-infrared spectra of this object in detail. 2MASS J2213$-$2136 was discovered by \cite{cruz2007} and is another near duplicate of 2MASS J0141$-$4633. All three of these are plotted in the right panel of Figure~\ref{lowg_L0} along with the spectrum of the L0 Pleiad dwarf Roque 25 (top, discovered by \citealt{martin1998}), and the late-M giant IRAS 14303$-$1042 (bottom).

The optical spectra of these objects reveal that the \ion{K}{1} and \ion{Na}{1} lines are weaker than normal as are the bands of CrH and FeH. Even weaker alkali lines and hydride bands, such as those seen in IRAS 14303$-$1042, are used as the gravity-dependent hallmarks of a giant spectrum. We interpret these spectral anomalies in 2MASS J0141$-$4633, 2MASS J0241$-$0326, and 2MASS J2213$-$2135 to be indicative of a surface gravity intermediate between that of a normal L dwarf and a late-M giant. Other peculiar features are the strengths of the VO bands, which are stronger than those seen in a normal M/L dwarf despite the fact that the TiO bandstrengths mimic those seen in an early-L dwarf. The oxides TiO and VO disappear from late-M and early-L dwarfs because of condensation. In the standard sequence (\citealt{kirkpatrick1999}), TiO disappears first and VO disappears at even later spectral types (i.e., cooler effective temperatures). Lower gravity leads to lower atmospheric pressure. This inhibits the formation of grains, pushing condensation to cooler temperatures (\citealt{lodders2002}). The limiting cases of this phenomemon are the giant stars. At their very low gravities of log(g)$\approx$0, late-M giants never reach temperatures low enough to trigger the formation of TiO and VO condensates, so gaseous TiO and VO still exist in their atmospheres and as absorbers in their optical spectra. (See spectrum of IRAS 14303$-$1042 in Figure~\ref{lowg_L0}.) 2MASS 0141$-$4633, 2MASS J0241$-$0326, and 2MASS J2213$-$2135 fall at values of T$_{eff}$ where the formation of titanium-bearing condensates has already begun but where vanadium condensation has not yet been triggered. At these temperatures, low-gravity L dwarfs are very easy to spot via their weak TiO bands and overly strong VO bands.

Given that the spectra of 2MASS J0141$-$4633, 2MASS J0241$-$0326, and 2MASS J2213$-$2135 are clearly more unusual and point to a lower gravity than that of Roque 25, we can conclude that their ages are significantly younger than the 100 Myr age assigned to this Pleiades member. In fact, fits of the optical and near-infrared spectra of 2MASS J0141$-$4633 from \cite{kirkpatrick2006} indicate 1 Myr $<$ age $<$ 50 Myr and log(g)$\approx$4.0, implying a mass of 6 $M_{Jupiter}$ $<$ M $<$ 25 $M_{Jupiter}$.

\subsubsection{2MASS J1022+5825 (L1 pec)}

This object was discovered by \cite{reid2008}. As seen in Figure~\ref{lowg_L1}, the best overall match is to a field L1 dwarf, but all of the alkali lines (\ion{K}{1}, \ion{Na}{1}, \ion{Rb}{1}, and \ion{Cs}{1}) are noticeably weaker in this object than in normal field L dwarfs, and the VO bands are more prominent than expected. In these respects its spectrum has a very strong resemblance to Roque 25, the L0 Pleiad, and suggests an age for 2MASS J1022+5825 of roughly 100 Myr.

Another unusual feature in the spectrum of 2MASS J1022+5825 is its strong H$\alpha$ emission line with a measured equivalent width of 128 \AA. As \cite{schmidt2007} show, the emission strength just two nights later had dropped to 26 \AA. Such variable emission has the potential to complicate interpretation of the gravity-sensitive features used in our analysis, especially when objects may have been spectroscopically observed on only one occasion. Such emission has two mechanisms that can alter the measured strengths of lines and bands. First, as noted in late-M and L dwarf flares (\citealt{liebert1999,martin1999b,schmidt2007}), emission in the cores of resonance lines can partially or totally fill in the absorption lines causing them to appear weaker than they really are. Second, flux from the superheated material above the photosphere adds continuum flux, suppressing the contrast between absorption lines and bands and thereby making the lines and bands appear weaker than normal (i.e., it veils the features). 

Fortunately such activity is not the cause of the features seen here. The first effect would amplify the appearance of low gravity via anemic alkali linestrengths, but the second effect would contradict it because it would cause the VO bandstrengths to weaken, not strengthen. Hence, low gravity is still the most likely explanation for the peculiarities seen in this spectrum. 
%As long as there is a rich spectrum of gravity diagnostics whose strengths are anti-correlated, disentangling the effects of gravity and emission/veiling caused by chromospheric activity, flaring, or the presence of accretion from a disk should be possible. This is easily done at late-M and early-L types because optical spectra contain both oxide bands and alkali lines which have the opposite behavior as gravity is lowered. At later types, VO has completely disappeared into condensates and the only gravity diagnostics that remain (the hydride bands and the alkali lines) all weaken at lower gravity. Fortunately, objects this late rarely show emission lines or flaring (\citealt{gizis2000,west2004}; see also Fig.\ 5 from \citealt{kirkpatrick2000}), so such an effect should only rarely confuse our interpretation.
In the case of 2MASS J1022+5825 we find that the emission appears to be confined to H$\alpha$ because the strengths of the alkali lines do not change appreciably between the two nights. Veiling also appears not to be an issue because the VO bandstrengths are unaffected between the observations. In this case we are further reassured that the noted spectral peculiarities are caused by low gravity and not by other effects.

\subsubsection{2MASS J0033$-$1521, G 196-3B, and 2MASS J2208+2921 (L2 pec)}

2MASS J0033$-$1521, discovered by \cite{gizis2003} and rediscovered as part of the southern L dwarf sample, is shown in the left panel of Figure~\ref{lowg_L2}. Its alkali lines are weaker than those of a standard early-L dwarf, the most obvious manifestation being the shallowness of the cores of the \ion{K}{1} doublet at 7665 and 7699 \AA. Hydride bandstrengths are also weaker than normal, and this is particularly obvious in the CaH band near 6950 \AA\ and the Wing-Ford FeH band at 9896 \AA. Two bands noticeably weak or absent from the spectrum are TiO and VO, both of which are still evident in the spectra of normal early-L dwarfs in the field; the 8432 \AA\ band of TiO and weak VO absorption near 7900 \AA\ usually sculpt the continuum of an early-L dwarf but this is not seen in 2MASS J0033$-$1521.

We can interpret the peculiarities in this spectrum as the result of low gravity due to the weakness of the alkali lines and hydride bands. However, we note that this object does not have the strong VO bands seen in the low-gravity L0 and L1 dwarfs discussed above. If this object is truly low-gravity, then it falls at a temperature cool enough for VO condensation to have already taken place. Unfortunately for low-gravity classification, the loss of gaseous oxides robs the optical spectrum of some of its most telltale low-gravity signatures.

There is an additional problem at these temperatures: there are currently no firmly established, low-gravity cluster members with optical types later than L0 to use as age comparisons. In the case of 2MASS J0033$-$1521, its unusual features are not that anomalous when compared to a normal field L2 dwarf, and it is likely that a lower gravity object of the same class would still exhibit some VO absorption. We thus surmise that 2MASS J0033$-$1521 must lie at the old end of the predicted age range where low-gravity signatures are evident, and this gives it an age of $\sim$100 Myr.

G 196-3B was discovered by \cite{rebolo1998} while performing a search for companions to young (X-ray active), nearby K and M dwarfs. The primary, G 196-3A, is an M2.5 dwarf believed to have an age between 20 Myr and 300 Myr (\citealt{kirkpatrick2001}). 2MASS J2208+2921 was discovered by \cite{kirkpatrick2000} and labeled as one of only two peculiar L dwarfs known at that time. At the time of that paper, the causes for its peculiarity were not understood but an analysis of the feature strengths on a newly acquired Subaru spectrum reveals that this object is probably also a low-gravity L dwarf. 

Spectra of G 196-3B and 2MASS J2208+2921 are shown in the right panel of Figure~\ref{lowg_L2}. In G 196-3B the cores of the \ion{K}{1} doublet are noticeably weaker than field L dwarfs of similar type. Lines of \ion{Na}{1}, \ion{Rb}{1}, and \ion{Cs}{1} over the 7500-9000 \AA\ range are very weak but present at this resolution. The hydride bands of CaH, CrH, and FeH are weaker than normal. These features point to low gravity (youth) as the cause, and the independent age estimate from the primary confirms this. In the spectrum of 2MASS J2208+2921 the alkali lines are all weak, most notably the cores of the \ion{K}{1} doublet. The hydride bands -- CaH, CrH, and FeH -- are all weak as well. Although the spectrum shows only very weak absorption by TiO at 8432 \AA, the VO bands near 7400 \AA\ and 7950 \AA\ are obvious and slightly stronger than those seen in G 196-3B. Nevertheless, this spectrum is very similar to that of G 196-3B, so we suspect that 2MASS J2208+2921 has a similar age of $\sim$100 Myr.

\subsubsection{2MASS J1615+4953 (L4 pec) and Gl 417B (L4.5)}

2MASS J1615+4953 was discovered by \cite{cruz2007}, who noted possible low-gravity signatures in its optical spectrum. Gl 417B was discovered by \cite{kirkpatrick2001}, who assigned an age of 80-300 Myr to it based on the activity, lithium abundance, and kinematics of its G dwarf primary, Gl 417A. \cite{bouy2003} find that Gl 417B is a binary with separation of $0{\farcs}070{\pm}0{\farcs}0028$ and magnitude differences in the HST/WFPC2 filters of $\Delta{m_{F814W}}= 1.07{\pm}0.11$ mag and $\Delta{m_{F1042M}} = 1.04{\pm}0.11$ mag. It can be shown that the relation between $M_I$ and optical spectral type suggests that the L4.5 composite spectrum is comprised of two spectra differing by roughly 2 or 2.5 spectral subclasses (e.g., an L4 + L6/L6.5 or an L4.5 + L6.5/L7). It can further be shown that such composite spectra differ little from that of the primary, so we can consider the peculiarities in the composite spectrum to be indicative of peculiarities in the primary spectrum alone.

The spectrum of 2MASS J1615+4953 in the left panel of Figure~\ref{lowg_L4} has low signal-to-noise but the overall morphology best fits that of an L4 dwarf. A comparison to an even earlier, L3 dwarf shows that 2MASS J1615+4953 is too blue in its overall slope. We interpret this as an absence of absorption from the wings of the \ion{Na}{1} ``D'' doublet at 5890 and 5896 \AA. A more detailed examination also shows that the \ion{K}{1} line cores are weaker along with all of the bands of CaH, CrH, and FeH. Low gravity is most likely the cause, and we tentatively assign this object an age estimate of $\sim$100 Myr based on the fact that it is more peculiar than Gl 417B ($\sim$80-300 Myr; see below). However, due to the poor signal-to-noise ratio of this spectrum we can not be completely certain that low gravity is the correct interpretation of its peculiarities. This object may prove to be an overly dusty L dwarf akin to those described in \cite{looper2008}.

The spectrum of Gl 417B in the right panel of Figure~\ref{lowg_L4} shows some slight deviations from a normal, old, field mid-L dwarf. Lines of \ion{Rb}{1} are somewhat weaker than in a field mid-L dwarf although other alkali lines of \ion{K}{1}, \ion{Na}{1}, and \ion{Cs}{1} aren't noticeably different in strength. The TiO band at 8200 \AA\ may also be somewhat weaker than normal. What is clear is that the spectrum of this fairly young object is not all that different from a much older field L4 or L5 dwarf. This supports the idea that peculiarities in the spectrum due to lower gravity are obvious only at ages $<$100-200 Myr, which is the age when these brown dwarfs are expected to have reached their final radius (\citealt{burrows1997}).

\subsubsection{2MASS J2244+2043 (L6.5)}

This is the final low-gravity spectrum illustrated here and it is the latest. The object was discovered as a byproduct of the 2MASS search for red Active Galactic Nuclei, led by B.\ O.\ Nelson and R.\ M.\ Cutri, and was first published in the trigonometric parallax paper of \cite{dahn2002}. The optical spectrum, typed as L6.5, is plotted in Figure~\ref{lowg_L6.5} along with the optical L6 and L7 standards from \cite{kirkpatrick1999}. The long-wavelength end of the spectrum is rather noisy, but it appears to show no Wing-Ford FeH band at 9896 \AA, which does not match its spectral type or the fact that the 8692 \AA\ band of FeH is clearly present. Other than that, however, there are no other obvious spectral oddities and no reason to label this spectrum as peculiar. The reason for discussing this object here is that its near-infrared spectrum is incredibly peculiar (\citealt{mclean2003,looper2008}) and almost certainly the byproduct of much lower gravity. Not only is the H-band portion of the spectrum markedly triangular, an indicator of low gravity, but the $J-K_s$ color of 2.48$\pm$0.15 makes it one of the reddest L dwarfs known.

So why, then, is the optical spectrum so ordinary in appearance? Unlike the earlier L dwarfs discussed above, TiO and VO are no longer available to use as temperature or condensation bellwethers, the condensation of the oxides having completely stripped them from the photosphere. In mid- to late-L dwarf spectra, the main shapers of the emergent flux in the optical are the pressure broadened \ion{Na}{1} and \ion{K}{1} lines. In atmospheres of lower gravity and pressure, these alkali absorptions will be less pronounced, giving them strengths comparable to a much warmer field object. In late-L dwarfs, the hydrides begin to disappear as well, but because this condensation will likely be inhibited in a lower gravity object just as it is for TiO and VO, these hydrides will appear in the spectrum at later types than normally seen. The combination of these two effects, then, conspire to create a spectrum which is virtually indistinguishable from an earlier type L dwarf of higher gravity. In other words, the lower gravity spectrum will have weaker alkalis and stronger hydrides than a high gravity L dwarf of comparable temperature. Hence there may be a degeneracy at mid- and late-L types whereby low-gravity objects are very difficult or impossible to distinguish in the optical. 2MASS J2244+2043 may be the first such object where this degeneracy is seen. 

\section{Analysis}

\subsection{Optical Spectral Type vs.\ Near-IR Color}

As suggested in \cite{kirkpatrick2006}, near-infrared colors may provide an independent means of identifying low-gravity L dwarfs. In the case of 2MASS J0141$-$4633, the emergent flux at $H$- and $K$-bands is significantly higher, relative to the $J$-band flux, than that of a normal dwarf of the same optical spectral type. This is evident in its 2MASS measured near-infrared color of J-K$_s$=1.735$\pm$0.054, which is far redder than the J-K$_s$=1.29 color (see Figure~\ref{shrimp}) of a normal L0 dwarf. The unusually red color is believed to be caused by reduced collision-induced absorption by H$_2$ as expected at lower gravity (lower pressure). (See \citealt{borysow1997} for opacity plots of CIA H$_2$ demonstrating that $H$-band and especially $K$-band should be the most affected.) Another way of stating this is that low gravity dwarfs should tend to have redder J-K$_s$ colors than higher gravity dwarfs of the same type. Optically determined spectral types should be independent of changes in the near-infrared CIA H$_2$ opacity, so plots of optical type vs.\ J-K$_s$ color may be useful in ferreting out other young objects. We can check this by making the plot shown in Figure~\ref{shrimp}, which highlights objects already known to have the spectroscopic hallmarks of lower gravity.

To construct this plot we have built three samples spanning the ultra-cool dwarf regime from M7 to L8:
  
\noindent $\bullet$ M7-M9.5: Any published objects classified as M7 to M9.5 based on optical spectra were eligible for inclusion. However, objects selected photometrically from near-infrared surveys were excluded as this might bias the distribution of J-K$_s$ colors. Primarily this sample is built from published spectroscopic follow-up of objects selected from proper motion studies or selected from optical photometric surveys. To ensure adequate statistics we built this list at random until the number of objects in each integral subtype bin (M7-7.5, M8-8.5, and M9-9.5) was roughly twenty-five.

\noindent $\bullet$ L0-L5.5: Finding L dwarfs free of near-infrared photometric selection bias would result in very small or null lists of objects, so we simply drew our sample from the lists given at http://www.DwarfArchives.org. Objects had to have well measured optical types (i.e., no uncertain types were included) and errors in the 2MASS J-K$_s$ color had to be $<$0.10 mag.

\noindent $\bullet$ L6-L8: The selection criteria were the same as for the L0-L5.5 sample
except that the restriction on J-K$_s$ color error was dropped to increase the sample size.

These three samples are graphically presented in Figure~\ref{shrimp}. Each bin along the y-axis represents a full integral subtype: ``M7'' includes M7 and M7.5 dwarfs, ``M8'' includes M8 and M8.5 dwarfs, etc. ``L8'' includes only type L8 because optical types of L8.5, L9, and L9.5 are not currently defined. In order to show more information than a standard histogram would allow, we have plotted the distributions as follows. For each bin, the median color in the group is plotted highest in the bin; colors falling further from the median are plotted progressively lower down the y-axis. This gives each bin a patterning of points that look approximately like an inverted ``V''. As each individual point contributing to the distribution is plotted on such a diagram, the error bars on each of the individual color measurements can also be shown. 

With these samples in hand, we then checked objects comprising these lists and marked those displaying signatures of low-gravity in their spectra. These low-gravity objects are labeled on Figure~\ref{shrimp}. As predicted, the low-gravity objects tend to have redder J-K$_s$ colors, sometimes dramatically redder, than the median color for their optical spectral class. However, there are also low-gravity dwarfs with J-K$_s$ colors that aren't unusually red.

Moreover, there appear to be some objects having very red J-K$_s$ colors for their type but no indication of lower gravity in their optical spectra. Two examples are the M9.5 dwarf SSSPM J2310$-$1759 (\citealt{lodieu2005}; J-K$_s$=1.407$\pm$0.041) and the L4.5 dwarf 2MASS J0512$-$2949 (J-K$_s$=2.178$\pm$0.071) from the southern L dwarf sample of Table~\ref{l_candidates} and Table~\ref{new_spec_obs}.  Two other examples are noted by \cite{looper2008}, who study two particularly red L dwarfs -- the optical L4.5 2MASS J18212815+1414010 (J-K$_s$=1.78$\pm$0.03) and the optical L6 2MASS J21481628+4003593 (J-K$_s$=2.38$\pm$0.04) -- the unusually red colors for which appear to be the result of an overabundance of atmospheric dust and are perhaps unrelated to low gravity. (On the blue side of the distribution, unusually blue objects are sometimes low-metallicity dwarfs [\citealt{kirkpatrick2008a}] in which collision induced absorption by H$_2$ is the dominant absorption mechanism in the near-infrared, whereas other blue objects may have atypically thin and/or large-grained clouds decks that may be unrelated or only partly the result of lower metallicity [\citealt{burgasser2008}].) Therefore, unusually red near-infrared colors alone can not be used to identify low-gravity dwarfs because higher metallicity and/or dustier L dwarfs can also exhibit redder colors for their optical type.

\subsection{Are We Finding Too Many Young L Dwarfs?}

As the number of low-gravity ultra-cool dwarfs in the published literature continues to grow (see Table~\ref{young_dwarfs}), we must ask whether their frequency of occurrence is consistent with expectations. If the frequency is much higher than predictions, then this would suggest that we may have assigned the wrong physical cause to peculiarities observed in the optical spectra above. There are also clear indications that these low-gravity dwarfs tend to be found in the southern hemisphere, where known, nearby ($<$100 pc) young stellar associations are concentrated. Because non-neglible color selection biases exist for late-M dwarfs in the studies discussed below, we restrict the subsequent discussion to L dwarfs alone.

The first two peculiar L dwarfs were noted by \cite{kirkpatrick2000} and we now believe that one of these, 2MASS J2208+2921, can be explained via low-gravity (youth). That original sample contained 92 L dwarfs with optical spectra and spectral type determinations. Three of these -- GD 165B, Gl 417B, and Gl 584C -- are not strictly ``field'' L dwarfs since they are companions to stars. As a result the percentage of low-gravity field L dwarfs in this sample is thus 1/89, or 1.1$\pm$1.1\%. It should be noted that this sample is comprised of 73 objects north of Dec=0$^\circ$ and 16 objects south of Dec=0$^\circ$. Of the latter group, only one is south of Dec=$-$30$^\circ$.

The southern L dwarf sample in the current paper is comprised of 24 L dwarfs, of which one (Gl 618.1B) is a companion object, and all have optical spectra and spectral classifications. The identification of three of these (2MASS J0033$-$1521, 2MASS J0141$-$4633, and DENIS-P J0357$-$4417) as young objects gives a low-gravity percentage of 3/23, or 13.0$\pm$7.5\%, for field L dwarfs. Despite its large error bar, this result appears at odds with the frequency found in the earlier Kirkpatrick sample. It is important to note, however, that this newer sample contains only objects located south of Dec=0$^\circ$.

The on-going sample of Cruz and collaborators expands markedly on the Kirkpatrick samples. All objects were photometrically selected using the same prescription regardless of hemisphere (\citealt{cruz2003,reid2008}), so the sample is uniform. In that collection there are 241 field L dwarfs with optical spectra and spectral classifications, 22 of which show low-gravity features\footnote{Some of these low-gravity L dwarfs remain unpublished but will be discussed in a future paper by Cruz and collaborators.}. That sample, comprised of a nearly equal split of objects at northern vs.\ southern declinations (116 in the north, 125 in the south), gives a low-gravity fraction of 22/241, or 9.1$\pm$1.9\%. 

Combining all of these samples and removing duplicates gives a total fraction of 23/303, or 7.6$\pm$1.6\%, for low-gravity field L dwarfs. Broken down by hemisphere, we find low-gravity fractions of 7/172 (4.1$\pm$1.5\%) for Dec$>0^\circ$ and 16/131 (12.2$\pm$3.1\%) for Dec$<0^\circ$, confirming the north/south dichotomy seen in the smaller Kirkpatrick samples alone. 

Are these empirical numbers in agreement with expectations? Both \cite{allen2005} and \cite{burgasser2004c} provide predictions for the makeup of the low-mass star and brown dwarf populations in the Solar Neighborhood. The former paper uses Bayesian statistical methods to compute the mass and age distribution that best constructs the observed field luminosity function, using the evolutionary models of \cite{burrows2001}. The latter work uses Monte Carlo simulations on a range of plausible star formation rates and initial mass functions along with theoretical evolutinary models from \cite{burrows1997} and \cite{baraffe2003} to predict an array of possible outcomes for the makeup of the Solar Neighorhood. Both works demonstrated that L dwarfs are expected to be, on average, younger than M dwarfs or T dwarfs. Kinematic evidence supports this claim. \cite{zapatero2007} measure $U,V,W$ space motions and deduce that nearby L and T dwarfs appear to comprise a kinematically younger population, with ages in the likely interval 0.5-4 Gyr, than that of stars with types G through early-M. 

The underlying reason for the relative youth of L dwarfs is the fact that they are comprised of two distinct populations. Some L dwarfs are actually low-mass stars, but theory predicts that the only ``stellar'' L dwarfs are those falling in a very restrictive mass range ($\sim$0.072-0.085 $M_\odot$ according to the \citealt{baraffe1998,baraffe2003} models; see figure 9.20 of \citealt{kirkpatrick2008}). The rest of the L dwarfs are brown dwarfs that never settle onto a main sequence and continue to cool with time. Hence, the transitory nature of the ``substellar'' L dwarfs means that the only brown dwarfs seen at these spectral types are those that have not yet had sufficient time to cool into the T dwarf class.

To compare our observational data to predictions, we use evidence from previous sections that our ability to label an L dwarf spectrum as low gravity is indicative of the fact that the object has an age of $\sim$100 Myr or younger. The \cite{allen2005} results predict that within a given volume of space there is a 3.1\% probability of a field M6 dwarf being younger than 100 Myr. At type L0 this probability increases to 5.6\%. At mid- to late-L this probability drops to 3.7\% and plummets further to 2.1\% for early-T and 1.6\% for late-T. \cite{burgasser2004c} plots the median age as a function of effective temperature and shows that the median age is substantially lower in the early-L through mid-T dwarf regime than it is for late-M dwarfs or late-T dwarfs.

Such numbers are applicable to a complete volume-limited sample. Unfortunately, the Kirkpatrick samples are primarily limited by the depths of 2MASS, and the Cruz samples, which were designed to provide a complete census of L dwarfs within 20 pc of the Sun using 2MASS data, used a sliding magnitude limit depending upon the J-K$_s$ color of the object in 2MASS. In both the Cruz and \cite{kirkpatrick2000} samples, late-L dwarfs were specifically targetted by using fainter magnitude limits for candidates with redder colors. As we have seen in \S4.3, this biases our searches in favor of young, earlier L dwarfs at fainter magnitudes (and presumably larger distances) since these low-gravity objects tend to have redder J-K$_s$ colors (see Figure~\ref{shrimp}). 

Very young objects are not expected to be well mixed with older constituents of the Milky Way, so our all-sky total is the best metric to use for comparison. Our overall rate of 7.6$\pm$1.6\% for L dwarfs with ages less than $\sim$100 Myr is in rough agreement with the \cite{allen2005} predicted rate of 3.7-5.6\% once the J-K$_s$ color selection bias is considered. The difference between hemispheres is likely attributable to the fact that the southern sky has a preponderance of young, nearby associations, and our selection technique is particularly well tuned to find examples beyond the intended distance cutoff. We discuss more about these associations in the following section.

\subsection{Are These Low-gravity Field Dwarfs Members of Known Young Associations?}

Table~\ref{young_dwarfs} lists the low-gravity field dwarfs discussed above along with other low-gravity ultracool field dwarfs from other published work. This list contains 20 objects total. (Several other new, low-gravity discoveries not listed here will be discussed in \citealt{cruz2008}.) The table gives the object name in column 1, the optical and near-infrared spectral type along with their references in columns 2-5, the 2MASS $J-K_s$ color and error in column 6, the discovery reference and reference pointing out the low-gravity nature of the object in columns 7-8, and notes on the diagnostics used for establishing low gravity in column 9. Given their implied youth and relative proximity to the Sun, could these objects be members of previously identified young stellar associations?

Figure~\ref{skymap} shows the locations of these 20 sources plotted on the celestial sphere. Plotted for comparison are the locations of members of the four young ($<$100 Myr), nearby ($<$60 pc) stellar associations currently recognized (\citealt{zuckerman2004}) -- the TW Hydrae Association ($\sim$8 Myr), the $\beta$ Pictoris Moving Group ($\sim$12 Myr), the Tucana-Horologium Association ($\sim$30 Myr), and the AB Doradus Moving Group ($\sim$50 Myr). We have coded the symbols for each of the 20 sources so that those believed to have ages near 100 Myr are shown as solid stars and those believed to be substantially younger ($\sim$10 Myr, as determined in \S4.2 or from references in the literature) are shown as open stars. 

This figure shows that some of these young, ultracool objects have sky locations and estimated ages similar to these known groups. The four known/suspected members of the TW Hydrae Association -- SSSPM J1102$-$3431, 2MASS J1139$-$3159, 2MASS J1207$-$3932, and DENIS J1245$-$4429 -- are obvious in the lower left portion of the figure. As noted in \cite{kirkpatrick2006}, 2MASS J0141$-$4633 has a sky location coincident with both the $\beta$ Pictoris Moving Group and the Tucana/Horologium Association along with an estimated age range in agreement with both groups. Other very young candiates -- 2MASS J0241$-$0326 and 2MASS 2213$-$2136 -- have sky locations and assumed ages consistent with known members of the $\beta$ Pictoris Moving Group.

Others appear to have sky locations and young ages inconsistent with membership in any of these known groups. Three northern hemisphere objects lying along an almost constant line of RA=10$^h$23$^m$ fall in a part of the sky well separated from the other groups. All five of the young candidates with $16^h<$RA$<24^h$ also lie in a part of the sky sparsely populated by known association members. Nonetheless, given the closeness of some of these nearby groups, our Solar System may fall near an edge or even within the true spatial volume of the association. In this case, members can fall almost anywhere on the sky and locations provide only weak constraints for individual objects. Kinematic and distance information is required to establish membership with certainty.

At the moment we have proper motions for only a few of these 20 objects. Trigonometric parallaxes and radial velocities are also needed so that distances and UVW space motions can be measured. Only then we can test for membership in known groups and see if there may be objects, particularly the northern ones, that may be hinting at the presence of heretofore unrecognized, young associations. Most intriguing is the possibility of uncovering young associations solely of low-mass stars, assuming such a mode of star formation can actually occur. We have begun some of this follow-up work and will report progress in forthcoming papers.

\section{Analyzing the Lithium Results}

A number of the newly acquired Keck-LRIS spectra from Figure~\ref{allspec_lin} and Figure~\ref{new_LT_lin} show the 6708-\AA\ line in absorption. Zooms of the 6300-7100 \AA\ region of these eight spectra are shown in Figure~\ref{lithium_spectra}. Equivalent width measurements or limits for all of the LRIS spectra are given in Table~\ref{new_spec_obs} through Table~\ref{new_spec_t}. As has been true in previous summaries, we do not detect the \ion{Li}{1} line in any the late-M or T dwarfs at the $\sim$10 \AA\ resolution we employ.

\subsection{Lithium Absorption Strength as a Function of Spectral Type}

Using the list of lithium detections along with the non-detections in Table~\ref{new_spec_obs} and Table~\ref{new_spec_ml}, we can update the lithium statistics plot first presented as Figure 7 of \cite{kirkpatrick2000}. This update is shown in Figure~\ref{li_stats}. In the upper panel we show the measured \ion{Li}{1} equivalent widths as a function of optical spectral type for all L dwarfs for which we have obtained LRIS spectra, now numbering 123 objects. The two panels below this encapsulate the statistical analysis of this distribution. In the middle panel is shown the percentage of L dwarfs with \ion{Li}{1} detections of 3 \AA\ or more as a function of optical type. Due to the increased sample of late-L dwarfs now available, we can now state confidently that there is a slight turnover at late-L in the fraction of objects with strong \ion{Li}{1} lines. In fact, none of the T dwarfs for which we have optical spectra in Table~\ref{new_spec_t} or those included in \cite{burgasser2003} show \ion{Li}{1}, although their detection limits are generally worse than 3 \AA\ equivalent width. This weakening of the \ion{Li}{1} line is further demonstrated in the bottom panel of Figure~\ref{li_stats}, which shows as a function of type the equivalent width of those lines which are detected. After growing steadily from early-L through mid-L, these \ion{Li}{1} strengths peak around L5.5-L6.5 then weaken for later types.

The peak in \ion{Li}{1} strength at $\sim$L6 corresponds roughly to the temperatures where monatomic lithium
is expected to disappear into molecular form (\citealt{lodders1999}). It should be kept in mind, however, that
our \ion{Li}{1} measurements are made against a relative continuum heavily influenced
by wings of the strong \ion{Na}{1} ``D'' doublet to the blue and strong \ion{K}{1} doublet to the red. That having been said, \cite{lodders1999} shows that both Na and K are
expected to disappear into molecular species (solid Na$_2$S and gaseous KCl) at temperatures much lower than this, so the abrupt change in \ion{Li}{1} strength
is probably not caused by modulations in the continuum opacity. Hence, the Li weakening can safely be attributed to the disappearance of monatomic lithium
from the atmosphere. \cite{lodders1999} predicts that the culprit is the formation of LiCl gas. It should also be noted that lithium is expected to disappear in monatomic form before either cesium or rubidium does, again in agreement with our observations.

\subsection{Lithium Absorption Strength as a Function of Age}

We now have in hand a sufficiently large sample of L dwarfs spanning ranges in temperature and age that we can examine in detail the ``second order'' effect of \ion{Li}{1} line strength versus age. To quantify this, we select four optical spectral types -- L0, L2, L4, and L6.5 -- where we have a good sampling across a wide span of ages. Objects in each of those bins can be divided into two sets: those with normal spectra and those with spectra that appear peculiar because of low-gravity effects, as discussed in \S4.2. Table~\ref{lithium_age_grid} lists objects by spectral type (column 1), name (columns 2, 4, 6, and 8) and their \ion{Li}{1} equivalent width measurements (columns 3, 5, 7, and 9). Dwarfs with normal, non-peculiar spectra are given in columns 2-5, with those showing no lithium in columns 2-3, and those with lithium in columns 4-5. Dwarfs exhibiting peculiar, low-gravity spectra are listed in columns 6-7, and those with very peculiar, low-gravity spectra are listed in columns 8-9. By dividing the columns in this way we are subdividing the spectra into empirical age bins. Normal spectra without detected \ion{Li}{1} are expected to be high mass and old. Normal spectra with \ion{Li}{1} are expected to be somewhat lower in mass, and because they have the same spectral type as the non-lithium dwarfs, must therefore be younger. Spectra exhibiting low-gravity features are expected to be younger still, and those showing the most peculiar features are expected to be of even lower gravity and the youngest of all. 

We analyze each grouping in the table:

\noindent{\it L0:} We have optical spectra of six objects with normal L0 spectral types. There are five where no lithium line was detected, and two of these had no detection down to 0.5 \AA\ equivalent width (EW). These are presumably the older, higher-mass L0's. The fifth normal L0 exhibits a \ion{Li}{1} line with 3 \AA\ EW. Presumably this object is younger and lower in mass than the others. We have optical spectra of an additional six L0's with unusual spectra. The three less peculiar ones, which we believe have an age near 100 Myr (in fact, one is a member of the Pleiades), show no \ion{Li}{1} line. Two of these show no line down to at least 3 \AA\ EW. The other objects, which we believe to have ages substantially less than 100 Myr based on discussion in \S4.2, show no lines down to 2 \AA\ EW or better. Figure~\ref{li_seq_L0} plots the highest-SNR spectrum from each of the four age bins. The inset is a zoom around the location of the 6708-\AA\ \ion{Li}{1} line to demonstrate how the Li strength varies as a function of age. Notice that the line when detectable is still very weak. 

\noindent{\it L2:} We have optical spectra of twelve objects with normal L2 spectral types. There are nine with no \ion{Li}{1} detection, with all but one of these having no measurable line down to 1 \AA\ EW. The three other normal L2s have \ion{Li}{1} detections of 1.7, 6, and 8 \AA\ EW and are presumably younger and lower mass than the first nine. The object with the 1.7 \AA\ EW detection is the binary Kelu-1 AB, for which \cite{liu2005} estimate an age of 300-800 Myr. This object may be somewhat older than the other objects in its age bin because the detected lithium is presumed to come from the lower mass secondary, not from the primary itself. We also note three L2s whose spectra appear peculiar due to low gravity. All of these are believed to be $\sim$100 Myr old, and they have \ion{Li}{1} EWs of 4 and 6 \AA. No L2 dwarfs have yet been identified in the ``Very Peculiar Dwarfs'' category of Table~\ref{lithium_age_grid}. Figure~\ref{li_seq_L2} shows an example spectrum from each age bin. Appreciable \ion{Li}{1} detections are seen in the two bins of lowest age.

\noindent{\it L4:} We have optical spectra of six objects with normal L4 spectral types. Of these, five have no measurable \ion{Li}{1} line, the most stringent limits being the 0.5 and 0.7 \AA\ measures in the two objects with the highest signal-to-noise. One of these, GD 165B, is believed to have an age of 1.2-5.5 Gyr based on age estimates of its white dwarf primary (\citealt{kirkpatrick1999b}). The other two normal L4s have measured EWs of 3.3 and 10.5 \AA\ and presumably are younger and lower mass than the others. Additionally we have optical spectra of one peculiar L4 and one peculiar L4.5, which are assumed to be lower gravity than the normal L4s. The slightly peculiar L4.5, Gl 417B, is believed to have an age of 80-300 Myr based on age indicators from its G dwarf primary (\citealt{kirkpatrick2001}), and it shows a \ion{Li}{1} EW of 11.5 \AA. Further complicating the interpretation of this spectrum is the fact that imaging by the Hubble Space Telescope hints that the B component itself is a double (\citealt{bouy2003}). The other peculiar L4, 2MASS J1615+4953, has no \ion{Li}{1} detection down to 8 \AA\ EW but the SNR is poor and this object may be unusual because of excess dust and not because of low gravity, as stated in \S4.2. No extremely peculiar L4 dwarfs are yet known, so the rightmost column in Table~\ref{lithium_age_grid} is again blank. Figure~\ref{li_seq_L4} shows a spectrum for each of the other three age bins. 

\noindent{\it L6.5:} We have optical spectra of six objects with normal L6.5 spectral types. Three of these do not show \ion{Li}{1}, and two of these have stringent limits of at least $<$2 \AA\ EW. The three other L6.5 spectra all show \ion{Li}{1}. Two of these show a strong line of at least $\sim$12 \AA\ EW. Although both appear to have normal spectra in the optical, 2MASS J2148+4003 has a strikingly unusual near-infrared spectrum whose oddities are likely due to excessive atmospheric dust (\citealt{looper2008}). The final L6.5 spectrum, 2MASS J2244+2043, shows a weaker \ion{Li}{1} line of only $\sim$5 \AA\ EW. Although its optical spectrum appears normal, we place it in the ``peculiar'' bin only because its near-infrared spectrum show odd features that are best attributed to low gravity (\citealt{mclean2003}). No L6.5 objects are known that might occupy the rightmost columns of the table. Figure~\ref{li_seq_L6.5} depicts these results graphically, with insets showing the range in \ion{Li}{1} EW.

To summarize, the true strength of \ion{Li}{1} line as a function of age is difficult to interpret from the extant data for several reasons: (1) Many of the age bins in Table~\ref{lithium_age_grid} are sparsely populated so conclusions are based on poor statistics; (2) both known and unrecognized binarity makes the interpretation of even well measured lines difficult as it is not known {\it a priori} what the relative contributions to the line strength are from the primary and the secondary; (3) some of these spectra have SNR levels that are too low at 6708 \AA\ to obtain accurate EW measurements. For now, we see evidence indicating that the \ion{Li}{1} line weakens markedly for the youngest ages, but the most important tests of this -- the very peculiar L dwarfs of lowest gravity -- have not yet been identified for types later than $\sim$L0. High signal-to-noise spectra of additional low-gravity field L dwarfs, such as those to be discussed in \citealt{cruz2008}, and confirmed L dwarf members of young clusters are needed to bolster the statistics. 

\subsection{Comparison to Theory}

Theory stipulates that objects below $\sim$60 M$_{Jup}$ will never reach the lithium fusion temperature of $\sim$2$\times$10$^6$K, so they will always retain their primordial complement of lithium. Objects of higher mass easily destroy their lithium, which is not produced as a byproduct of normal thermonuclear reactions in the cores of stars and brown dwarfs. Therefore, the absence of lithium can be used as a direct test of substellarity, as first proposed by \cite{rebolo1992}. \cite{chabrier1996} calculated that brown dwarfs with masses of 60 M$_{Jup}$ will have depleted their stores of lithium by a factor of two in 250 Myr and by a factor of one hundred in 1 Gyr. Higher mass objects burn up the lithium more quickly, a 70 M$_{Jup}$ object reducing its store by a factor of two in 130 Myr and by a factor of one hundred in 220 Myr. An 80 M$_{Jup}$ object accomplishes these same feats in 90 and 140 Myr, respectively.

Complete lithium destruction occurs because these objects spend the first portion of their lives as M dwarfs, where they are fully convective and fully mixed. Therefore, material in the upper atmosphere of the object eventually circulates to the core where lithium burning can take place. All brown dwarfs with masses above 60 M$_{Jup}$ are expected to go through the fully convective mid- to late-M dwarf regime (see evolutionary tracks in Figures 8 and 9 of \citealt{burrows1997}), residing there for at least the first $\sim$400 Myr of their lives, then they evolve to lower temperatures where they may no longer be convective. This allows plenty of time for the depletion to have occurred, according to the models.

However, retaining a primordial abundance of lithium and being able to detect that lithium observationally in the spectrum are two different issues. There are two regimes where the lithium test may have to be amended. The first concerns objects of very cool temperature. We have already observed that the equivalent width of the \ion{Li}{1} line at 6708 \AA\ diminishes on average at types later than $\sim$L6, presumably due to the formation of lithium-bearing molecules at colder temperatures. The second concerns objects of very low gravity. As first mentioned in \cite{kirkpatrick2006} the lithium test should also be carefully considered before being applied to brown dwarfs of very young age. The reason for this is the expected (and observed) weakening of the other alkali lines as gravity is diminished; the \ion{Li}{1} line is expected to weaken too, thus making the lithium test more difficult to use in this regime. We derived a test of this using empirical data in \S5.2, but results were inconclusive. 

Do atmospheric models of L dwarfs predict a weakening of the \ion{Li}{1} line?
Figure~\ref{models} shows the behavior of 
models\footnote{These synthetic spectra, which are updated versions of those
from \cite{allard2001}, have been computed with
version 15 of the {\tt PHOENIX} model atmosphere code.
These models incorporate numerous updates including but not limited to
the use of the new \cite{barber2006} water line list
and the hydride line lists from \cite{dulick2003} and \cite{burrows2002}.
The lithium abundance was set to the \cite{lodders2003} recommended
value of A(Li) = 3.28 (where A(H) = 12.0); all other elemental abundances
where set to those of \cite{asplund2005}. 
Cloud and cloud-free scenarios were modeled in the standard
``Cond'' and ``Dusty'' approaches of \citealt{allard2003}.} at three different temperatures as gravity is lowered. It should be noted that interior physics, such as the destruction of lithium by convection toward the core, is not included in these simulations; the total lithium abundance is held fixed so that the behavior of the line is due solely to atmospheric chemistry and physics. In the 2400K models, corresponding roughly to an optical type of L0 (see \S9.4.1 of \citealt{kirkpatrick2008}), the \ion{Li}{1} line is seen only at the highest gravities and disappears entirely by log(g)=5.0 at the resolution (10 \AA) shown here. At 2100K, corresponding to an optical type of L2, the \ion{Li}{1} line weakens and then vanishes around log(g)=4.0. At 1800K, corresponding to types L4-4.5, the line weakens throughout the gravity sequence and vanishes only at log(g)=2.5.

In conclusion, we find that theory predicts the weakening of the \ion{Li}{1} line at lower gravities, an effect supported (albeit weakly) by current observational data. Although additional optical spectra of L dwarfs covering a wide range of gravities/ages are needed to bolster the empirical evidence, predictions nonetheless bear out the caveat addressed by \cite{kirkpatrick2006} in their study of 2MASS J0141$-$4633. Namely, they suggested that researchers apply the lithium test with caution for very young brown dwarfs because the \ion{Li}{1} line is expected to be considerably weaker and may, in fact, be undetectable at moderate resolution in those objects.

\section{Summary}

Using a large set of optical specta, we have shown that a subset of objects with peculiar spectra can be identified as low-gravity brown dwarfs. Low gravity is a hallmark of youth, as these objects have lower masses and more extended atmospheres than higher mass, older field dwarfs. We find that the observed percentage of young ($<$100 Myr old) L dwarfs in our sample -- 7.6$\pm$1.6\% -- is consistent with theoretical predictions once selection biases are considered. Even though low gravity enables these young objects to be identified readily via spectroscopy, it complicates the brown dwarf ``lithium test'' because it weakens the observable \ion{Li}{1} line for a given lithium abundance and spectral type. Our sample of young brown dwarfs is found to lie primarily in the southern hemisphere, which (probably not coincidentally) is the location of known, nearby young associations. Further work on these young objects is needed to establish or refute membership in these groups.

\acknowledgements

\begin{figure}
\epsscale{0.75}
\plotone{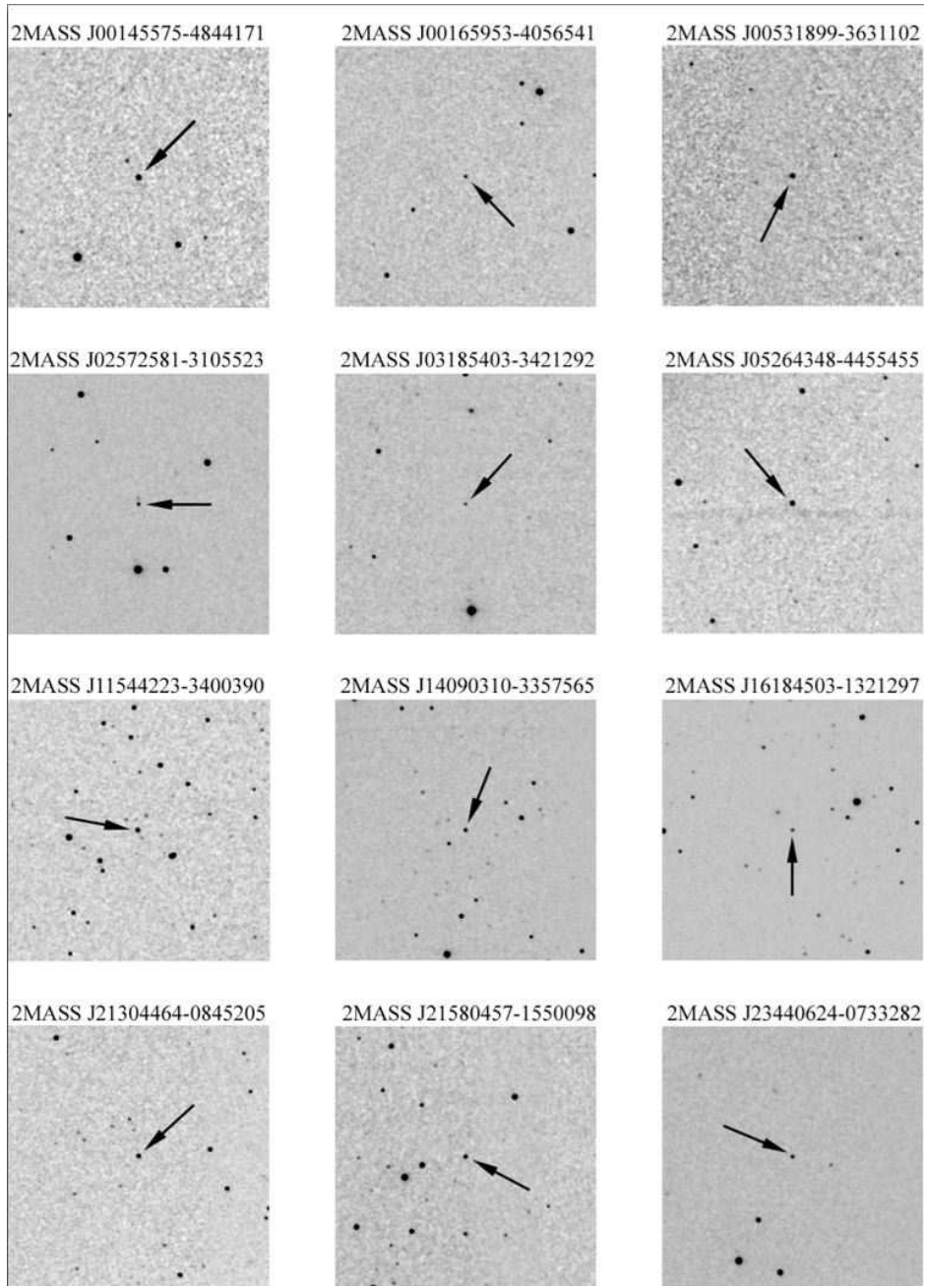}
\caption{2MASS $J$-band finder charts for the new objects identified in
Table~\ref{l_candidates}. All images are five arcminutes square with north up and east to the left. The positions of the new L and late-M dwarfs are marked with arrows. Note that 
the fainter object a few arcseconds north of 2MASS J02572581$-$3105523 is 
merely an artifact -- a latent image persisting from the previous 2MASS frame 
and caused by the brighter star due south. \label{finders}}
\end{figure}

\begin{figure}
\epsscale{0.9}
\plotone{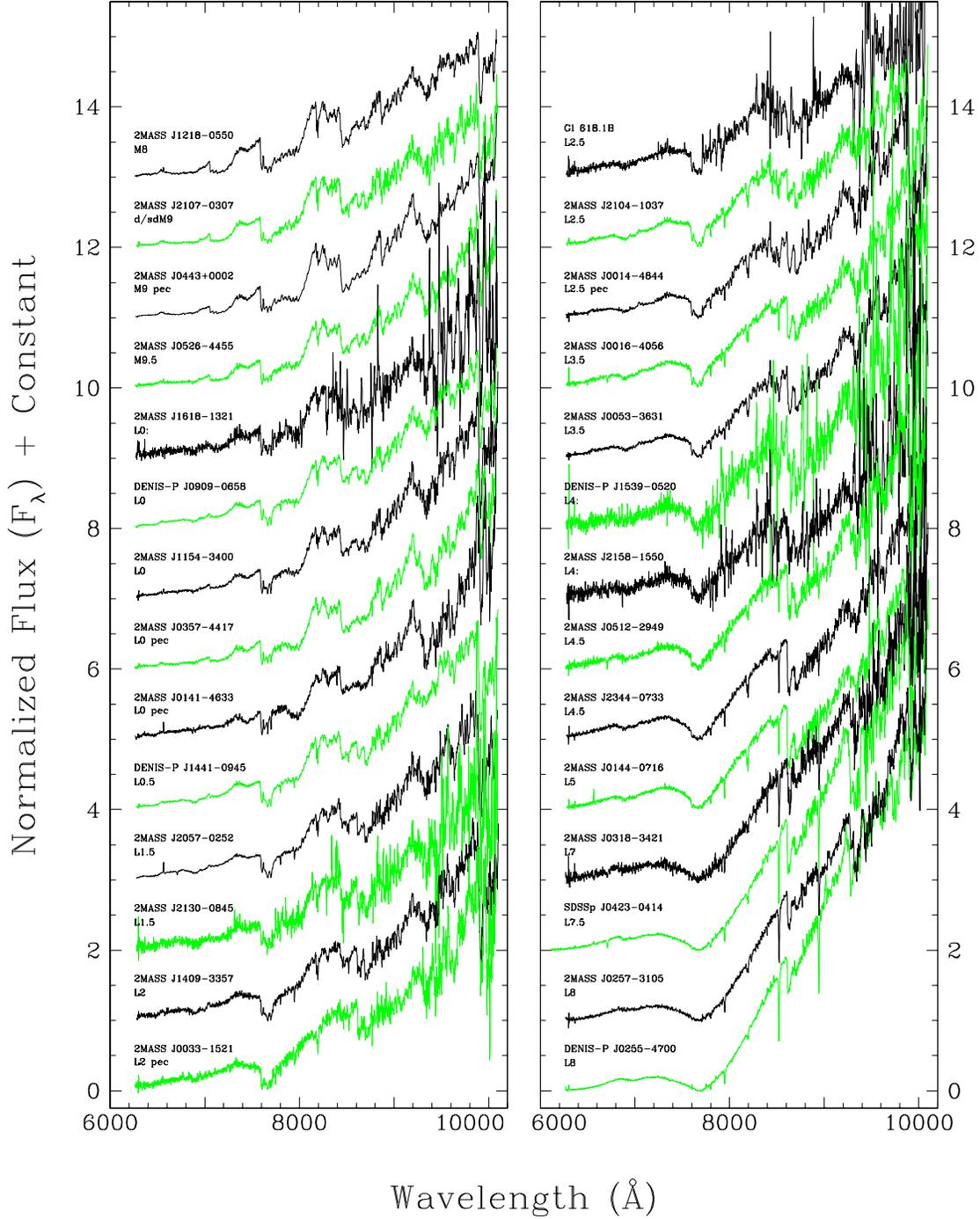}
\caption{Keck/LRIS spectra of the twenty-eight objects listed in Table~\ref{new_spec_obs}.
Each spectrum is normalized to one at 8250 \AA\ and an integral offset is
added to this normalized flux so that the spectra are clearly separated 
along the $y$-axis. Spectra are ordered from earliest to latest with M8 through
L2 shown in the left panel and L2.5 through L8 shown in the right. The spectra 
shown here are not corrected for telluric absorption. Colors of the spectra in Figure~\ref{allspec_lin}
through Figure~\ref{new_LT_log} alternate between black and green to aid in presentation.
\label{allspec_lin}}
\end{figure}

\begin{figure}
\epsscale{0.9}
\plotone{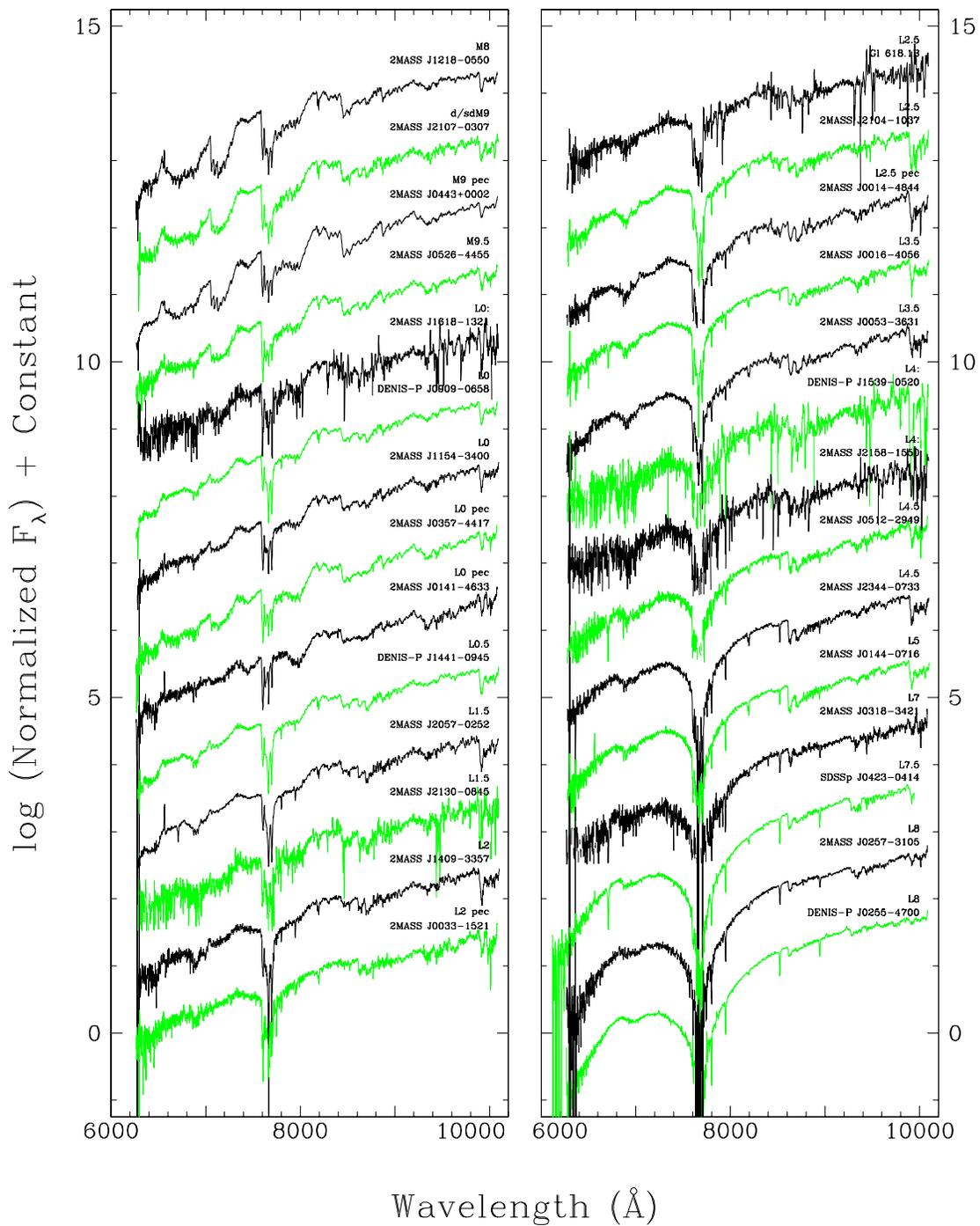}
\caption{The same data shown in Fig.\ 2 except that the $y$-axis is scaled
logarithmically to emphasize features at shorter wavelengths. 
\label{allspec_log}}
\end{figure}

\begin{figure}
\epsscale{0.9}
\plotone{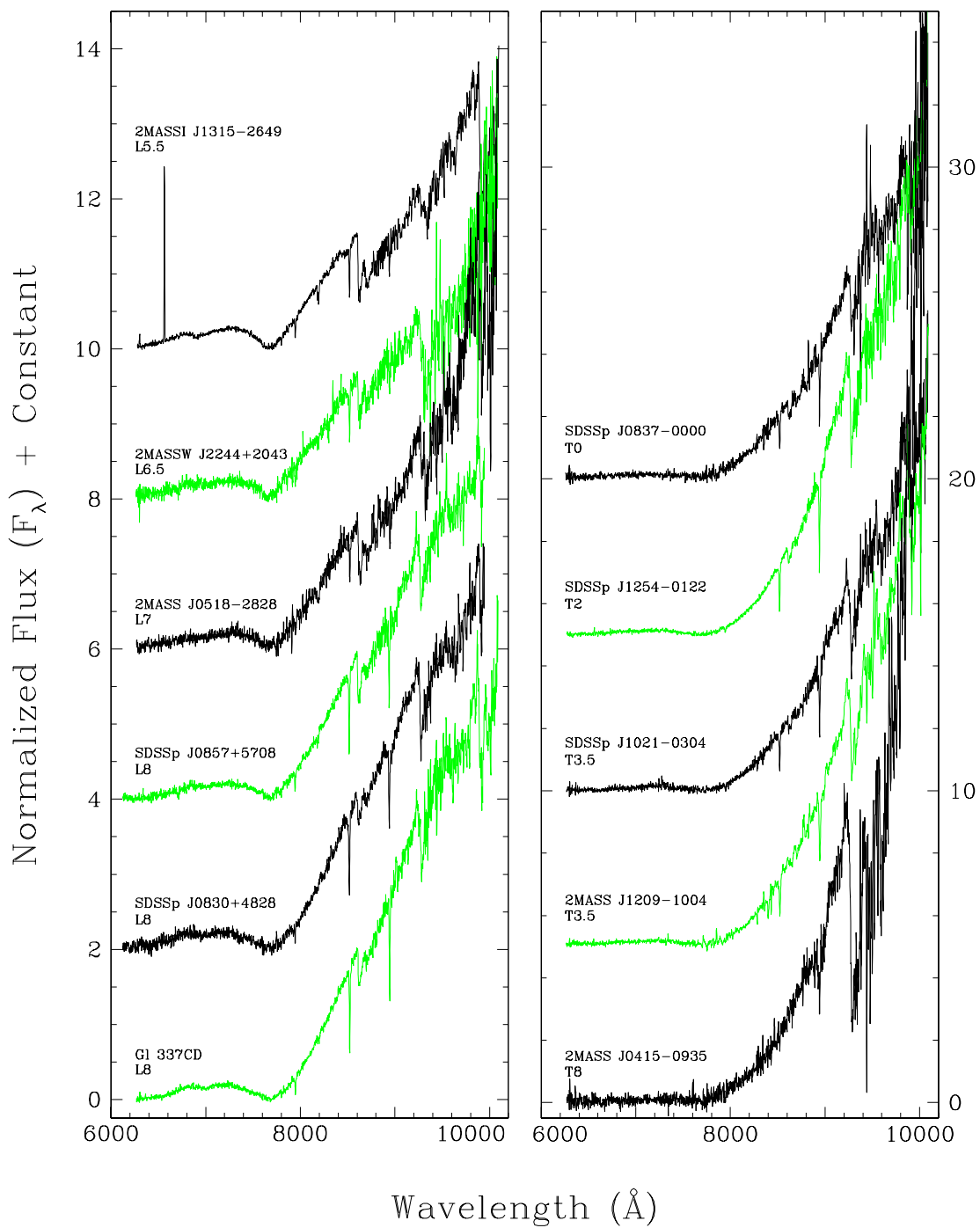}
\caption{Keck/LRIS spectra of the eleven objects in Table~\ref{new_spec_ml} and 
Table~\ref{new_spec_t}. All spectra have been normalized to unity at 8250 \AA.
Integral offsets have been added along the y-axis to separate the 
spectra vertically -- offsets are 2, 4, 6, 8, and 10 for 
the left panel and 5, 10, 15, and 20 for the right. All of these spectra have 
been corrected for telluric absorption except for 2MASS J2244+2043 and 2MASS
J0518$-$2828.
\label{new_LT_lin}}
\end{figure}

\begin{figure}
\epsscale{0.9}
\plotone{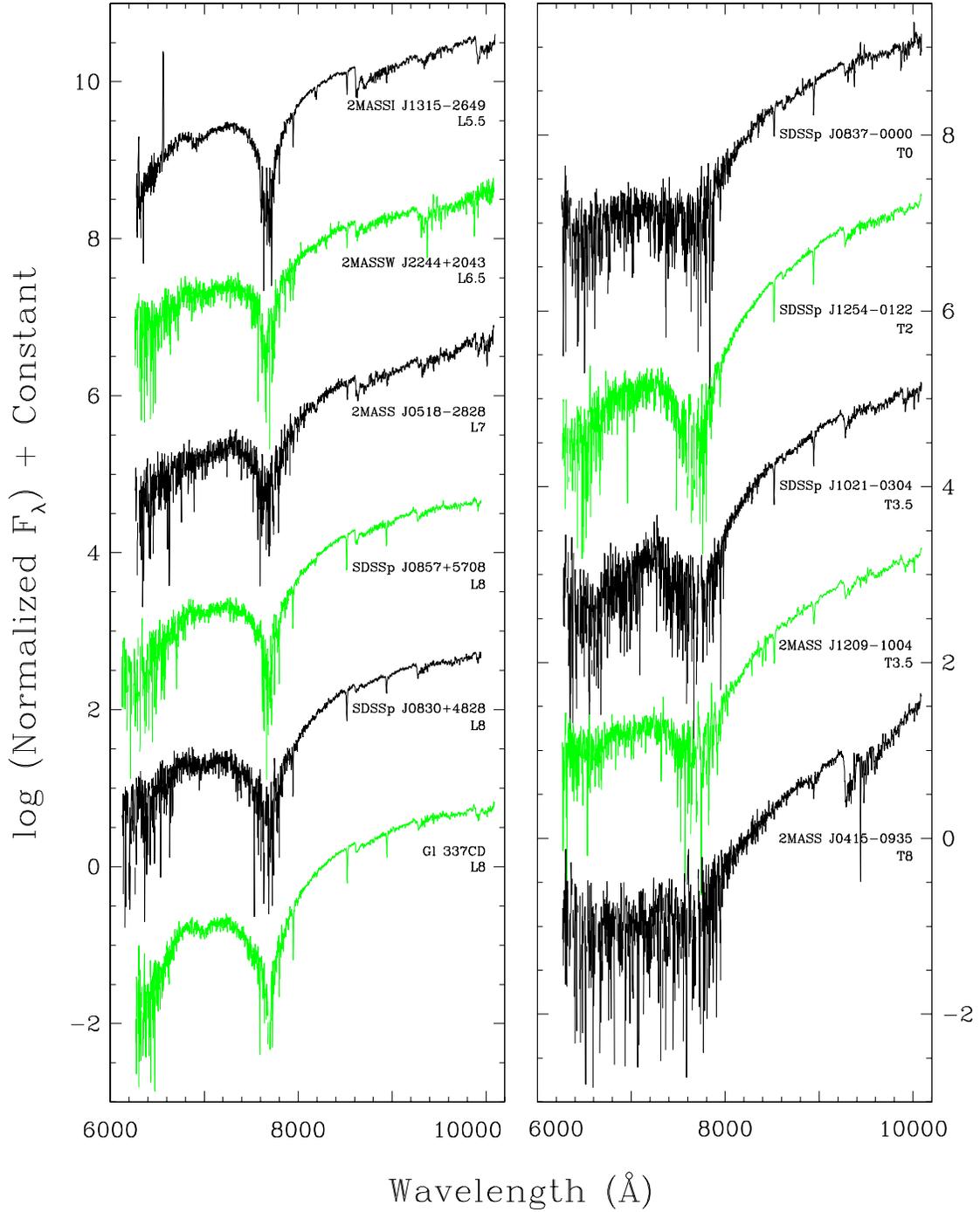}
\caption{The same as Figure~\ref{new_LT_lin} except that the objects 
are plotted on a logarithmic, rather than a linear, scale to 
emphasize features at the shortest wavelengths and to avoid overlap 
at the red end of the spectra. Vertical offsets are integral multiples
of two in both panels.
\label{new_LT_log}}
\end{figure}

\begin{figure}
\epsscale{0.9}
\includegraphics[scale=0.65,angle=270]{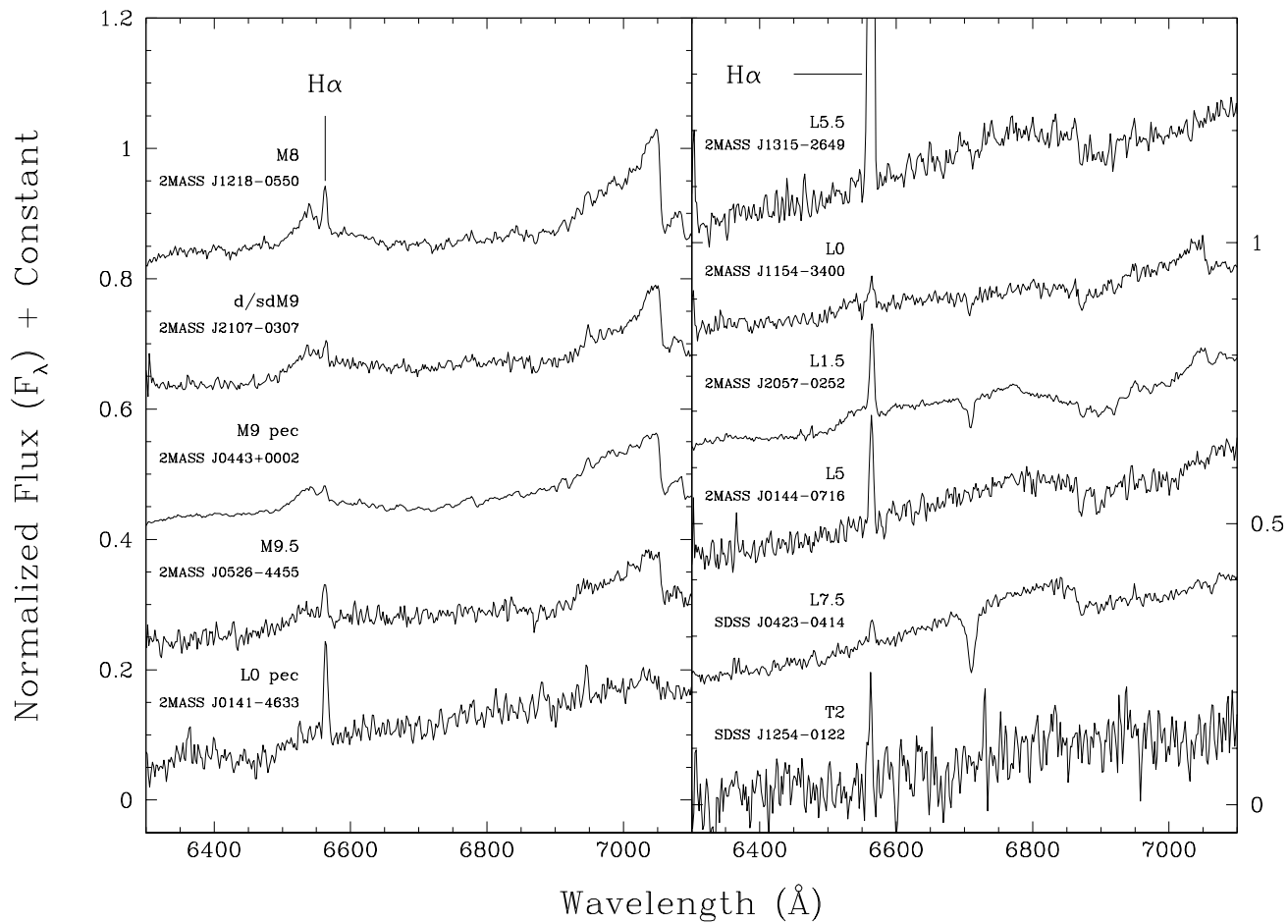}
\caption{Zooms of the Keck/LRIS spectra in the 6300-7100 \AA\ region for
those objects in Table~\ref{new_spec_obs} exhibiting H$\alpha$ emission. 
Note that three of the objects (2MASS J1154$-$3400, 
2MASS J2057$-$0252, and SDSS J0423$-$0414) exhibit both \ion{Li}{1} absorption
at 6708 \AA\ and H$\alpha$ emission. 
\label{halpha}}
\end{figure}

\begin{figure}
\epsscale{0.65}
\plotone{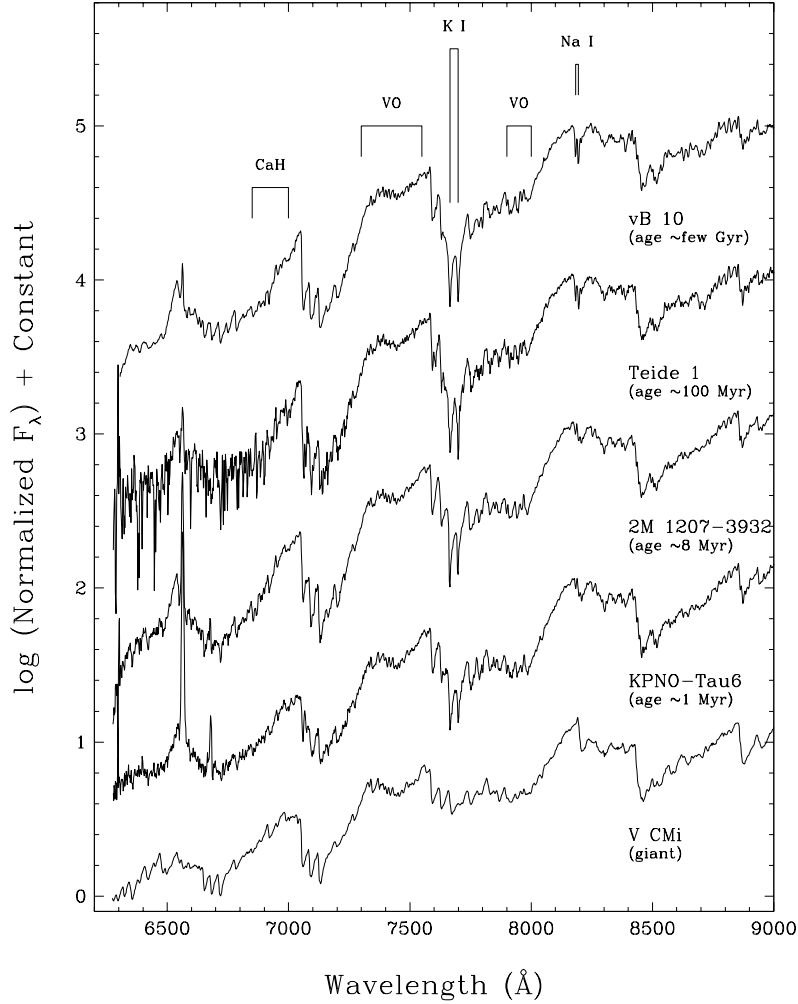}
\epsscale{0.75}
\caption{An optical spectral sequence showing feature changes as a function of age (gravity) at M8-M8.5. Each object is labeled with its age as deduced from membership in a cluster (either the Pleiades at $\sim$100 Myr, the TW Hydrae Association at $\sim$8 Myr, or the Taurus Molecular Cloud at $\sim$1 Myr) or from association with a higher mass primary (in the case of vB 10). Note the weakening of alkali lines (\ion{K}{1} and \ion{Na}{1}) and the CaH band as well as the strengthening of the VO bands from oldest (vB 10) to youngest (KPNO-Tau6). Although the change in spectral morphology is subtle from object to object, the plot demonstrates that gravity differences can be discerned at the $\sim$1 dex level in log(age). The spectrum of a late-M giant with log(g)$\approx$0 is shown at the bottom to illustrate the behavior of these features at even lower gravities. All spectra have been corrected for telluric absorption.
\label{lateM_gravity_sequence}}
\end{figure}

\begin{figure}
\plotone{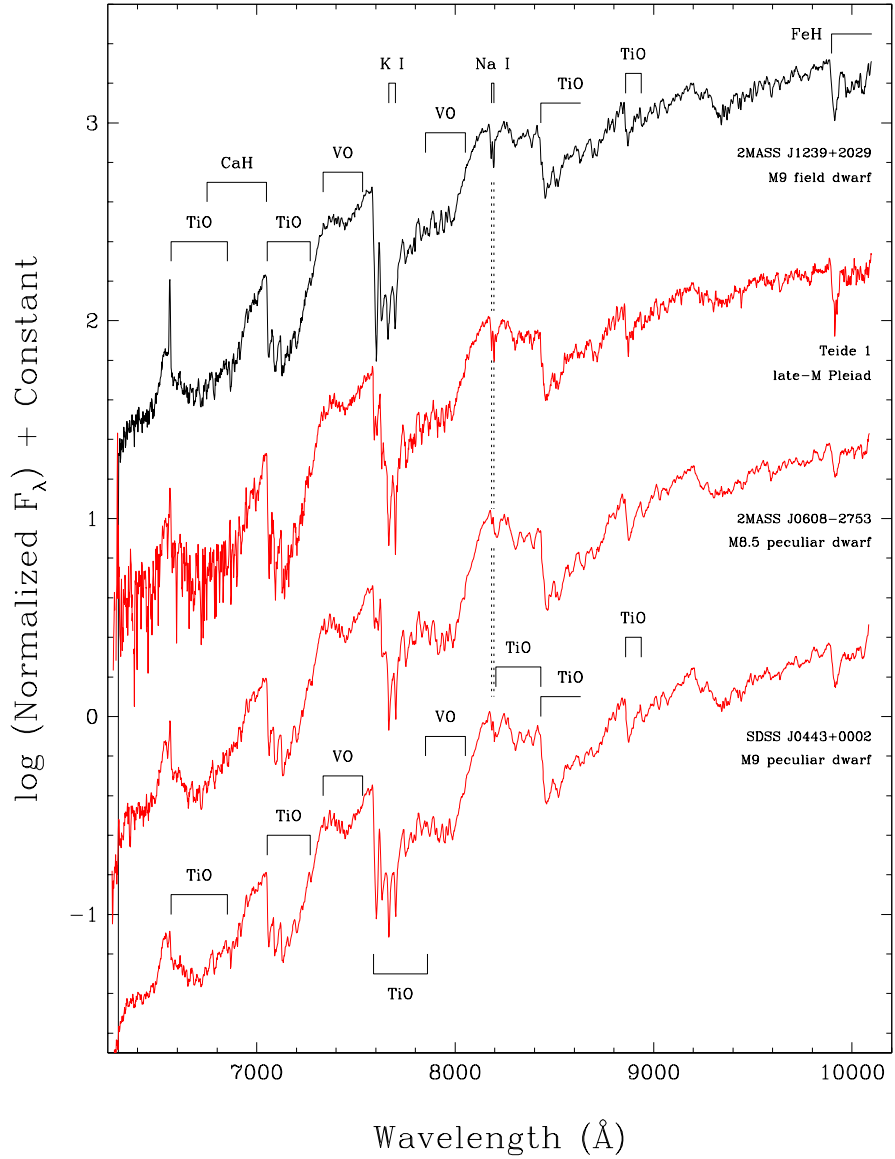}
\caption{The peculiar M8.5 dwarf 2MASS J0608$-$2753 and the peculiar M9 dwarf SDSS J0443+0002 compared to an M9 field dwarf (2MASS J1239+2029) and a late-M member of the Pleiades (Teide 1). Prominent spectral features are marked. Spectra have been normalized to unity at 8250 \AA\ and offset along the y-axis to separate them vertically.
\label{lowg_lateM}}
\end{figure}

\begin{figure}
\includegraphics[scale=0.65,angle=270]{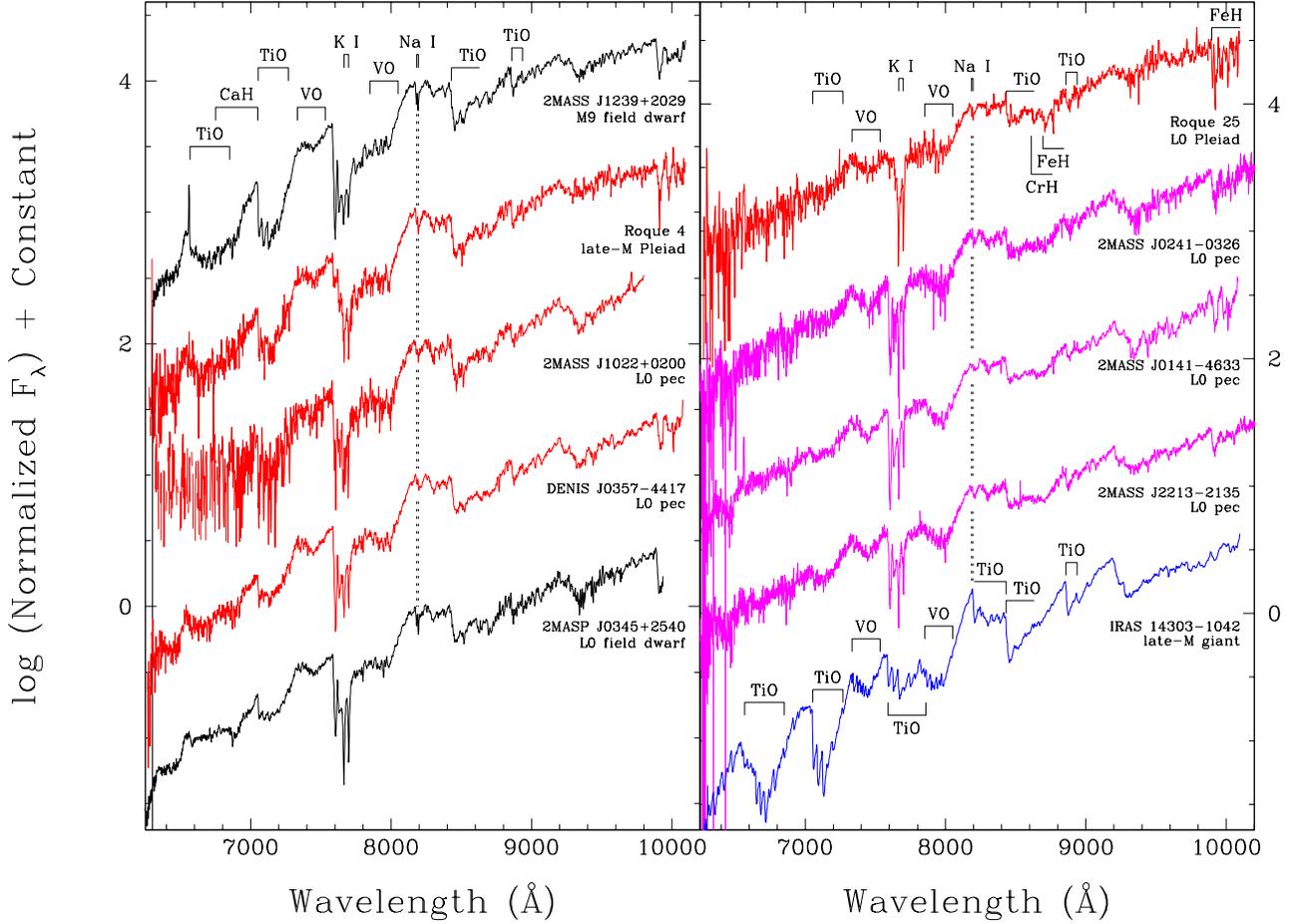}
\caption{(Left) The slightly peculiar L0 dwarfs 2MASS J1022+0200 and SDSS J0357$-$4417 compared to a normal field M9 (2MASS J1239+2029), an M9 dwarf in the Pleiades (Roque 4), and the standard L0 dwarf 2MASP J0345+2540 from \cite{kirkpatrick1999}. Prominent features are marked. (Right) The more peculiar L0 dwarfs 2MASS J0241$-$0326, 2MASS J0141$-$4633, and 2MASS J2213$-$2135 compared to an L0 member of the Pleiades (Roque 25) and a late-M giant (IRAS 14303$-$1042). In both panels spectra have been normalized to unity at 8250 \AA\ and offset along the y-axis to separate them vertically.
\label{lowg_L0}}
\end{figure}

\begin{figure}
\plotone{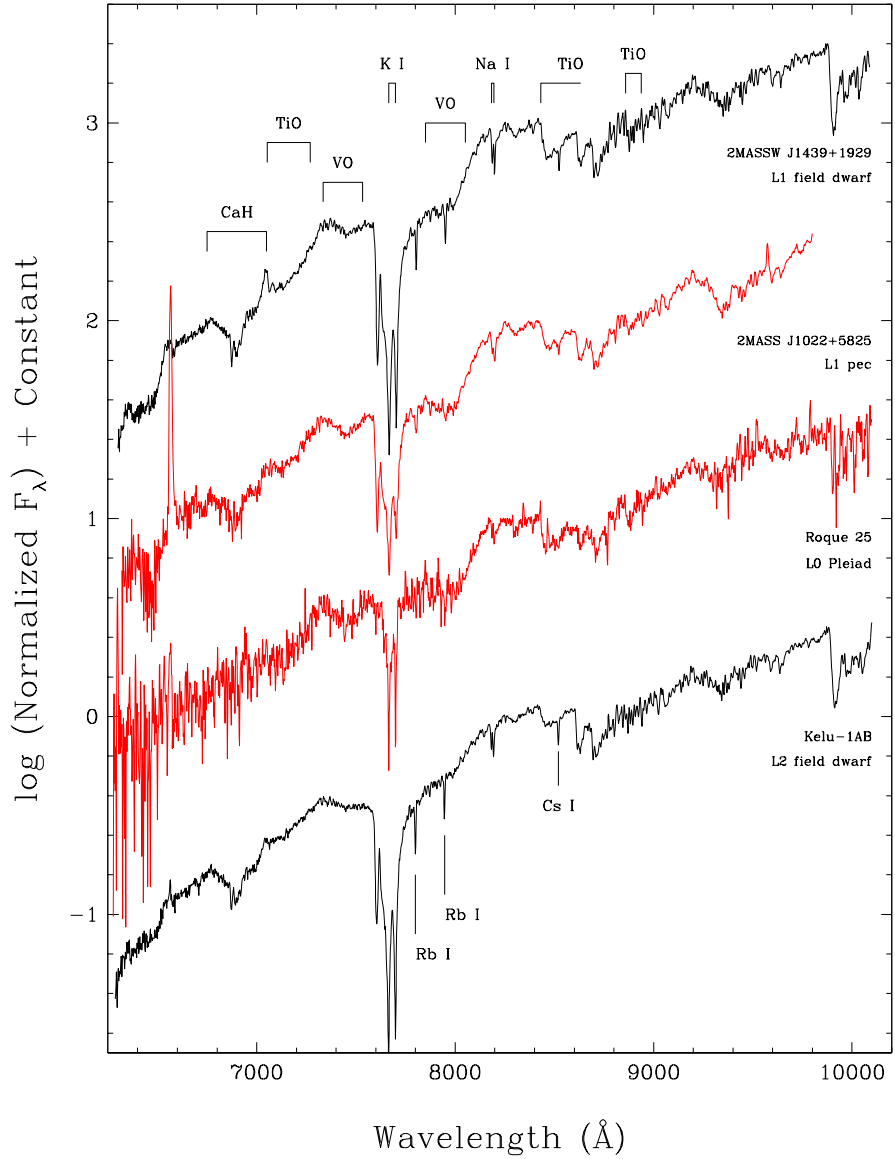}
\caption{The peculiar L1 dwarf 2MASS J1022+5828 compared to the standard L1 dwarf 2MASSW J1439+1929 and an L0 dwarf (Roque 25) in the Pleiades. Prominent features are marked. Spectra have been normalized to unity at 8250 \AA\ and have been offset along the y-axis to separate the spectra vertically.
\label{lowg_L1}}
\end{figure}

\begin{figure}
\includegraphics[scale=0.65,angle=270]{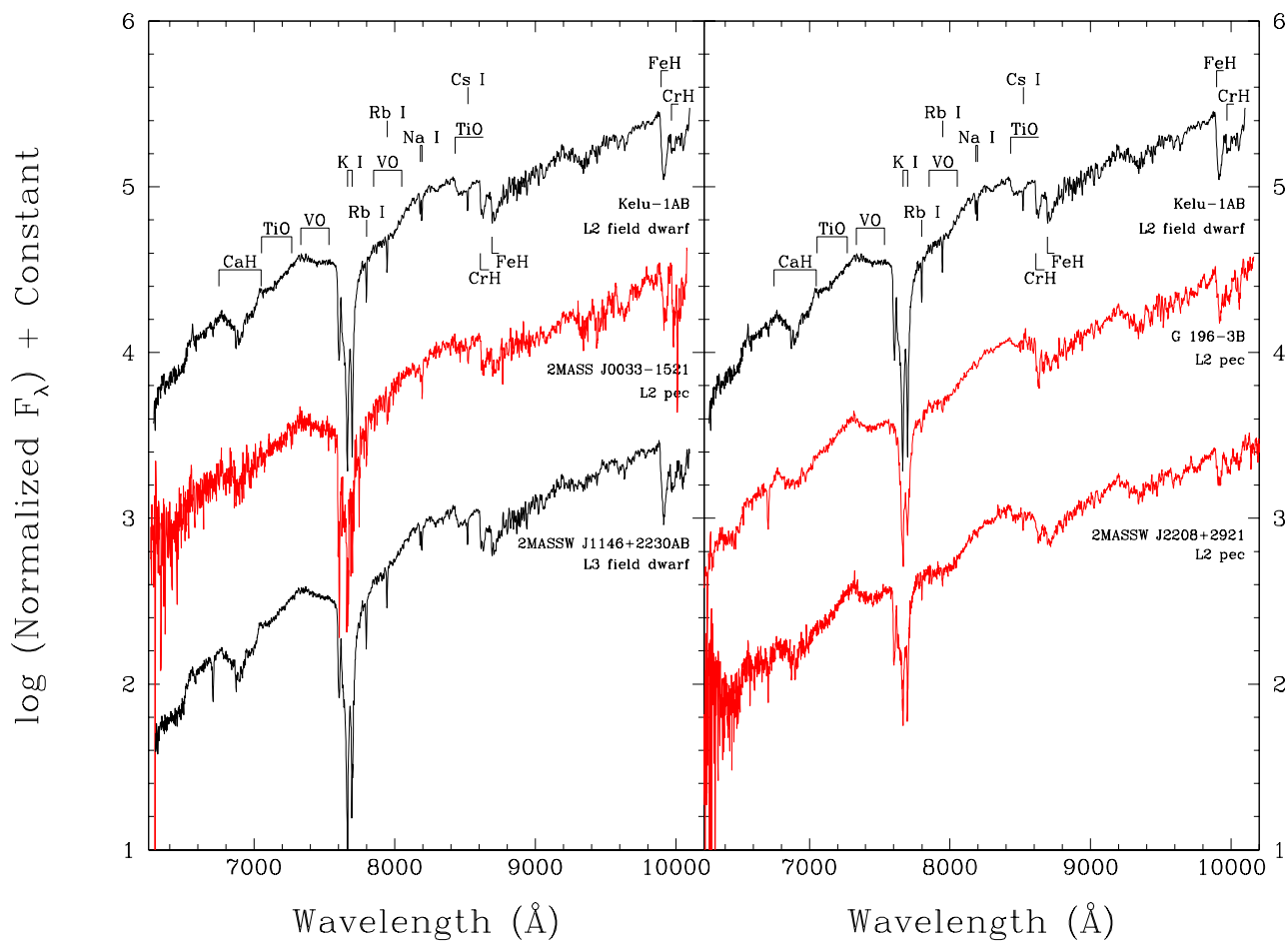}
\caption{(Left) The peculiar L2 dwarf 2MASS J0033$-$1521 compared to the standard L2 dwarf Kelu-1 and the standard L3 dwarf 2MASS J1146+2230. (Right) The peculiar L2 dwarfs G 196-3B and 2MASS J2208+2921 compared to the same L2 standard shown in the left panel. In both panels the spectra have been normalized to unity at 8250 \AA\ and a constant offset added to separate the spectra vertically.
\label{lowg_L2}}
\end{figure}

\begin{figure}
\includegraphics[scale=0.65,angle=270]{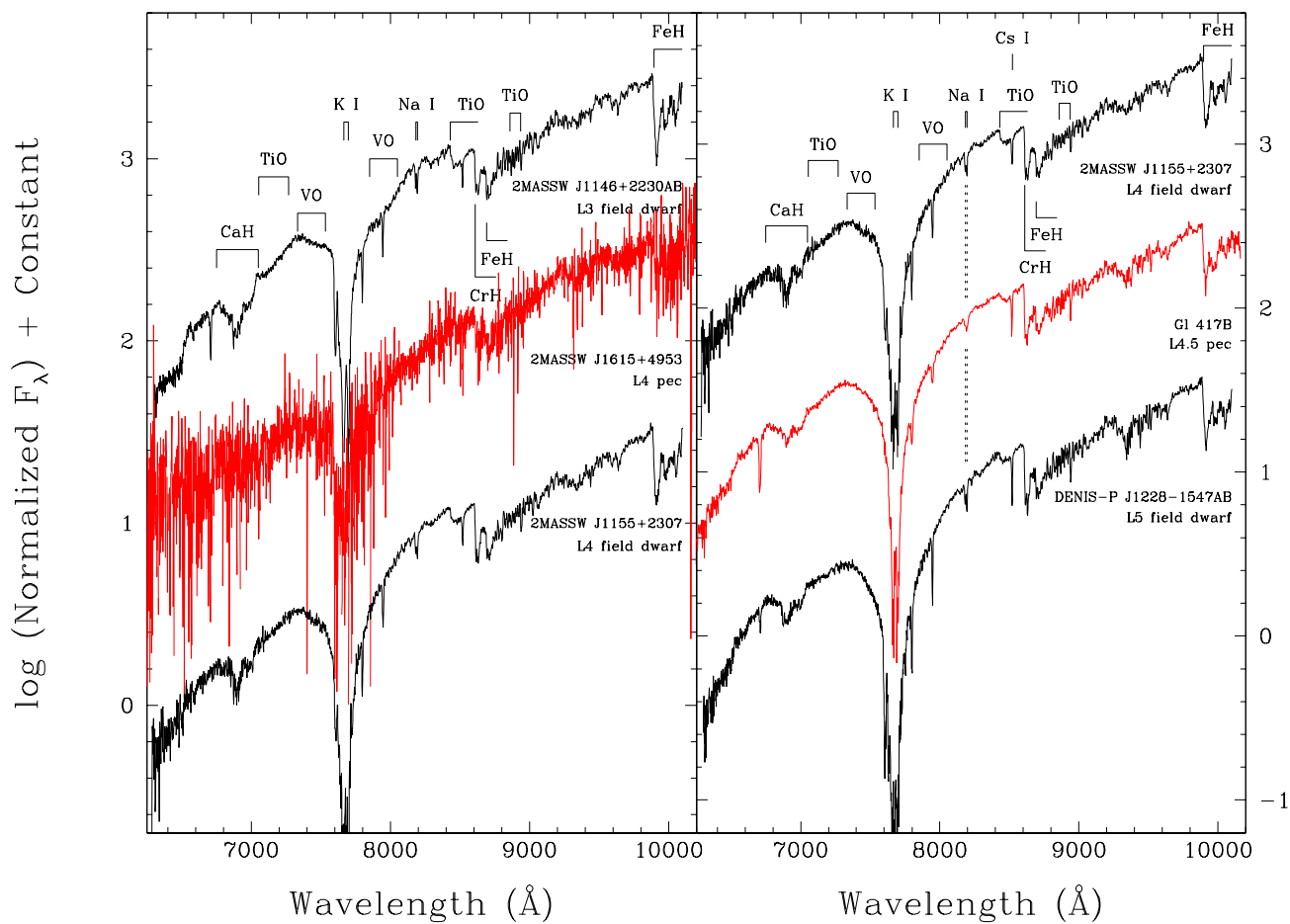}
\caption{(Left) The peculiar L4 dwarf 2MASS J1615+4953 compared to the standard L3 dwarf 2MASS J1146+2230 and the standard L4 dwarf 2MASS J1155+2307. (Right) The peculiar L4.5 dwarf Gl 417B compared to the same L4 standard shown in the left panel and the L5 standard DENIS J1228-1547. In both panels the spectra have been normalized to unity at 8250 \AA\ and a constant offset added to separate the spectra vertically.
\label{lowg_L4}}
\end{figure}

\begin{figure}
\plotone{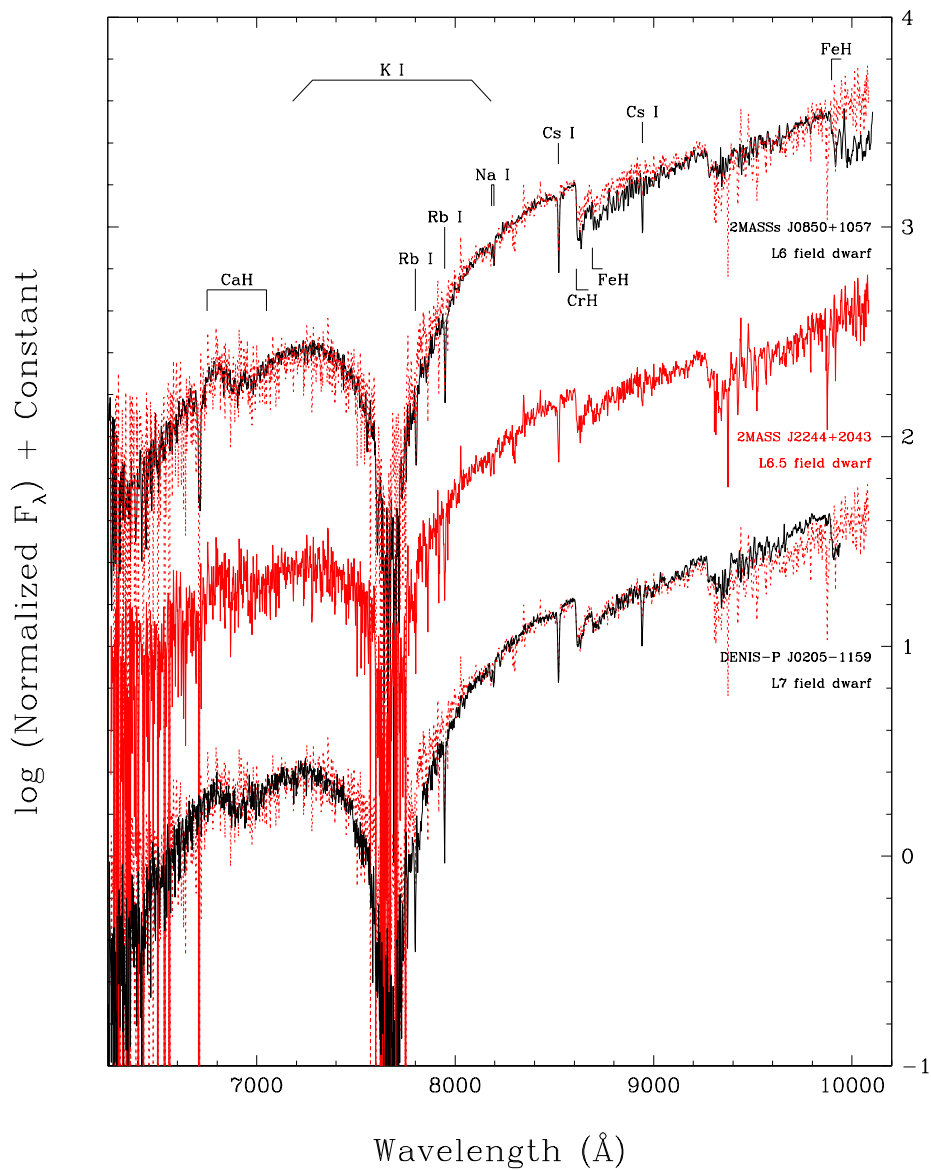}
\caption{2MASS J2244+2043 (red) compared to the L6 and L7 optical standards (black) 2MASS J0850+1057 and DENIS J0205$-$1159. Even though the near-infrared spectrum of this object is extremely peculiar with respect to normal late-L dwarfs, the optical spectrum differs only subtly from a normal L6.5 dwarf. In both panels the spectra have been normalized to unity at 8250 \AA\ and a constant offset added to separate the spectra vertically. Flux is plotted in logarithmic units so that features across all wavelengths can be more easily distinguished.
\label{lowg_L6.5}}
\end{figure}

\begin{figure}
\epsscale{0.75}
\plotone{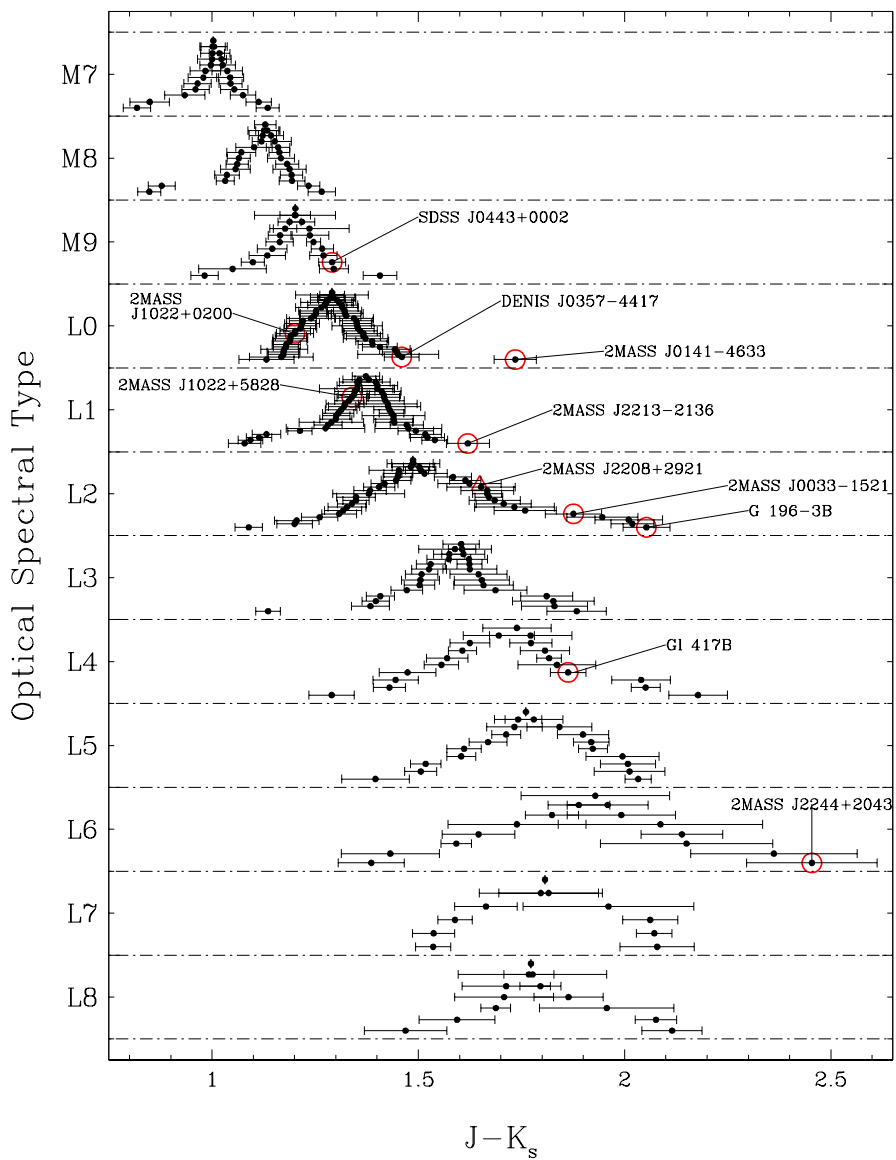}
\caption{J-K$_s$ colors for a collection of M7-L8 dwarfs with optically 
determined spectral types. Each bin represents a full integral subtype: 
``M7'' includes M7 and M7.5 dwarfs, ``M8'' includes M8 and M8.5 dwarfs, etc. 
For each group of objects, the median color in the group is plotted highest 
in the bin; colors falling farther from the median are plotted progressively 
farther down the y-axis. Sample selections are described in the text.
We use red circles to mark eleven objects we spectroscopically identify as low-gravity.
We also mark with a red triangle the color location of a peculiar L dwarf, 2MASS J2208+2921, from \cite{kirkpatrick2000} that did not meet the selection criteria for the plot.
\label{shrimp}}
\end{figure}

\begin{figure}
\epsscale{0.9}
\plotone{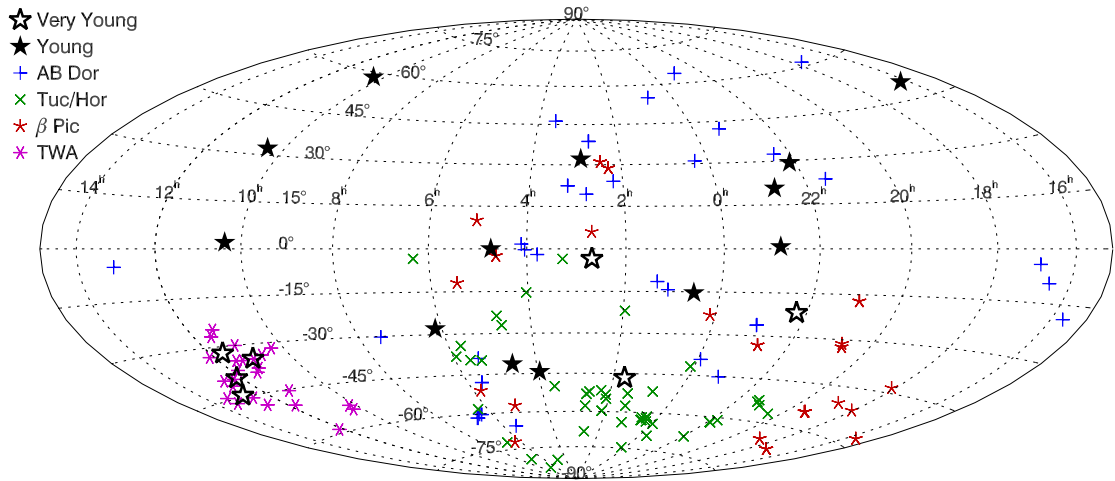}
\caption{Hammer (equal-area) projection of the sky in J2000 equatorial coordinates showing members of the AB Doradus Moving Group (blue plus signs), the Tucana-Horologium Association (green x's), $\beta$ Pictoris Moving Group (red asterisks), and the TW Hydrae Association (magenta snowflakes) as listed in \cite{zuckerman2004}. Locations of the 20 low-gravity field dwarfs of Table~\ref{young_dwarfs} are shown by solid stars for objects with age estimates $\sim$100 Myr and by open stars for objects with age estimates $\sim$10 Myr.
\label{skymap}}
\end{figure}

\begin{figure}
\epsscale{0.9}
\includegraphics[scale=0.65,angle=270]{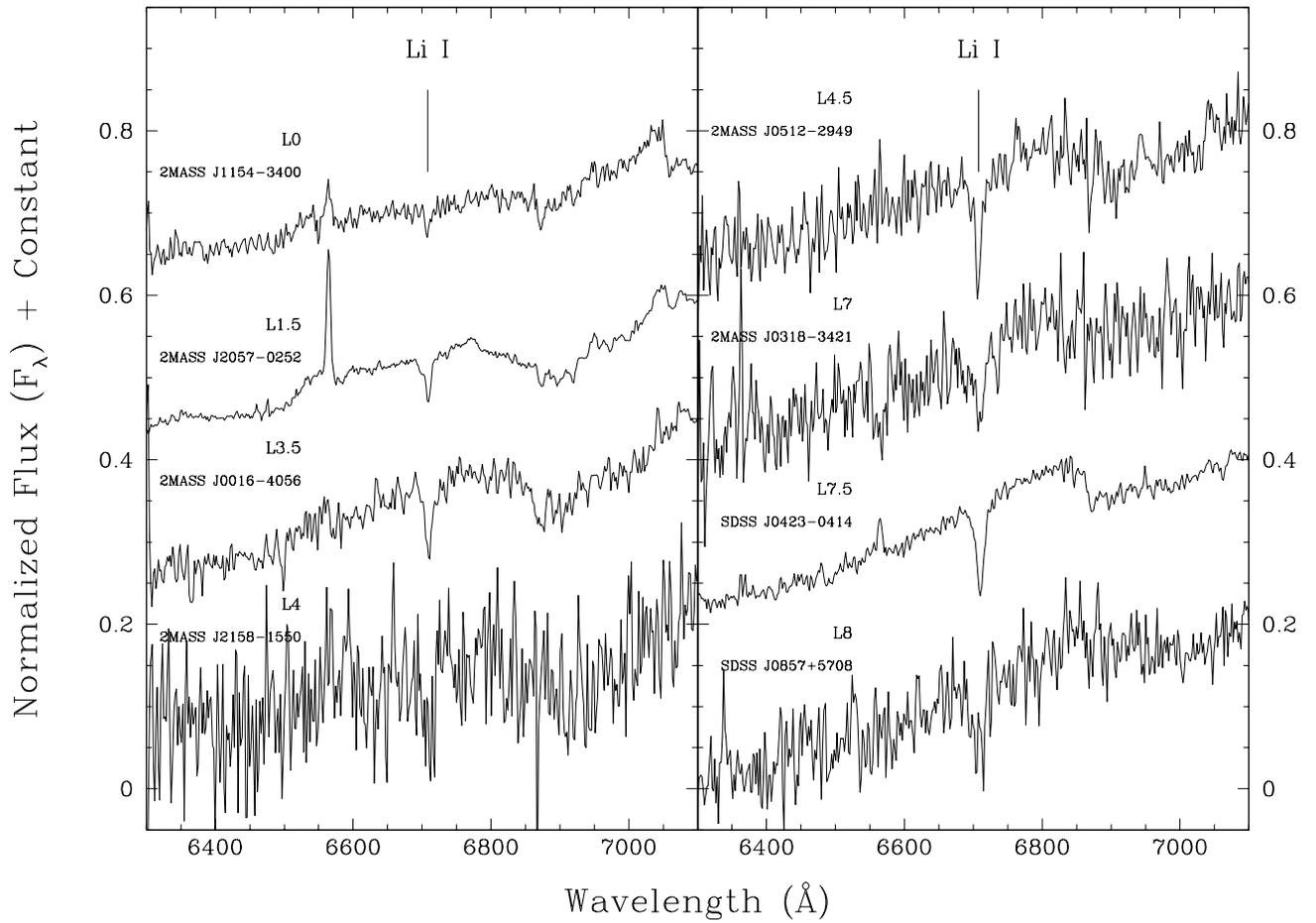}
\caption{Zooms of the Keck/LRIS spectra in the 6300-7100 \AA\ region for
those objects from the southern L dwarf sample and literature sample exhibiting \ion{Li}{1} absorption. 
\label{lithium_spectra}}
\end{figure}

\begin{figure}
\epsscale{0.7}
\plotone{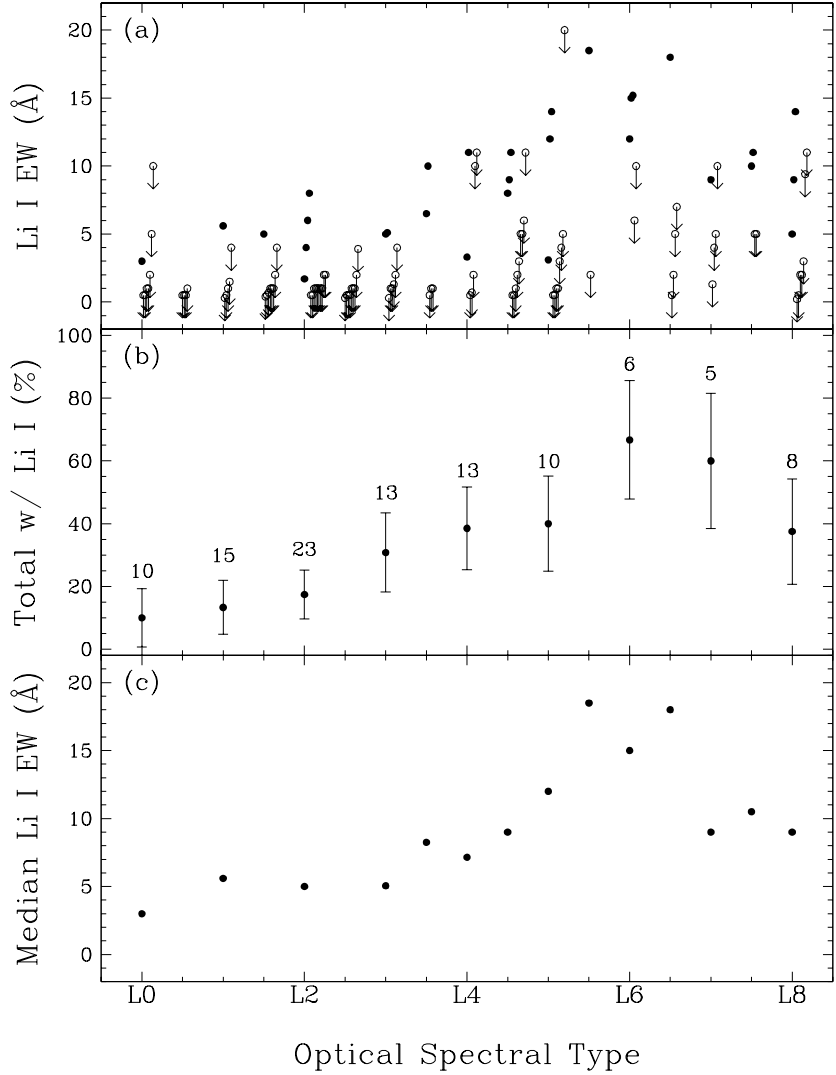}
\caption{An update of Fig.\ 7 from \cite{kirkpatrick2000}. (a) Li I equivalent widths as a function of spectral subclass for all L dwarfs for which we have ever obtained Keck-LRIS spectra. Solid circles denote objects having a detected Li I absorption line. Open circles with downward arrows denote upper limits to the Li I EW for those objects for which no line was detected. To avoid the overlapping of data caused by the quantization of spectral types, some of the points have been given slight offsets along the x-axis. (b) Percentage of L dwarfs showing Li I absorption as a function of spectral subclass. The only objects used in this computation are those for which a Li I equivalent width of 3 \AA\ or more would be detectable. Points have been binned into integer subtypes in which L0 and L0.5 dwarfs have been combined in the L0 bin, L1 and L1.5 dwarfs have been combined in the L1 bin, etc. The total number of objects in each bin is given above each data point. (c) Median Li strength as a function of spectral class for those objects for which lithium absorption was detected. \label{li_stats}}
\end{figure}

\begin{figure}
\epsscale{0.9}
\plotone{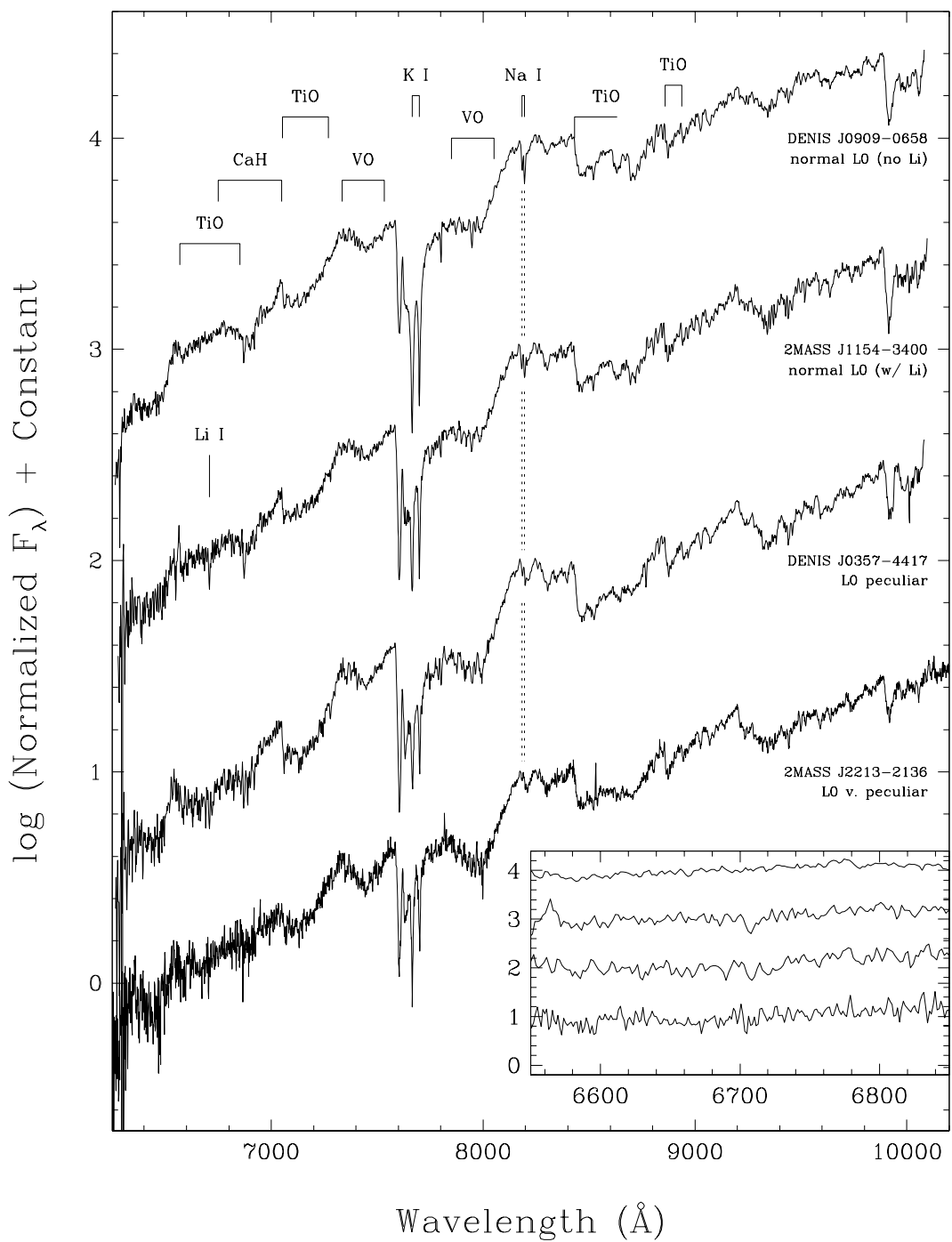}
\caption{Sample spectra of optical L0 dwarfs from each of the four age bins of Table~\ref{lithium_age_grid}. These are displayed on a logarithmic scale, normalized to one at 8250 \AA, with integers offsets added in the $y$-direction to separate spectra vertically. The spectrum with the best signal-to-noise ratio in each bin is shown. The inset shows a zoom-in around the 6708 \AA\ \ion{Li}{1} line for each of the four. In the inset, spectra are shown on a linear scale, normalized to one near 6750 \AA, and again offset by integers.
\label{li_seq_L0}}
\end{figure}

\begin{figure}
\epsscale{0.9}
\plotone{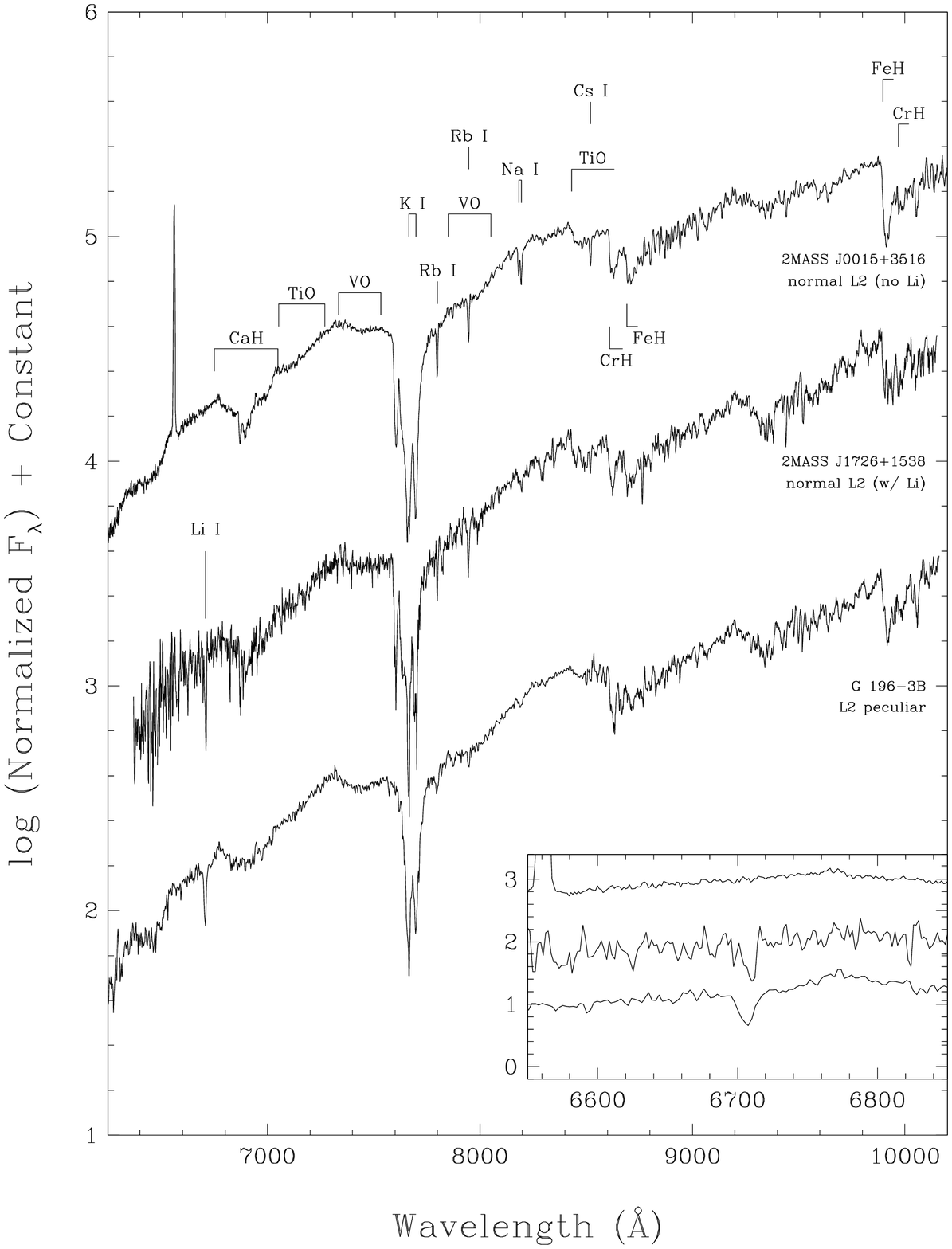}
\caption{Sample spectra of optical L2 dwarfs from each of the three populated age bins of Table~\ref{lithium_age_grid}. The inset shows a zoom-in around the \ion{Li}{1} line for each of the three. Normalizations and scalings are the same as in Figure~\ref{li_seq_L0}.
\label{li_seq_L2}}
\end{figure}

\begin{figure}
\epsscale{0.9}
\plotone{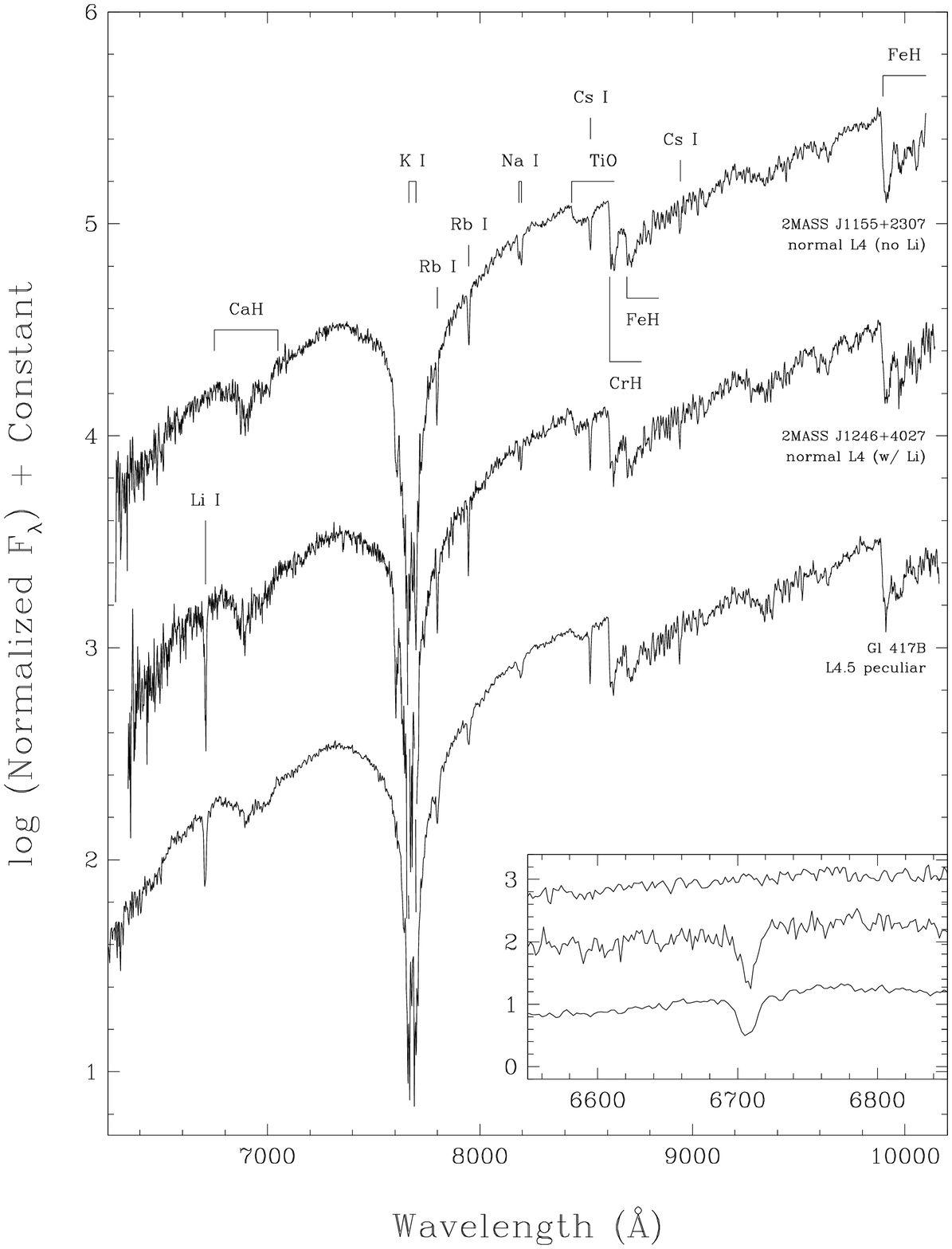}
\caption{Sample spectra of optical L4 dwarfs from each of the three populated age bins of Table~\ref{lithium_age_grid}. The inset shows a zoom-in around the \ion{Li}{1} line for each of the three. Normalizations and scalings are the same as in Figure~\ref{li_seq_L0}.
\label{li_seq_L4}}
\end{figure}

\begin{figure}
\epsscale{0.9}
\plotone{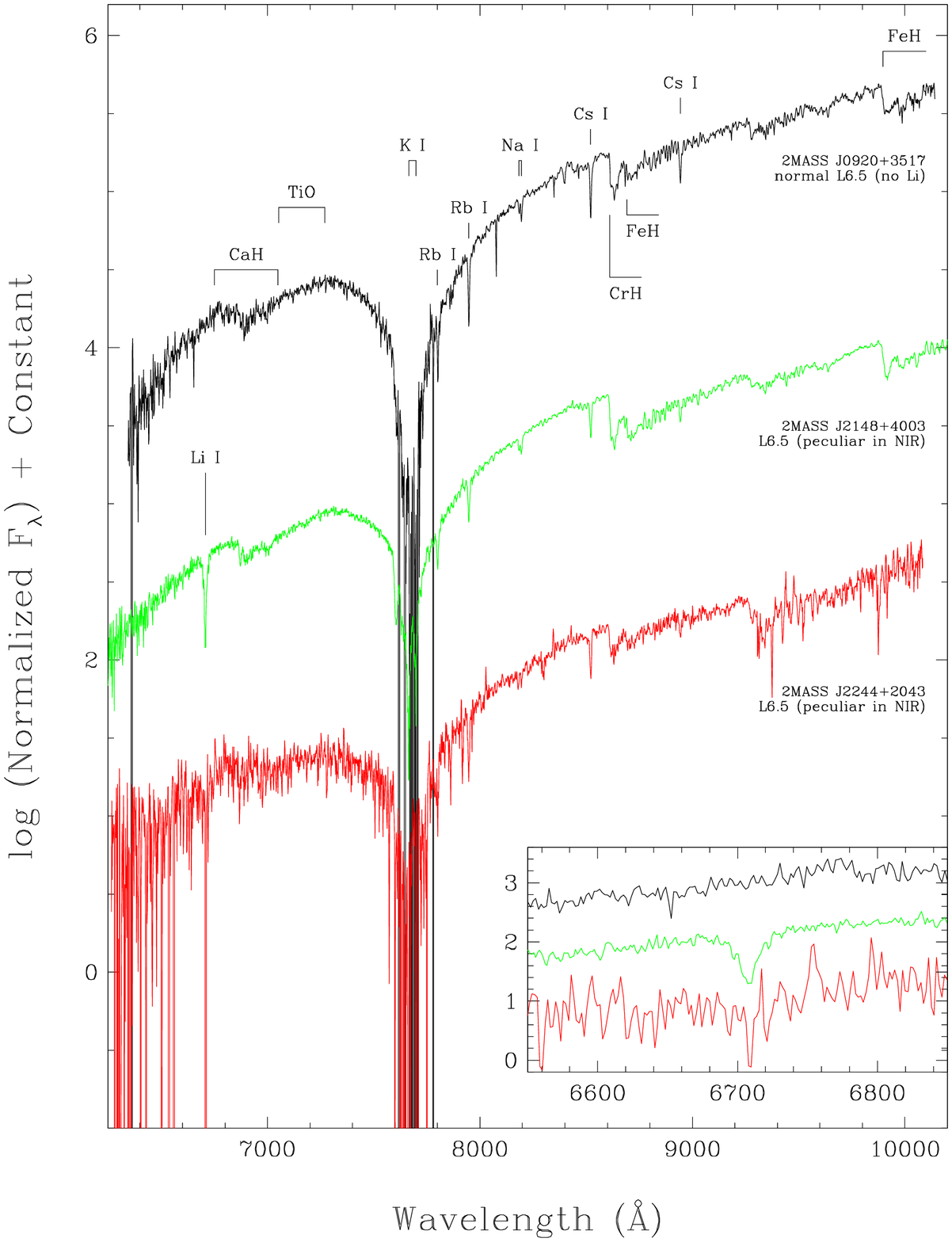}
\caption{Sample spectra of optical L6.5 dwarfs from each of the three populated age bins of Table~\ref{lithium_age_grid}. The inset shows a zoom-in around the \ion{Li}{1} line for each of the three. Normalizations and scalings are the same as in Figure~\ref{li_seq_L0}.
\label{li_seq_L6.5}}
\end{figure}

\begin{figure}
\epsscale{0.9}
\includegraphics[scale=0.65,angle=270]{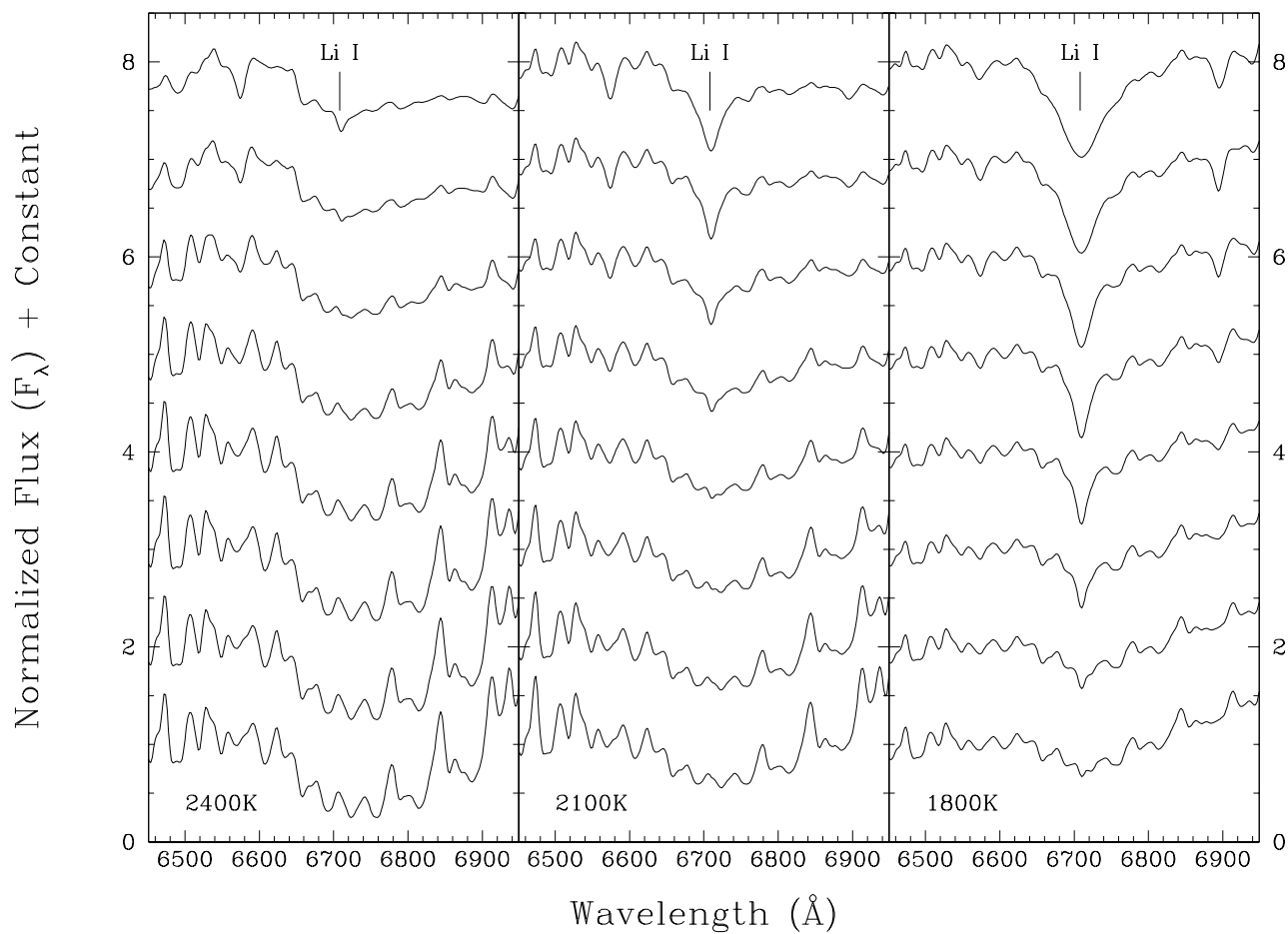}
\caption{Synthetic spectral models at effective temperatures of 2400K (left panel), 2100K (middle panel), and 1800K (right panel) for objects of solar abundance. Each panel illustrates the behavior of the spectra over a range of gravities, decreasing in half-dex increments from log(g)=6.0 at the top of each panel to log(g)=2.5 at the bottom. Note the universal weakening of the \ion{Li}{1} line as gravity is lowered. The spectra displayed here have a resolution of 10 \AA, identical to that of our empirical spectra.
\label{models}}
\end{figure}

We would like to thank Peter Allen for useful discussions on his 2005 paper. We would like to thank the staff at the W.\ M.\ Keck Observatory for their help in acquiring the LRIS data: observing assistants Joel Aycock, Gary Puniwai, Julie Rivera, Gabrelle Saurage, and Cynthia Wilburn as well as instrument specialists Paola Amico, Randy Campbell, Bob Goodrich, Grant Hill, Marc Kassis, and Greg Wirth. We would also like to thank the staff at the Subaru Telescope for their guidance in taking and reducing the FOCAS spectra: telescope operators Alan Hatakeyama and Robert Potter along with FOCAS support astronomer Takashi Hattori. We further wish to recognize and acknowledge the very significant cultural role and reverence that the summit of Mauna Kea has always had within the indigenous Hawaiian community.  We are most fortunate to have the opportunity to conduct observations from this mountain. This publication makes use of data products from the Two Micron All Sky Survey, which is a joint project of the University of Massachusetts and the Infrared Processing and Analysis Center/California Institute of Technology, funded by the National Aeronautics and Space Administration and the National Science Foundation. This research has made use of the NASA/IPAC Infrared Science Archive, which is operated by the Jet Propulsion Laboratory, California Institute of Technology, under contract with the National Aeronautics and Space Administration. The finder charts in Figure~\ref{finders} were generating using the mosaicking service available at http://hachi.ipac.caltech.edu:8080/montage/. Our research has benefitted from the M, L, and T dwarf compendium housed at DwarfArchives.org whose server was funded by a NASA Small Research Grant, administered by the American Astronomical Society.

\clearpage

\pagestyle{empty}

\begin{deluxetable}{lccccllll}
\tabletypesize{\scriptsize}
\rotate
\tablewidth{9in}
\tablecaption{List of L Dwarf Candidates\label{l_candidates}}
\tablehead{
\colhead{2MASS} & 
\colhead{$J$} &
\colhead{$H$} &
\colhead{$K_s$} &
\colhead{$J-K_s$} &
\colhead{Opt.\ Sp.} &
\colhead{Type} &
\colhead{Discovery} &
\colhead{Other} \\
\colhead{designation\tablenotemark{c}} & 
\colhead{mag} &
\colhead{mag} &
\colhead{mag} &
\colhead{color} &
\colhead{Type} &
\colhead{Ref.} &
\colhead{Ref.} &
\colhead{designation} \\
\colhead{(1)} &
\colhead{(2)} &
\colhead{(3)} &
\colhead{(4)} &
\colhead{(5)} &
\colhead{(6)} &
\colhead{(7)} &
\colhead{(8)} &
\colhead{(9)} 
}
\startdata 
\multicolumn{9}{c}{Confirmed Late-M and L Dwarfs} \\
\\
2MASS J00145575$-$4844171 & 14.050$\pm$0.035 & 13.107$\pm$0.036 & 12.723$\pm$0.030 &  1.327$\pm$0.046 & L2.5 pec& 1 & 1 & \\
2MASS J00165953$-$4056541 & 15.316$\pm$0.061 & 14.206$\pm$0.048 & 13.432$\pm$0.038 &  1.884$\pm$0.072 & L3.5    & 1 & 1 & \\
2MASS J00242463$-$0158201 & 11.992$\pm$0.035 & 11.084$\pm$0.022 & 10.539$\pm$0.023 &  1.453$\pm$0.042 & M9.5    & 2 & 20, 21& BRI 0021$-$0214 \\
2MASS J00332386$-$1521309 & 15.286$\pm$0.056 & 14.208$\pm$0.051 & 13.410$\pm$0.039 &  1.876$\pm$0.068 & L2 pec  & 1 & 3 & \\
2MASS J00511078$-$1544169 & 15.277$\pm$0.050 & 14.164$\pm$0.048 & 13.466$\pm$0.039 &  1.811$\pm$0.063 & L3.5    & 4 & 4 & \\
2MASS J00531899$-$3631102 & 14.445$\pm$0.026 & 13.480$\pm$0.031 & 12.937$\pm$0.029 &  1.508$\pm$0.039 & L3.5    & 1 & 1 & \\
2MASS J00584253$-$0651239 & 14.311$\pm$0.026 & 13.444$\pm$0.030 & 12.904$\pm$0.033 &  1.407$\pm$0.042 & L0      & 4 & 4 & \\
2MASS J01174748$-$3403258 & 15.178$\pm$0.036 & 14.209$\pm$0.039 & 13.489$\pm$0.037 &  1.689$\pm$0.052 & L2:     & 5 & 5 & \\
2MASS J01415823$-$4633574 & 14.832$\pm$0.043 & 13.875$\pm$0.026 & 13.097$\pm$0.032 &  1.735$\pm$0.054 & L0 pec  & 24& 24& \\
2MASS J01443536$-$0716142 & 14.191$\pm$0.026 & 13.008$\pm$0.029 & 12.268$\pm$0.023 &  1.923$\pm$0.035 & L5      & 1 & 6 & \\
2MASS J02052940$-$1159296 & 14.587$\pm$0.030 & 13.568$\pm$0.037 & 12.998$\pm$0.030 &  1.589$\pm$0.042 & L7      & 2 & 7 & DENIS$-$P J0205.4$-$1159 \\
2MASS J02511490$-$0352459 & 13.059$\pm$0.027 & 12.254$\pm$0.024 & 11.662$\pm$0.019 &  1.397$\pm$0.033 & L3      & 5 & 5 & \\
2MASS J02550357$-$4700509 & 13.246$\pm$0.027 & 12.204$\pm$0.024 & 11.558$\pm$0.024 &  1.688$\pm$0.036 & L8      & 1 & 8 & DENIS$-$P J0255$-$4700 \\
2MASS J02572581$-$3105523 & 14.672$\pm$0.039 & 13.518$\pm$0.032 & 12.876$\pm$0.032 &  1.796$\pm$0.050 & L8      & 1 & 1 & \\
2MASS J03185403$-$3421292 & 15.569$\pm$0.055 & 14.346$\pm$0.044 & 13.507$\pm$0.039 &  2.062$\pm$0.067 & L7      & 1 & 1 & \\
2MASS J03370359$-$1758079 & 15.621$\pm$0.058 & 14.412$\pm$0.050 & 13.581$\pm$0.041 &  2.040$\pm$0.071 & L4.5    & 4 & 4 & \\
2MASS J03572695$-$4417305 & 14.367$\pm$0.032 & 13.531$\pm$0.026 & 12.907$\pm$0.027 &  1.460$\pm$0.042 & L0 pec  & 1 & 9 & DENIS-P J035726.9$-$441730 \\
2MASS J04082905$-$1450334 & 14.222$\pm$0.030 & 13.337$\pm$0.030 & 12.817$\pm$0.023 &  1.405$\pm$0.038 & L2      & 5 & 26 & \\
2MASS J04234858$-$0414035 & 14.465$\pm$0.027 & 13.463$\pm$0.035 & 12.929$\pm$0.034 &  1.536$\pm$0.043 & L7.5    & 1 & 10, 23& SDSSp J042348.57$-$041403.5 \\
2MASS J04285096$-$2253227 & 13.507$\pm$0.023 & 12.668$\pm$0.027 & 12.118$\pm$0.026 &  1.389$\pm$0.035 & L0.5    & 11& 11& \\
2MASS J04390101$-$2353083 & 14.408$\pm$0.029 & 13.409$\pm$0.029 & 12.816$\pm$0.023 &  1.592$\pm$0.037 & L6.5    & 5 & 5 & \\
2MASS J04433761+0002051   & 12.507$\pm$0.026 & 11.804$\pm$0.024 & 11.216$\pm$0.021 &  1.291$\pm$0.033 & M9 pec  & 1 & 25& SDSS J044337.61+000205.1\\
2MASS J04455387$-$3048204 & 13.393$\pm$0.026 & 12.580$\pm$0.024 & 11.975$\pm$0.021 &  1.418$\pm$0.033 & L2      & 5 & 5 & \\
2MASS J04532647$-$1751543 & 15.142$\pm$0.035 & 14.060$\pm$0.035 & 13.466$\pm$0.035 &  1.676$\pm$0.049 & L3:     & 5 & 5 & \\
2MASS J05120636$-$2949540 & 15.463$\pm$0.057 & 14.156$\pm$0.048 & 13.285$\pm$0.042 &  2.178$\pm$0.071 & L4.5    & 1 & 5 & \\
2MASS J05233822$-$1403022 & 13.084$\pm$0.024 & 12.220$\pm$0.021 & 11.638$\pm$0.027 &  1.446$\pm$0.036 & L2.5    & 5 & 5 & \\
2MASS J05264348$-$4455455 & 14.082$\pm$0.033 & 13.307$\pm$0.028 & 12.705$\pm$0.027 &  1.377$\pm$0.043 & M9.5    & 1 & 1 & \\
2MASS J09095749$-$0658186 & 13.890$\pm$0.024 & 13.090$\pm$0.021 & 12.539$\pm$0.026 &  1.351$\pm$0.035 & L0      & 1 & 12& DENIS$-$P J0909$-$0658\\
2MASS J09532126$-$1014205 & 13.469$\pm$0.028 & 12.644$\pm$0.027 & 12.142$\pm$0.022 &  1.327$\pm$0.036 & L0:     & 13& 13& \\
2MASS J10101480$-$0406499 & 15.508$\pm$0.059 & 14.385$\pm$0.037 & 13.619$\pm$0.046 &  1.889$\pm$0.075 & L6      & 5 & 5 & \\
2MASS J10452400$-$0149576 & 13.160$\pm$0.024 & 12.352$\pm$0.025 & 11.780$\pm$0.023 &  1.380$\pm$0.033 & L1      & 14& 14& \\
2MASS J10584787$-$1548172 & 14.155$\pm$0.035 & 13.226$\pm$0.025 & 12.532$\pm$0.029 &  1.623$\pm$0.045 & L3      & 2 & 7 & DENIS$-$P J1058.7$-$1548 \\
2MASS J11544223$-$3400390 & 14.195$\pm$0.033 & 13.331$\pm$0.028 & 12.851$\pm$0.033 &  1.344$\pm$0.047 & L0      & 1 & 1 & \\
2MASS J12130336$-$0432437 & 14.683$\pm$0.035 & 13.648$\pm$0.025 & 13.014$\pm$0.030 &  1.669$\pm$0.046 & L5      & 5 & 5 & \\
2MASS J12185957$-$0550282 & 14.050$\pm$0.027 & 13.327$\pm$0.024 & 12.780$\pm$0.030 &  1.270$\pm$0.040 & M8      & 1 & 5 & \\
2MASS J12281523$-$1547342 & 14.378$\pm$0.030 & 13.347$\pm$0.032 & 12.767$\pm$0.030 &  1.611$\pm$0.042 & L5      & 2 & 7 & DENIS$-$P J1228.2$-$1547 \\
2MASS J13054019$-$2541059 & 13.414$\pm$0.026 & 12.392$\pm$0.025 & 11.747$\pm$0.023 &  1.667$\pm$0.035 & L2      & 2 & 15& Kelu$-$1 \\
2MASS J14090310$-$3357565 & 14.248$\pm$0.026 & 13.424$\pm$0.033 & 12.865$\pm$0.029 &  1.383$\pm$0.039 & L2      & 1 & 1 & \\
2MASS J14413716$-$0945590 & 14.020$\pm$0.029 & 13.190$\pm$0.031 & 12.661$\pm$0.030 &  1.359$\pm$0.042 & L0.5    & 1 & 8 & DENIS-P J1441$-$0945; G 124-62B\tablenotemark{a}  \\
2MASS J15074769$-$1627386 & 12.830$\pm$0.027 & 11.895$\pm$0.024 & 11.312$\pm$0.026 &  1.518$\pm$0.037 & L5      & 16& 16& \\
2MASS J15394189$-$0520428 & 13.922$\pm$0.029 & 13.060$\pm$0.026 & 12.575$\pm$0.029 &  1.347$\pm$0.041 & L4:\tablenotemark{b}     & 1 & 22& DENIS-P J153941.96$-$052042.4\\
2MASS J16184503$-$1321297 & 14.247$\pm$0.024 & 13.402$\pm$0.026 & 12.920$\pm$0.026 &  1.327$\pm$0.035 & L0:     & 1 & 1 & \\
2MASS J16202614$-$0416315 & 15.283$\pm$0.049 & 14.348$\pm$0.040 & 13.598$\pm$0.038 &  1.685$\pm$0.062 & L2.5    & 17& 17& Gl 618.1B \\
2MASS J20575409$-$0252302 & 13.121$\pm$0.024 & 12.268$\pm$0.024 & 11.724$\pm$0.025 &  1.397$\pm$0.035 & L1.5    & 1 & 5 & \\
2MASS J21041491$-$1037369 & 13.841$\pm$0.029 & 12.975$\pm$0.025 & 12.369$\pm$0.024 &  1.472$\pm$0.038 & L2.5    & 1 & 5 & \\
2MASS J21073169$-$0307337 & 14.200$\pm$0.032 & 13.443$\pm$0.031 & 12.878$\pm$0.030 &  1.322$\pm$0.044 & sd:M9   & 1 & 5 & \\
2MASS J21304464$-$0845205 & 14.137$\pm$0.032 & 13.334$\pm$0.032 & 12.815$\pm$0.033 &  1.322$\pm$0.046 & L1.5    & 1 & 1 & \\
2MASS J21580457$-$1550098 & 15.040$\pm$0.040 & 13.867$\pm$0.033 & 13.185$\pm$0.036 &  1.855$\pm$0.054 & L4:     & 1 & 1 & \\
2MASS J22064498$-$4217208 & 15.555$\pm$0.066 & 14.447$\pm$0.061 & 13.609$\pm$0.055 &  1.946$\pm$0.086 & L2      & 4 & 4 & \\
2MASS J22244381$-$0158521 & 14.073$\pm$0.027 & 12.818$\pm$0.026 & 12.022$\pm$0.023 &  2.051$\pm$0.035 & L4.5    & 4 & 4 & \\
2MASS J23440624$-$0733282 & 14.802$\pm$0.037 & 13.846$\pm$0.035 & 13.232$\pm$0.033 &  1.570$\pm$0.050 & L4.5    & 1 & 1 & \\
\\
\multicolumn{9}{c}{Objects Confirmed as Other Types} \\
\\
%2MASS J00323659$-$0910261& 13.947$\pm$0.037 & 13.080$\pm$0.037 & 11.955$\pm$0.032 &  1.992$\pm$0.049 & ---     & - & - & has bright opt. counterpart (AX?)\\ 
2MASS J01241236$-$4537057 & 13.410$\pm$0.029 & 12.493$\pm$0.026 & 12.076$\pm$0.026 &  1.334$\pm$0.039 & M giant & 1 & - & \\
2MASS J01343566$-$0931030 & 16.187$\pm$0.131 & 14.793$\pm$0.074 & 13.579$\pm$0.055 &  2.608$\pm$0.142 & carbon star & 1 & - & \\
%2MASS J02352800+0000402  & 11.949$\pm$0.024 & 10.709$\pm$0.024 & 10.248$\pm$0.021 &  1.701$\pm$0.032 & ---     & - & - & has bright opt. counterpart (AX?)\\
2MASS J04475750$-$0553241 & 13.789$\pm$0.029 & 12.441$\pm$0.032 & 11.734$\pm$0.019 &  2.055$\pm$0.035 & reddened& 1 & - & (early type, reddened star) \\
%2MASS J05013923$-$0852224& 15.941$\pm$0.079 & 14.203$\pm$0.037 & 13.427$\pm$0.037 &  2.514$\pm$0.087 & ---     & - & - & in nebular region (AX?)\\
2MASS J10152592$-$0204318 & 14.050$\pm$0.027 & 12.866$\pm$0.025 & 11.946$\pm$0.026 &  2.104$\pm$0.037 & carbon star & 1 & - & \\
2MASS J12274004$-$0027506 & 12.757$\pm$0.024 & 11.513$\pm$0.021 & 10.545$\pm$0.023 &  2.212$\pm$0.033 & carbon star & 18& - & FASTT 542 \\
2MASS J12562145$-$0811144 & 11.317$\pm$0.024 & 10.426$\pm$0.024 &  9.994$\pm$0.021 &  1.323$\pm$0.032 & M giant & 1 & - & \\
2MASS J13414737$-$0813470 & 12.872$\pm$0.024 & 11.813$\pm$0.021 & 11.444$\pm$0.026 &  1.428$\pm$0.035 & M giant & 1 & - & \\
2MASS J13451789$-$0829573 & 15.622$\pm$0.068 & 14.165$\pm$0.039 & 12.822$\pm$0.027 &  2.800$\pm$0.073 & QSO     & 1 & 1 & (redshift z=0.57) \\
2MASS J13592063$-$3023395 & 14.613$\pm$0.033 & 13.081$\pm$0.028 & 11.820$\pm$0.025 &  2.793$\pm$0.041 & carbon star & 14& - & \\
2MASS J14322874$-$0531178 & 13.989$\pm$0.028 & 12.531$\pm$0.023 & 11.240$\pm$0.021 &  2.749$\pm$0.035 & carbon star & 18& - & \\
2MASS J15010693$-$0531388 & 13.588$\pm$0.026 & 12.391$\pm$0.021 & 11.498$\pm$0.021 &  2.090$\pm$0.033 & carbon star & 5 & - & \\
2MASS J15151106$-$1332278 & 12.599$\pm$0.026 & 11.511$\pm$0.023 & 10.795$\pm$0.021 &  1.804$\pm$0.033 & carbon star & 14& - & \\
2MASS J15514921$-$0750489 & 16.867$\pm$0.181 & 15.080$\pm$0.072 & 13.533$\pm$0.042 &  3.334$\pm$0.186 & carbon star & 1 & - & \\
2MASS J16014265$-$1249447 & 14.416$\pm$0.029 & 13.105$\pm$0.029 & 12.382$\pm$0.029 &  2.034$\pm$0.041 & carbon star & 1 & - & \\
2MASS J20135152$-$2806020 & 14.242$\pm$0.030 & 13.461$\pm$0.028 & 12.944$\pm$0.027 &  1.298$\pm$0.040 & M8-9 III& 1 & - & \\
2MASS J21001879$-$0606550 & 15.306$\pm$0.045 & 13.557$\pm$0.025 & 12.073$\pm$0.021 &  3.233$\pm$0.050 & carbon star & 1 & - & \\
%2MASS J22120797$-$2739313& 15.384$\pm$0.072 & 14.729$\pm$0.070 & 13.493$\pm$0.055 &  1.891$\pm$0.091 & ---     & - & - & has bright opt. counterpart (AX?)\\
2MASS J22351322$-$4835588 & 13.809$\pm$0.029 & 12.949$\pm$0.029 & 12.124$\pm$0.026 &  1.685$\pm$0.039 & QSO     & - & 19& \\
\enddata
\tablecomments{
Spectral type and discovery references are (1) this paper, (2) \citealt{kirkpatrick1999}, 
(3) \citealt{gizis2003}, (4) \citealt{kirkpatrick2000}, (5) \citealt{cruz2003}, (6) \citealt{liebert2003}, 
(7) \citealt{delfosse1997}, (8) \citealt{martin1999}, (9) \citealt{bouy2003}, 
(10) \citealt{geballe2002}, (11) \citealt{kendall2003}, 
(12) \citealt{delfosse1999}, (13) \citealt{cruz2007}, (14) \citealt{gizis2002}, (15) \citealt{ruiz1997}, 
(16) \citealt{reid2000}, (17) \citealt{wilson2001}, (18) \citealt{totten1998}, (19) \citealt{savage1976}, 
(20) \citealt{luyten1980}, (21) \citealt{irwin1991}, (22) \citealt{kendall2004}, (23) \citealt{schneider2002},
(24) \citealt{kirkpatrick2006}, (25) \citealt{hawley2002}, (26) \citealt{wilson2003}.}
\tablenotetext{a}{Common proper motion for DENIS-P J1441$-$0945 and G 124-62A confirmed by
\cite{seifahrt2005}.}
\tablenotetext{b}{DENIS J1539$-$0520 is typed in the optical as L3.5 by \cite{reid2008}.}
\tablenotetext{c}{Source designations for 2MASS discoveries are given as ``2MASx Jhhmmss[.]ss$\pm$ddmmss[.]s''. The ``x'' in the prefix will vary depending upon the catalog from which the object was discovered: ``S'' is used for objects from the 2MASS All-Sky Point Source Catalog, ``SW'' is used for objects taken from the Survey Point Source Working Database, ``Ss'' is used for objects taken from the 2MASS Sampler Point Source Catalog, ``P'' is used for objects discovered in the prototype data. The suffix is the sexigesimal Right Ascension and Declination at J2000 equinox.}
\end{deluxetable}

\clearpage

\pagestyle{plaintop}

\begin{deluxetable}{llll}
\tabletypesize{\small}
\tablecaption{Nights of Observation at Keck\label{nights}}
\tablehead{
\colhead{Obs.\ Date} &
\colhead{Principal} &
\colhead{Other Observer} &
\colhead{Sky Conditions} \\
\colhead{(UT)} &
\colhead{Investigator} &
\colhead{Assisting} &
\colhead{} \\
\colhead{(1)} &
\colhead{(2)} &
\colhead{(3)} &
\colhead{(4)} 
}
\startdata 
2000 Aug 23 & Kirkpatrick & (none)      & cirrus throughout night \\
2000 Dec 26 & Kirkpatrick & Liebert     & clear \\
2000 Dec 27 & Carpenter   & Hillenbrand & clear \\
2000 Dec 28 & Carpenter   & Hillenbrand & clear, seeing poorer than average \\
2001 Feb 20 & Kirkpatrick & Liebert     & mostly clear, light clouds late in night \\
2001 Nov 13 & Stauffer    & Kirkpatrick & spotty clouds \\
2002 Jan 01 & Carpenter   & (none)      & clear \\
2002 Jan 02 & Carpenter   & (none)      & clear \\
2002 Jan 03 & Carpenter   & (none)      & clear first half, then fog forced closure \\
2002 Feb 19 & Kirkpatrick & Lowrance    & never opened (fog and snow)\\
2002 Feb 20 & Kirkpatrick & Lowrance    & never opened (fog)\\
2003 Jan 02 & Kirkpatrick & Lowrance    & clear \\
2003 Jan 03 & Kirkpatrick & Lowrance    & clear \\
2003 Dec 22 & Kirkpatrick & Lowrance    & clear \\
2003 Dec 23 & Kirkpatrick & Lowrance    & clear \\
2003 Dec 24 & Kirkpatrick & Lowrance    & clear \\
\enddata
\end{deluxetable}

\clearpage

\begin{deluxetable}{lcclllllllccc}
\tabletypesize{\scriptsize}
\rotate
\tablewidth{9in}
\tablecaption{New Spectroscopic Observations of Objects from Table~\ref{l_candidates}\label{new_spec_obs}}
\tablehead{
\colhead{Object} & 
\colhead{Obs.\ Date} &
\colhead{Int.} &
\colhead{CrH-a} &
\colhead{Rb-b/} &
\colhead{Cs-a/} &
\colhead{Color-d} &
\colhead{KI} &
\colhead{Oxide} &
\colhead{Final opt.} &
\colhead{Dist.\tablenotemark{d}} &
\colhead{H$\alpha$} &
\colhead{Li I} \\
\colhead{name} & 
\colhead{(UT)} &
\colhead{(s)} &
\colhead{} &
\colhead{TiO-b} &
\colhead{VO-b} &
\colhead{} &
\colhead{fit} &
\colhead{fit} &
\colhead{type} &
\colhead{(pc)} &
\colhead{EW (\AA)} &
\colhead{EW (\AA)} \\
\colhead{(1)} &
\colhead{(2)} &
\colhead{(3)} &
\colhead{(4)} &
\colhead{(5)} &
\colhead{(6)} &
\colhead{(7)} &
\colhead{(8)} &
\colhead{(9)} &
\colhead{(10)} &
\colhead{(11)} &
\colhead{(12)} &
\colhead{(13)}
}
\startdata 
2MASS J00145575$-$4844171     & 2003 Dec 23 &  600 & 1.57(2)    & ---        & ---        & 8.97(-)     & (3-4)     & (2-3)     & L2.5\tablenotemark{e}& 20: & $<$1 & $<$0.5 \\
2MASS J00165953$-$4056541     & 2003 Jan 02 & 1200 & 1.73(3-4)  & 1.30(4)    & 1.22(3-4)  & 7.83(-)     & (3)       & ---       & L3.5    & 29  & $<$1 & 6.5    \\
2MASS J00332386$-$1521309     & 2003 Dec 24 & 1200 & 1.50(2)    & ---        & ---        & 7.25(-)     & (2)       & (4)       & L2 pec  & 39: & $<$3 & $<$2   \\
2MASS J00531899$-$3631102     & 2003 Dec 23 &  600 & 1.69(3)    & 1.31(4)    & 1.16(3)    & 9.17(-)     & (4)       & ---       & L3.5    & 20  & $<$1 & $\le$1 \\
2MASS J01415823$-$4633574     & 2003 Dec 23 & 2400 & 0.97($<$0) & 0.68(0)    & 0.69($<$0) & 8.28(-)     & (pec)     & ---       & L0 pec  & 39: & 10.5 & $\le$1 \\
2MASS J01443536$-$0716142     & 2001 Feb 20 &  960 & 2.20(5)    & 1.75(6)    & 1.31(4)    & 11.40(5)    & (4-5)     & ---       & L5      & 13  & 13\tablenotemark{b}
   & $<$0.5 \\
DENIS-P J0255$-$4700          & 2000 Dec 26 & 2400 & 1.29(8)    & 2.14(7)    & 1.67(8)    & 36.54($>$8) & ---       & (8)       & L8      & {\it 5}   & $<$1 & $<$0.2 \\
2MASS J02572581$-$3105523     & 2003 Dec 24 & 1200 & 1.32(8)    & ---        & ---        & 29.63(8)    & ---       & (8)       & L8      & 9   & $<$2 & $<$0.5 \\
2MASS J03185403$-$3421292     & 2003 Dec 24 & 1800 & 1.49(7)    & ---        & ---        & 21.90(7)    & ---       & (7)       & L7      & 16  & $<$10\tablenotemark{c}
   & 9: \\
DENIS-P J035729.6$-$441730    & 2003 Dec 24 & 1200 & 1.02($<$0) & 0.69(0)    & 0.73($<$0) & 7.74(-)     & (0)       & ---       & L0 pec  & 32: & $<$2 & $\le$2 \\
SDSSp J042348.57$-$041403.5   & multi dates\tablenotemark{a}
                                            & 5100 & 1.56(7)    & 2.07(7)    & 1.72(8+)   & 26.81(7-8)  & ---       & ---       & L7.5    & {\it 15}& 3    & 11     \\
SDSS J044337.61+000205.1      & 2003 Dec 24 &  600 & 0.97($<$0) & 0.49($<$0) & 0.68($<$0) & 6.53(-)     & ($\sim$0) & ---       & M9 pec  & 15: & 2.5  & $<$2   \\
2MASS J05120636$-$2949540     & 2003 Dec 24 & 1200 & 1.78(3-4)  & ---        & ---        & 11.29(5)    & (4)       & (5)       & L4.5    & 26  & $\le$2 & 11   \\
2MASS J05264348$-$4455455     & 2003 Dec 24 &  600 & 1.15(0)    & 0.63($<$0) & 0.80(0)    & 6.42(-)     & ($\sim$0) & ---       & M9.5    & 30  & 4    & $<$1   \\
DENIS-P J0909$-$0658          & 2003 Dec 22 &  600 & 1.21(0)    & 0.79(1)    & 0.79(0)    & 6.45(-)     & (0)       & ---       & L0      & 25  & $\le$1 & $<$0.5 \\
2MASS J11544223$-$3400390     & 2003 Jan 02 &  600 & 1.17(0)    & 0.72(0)    & 0.78(0)    & 7.20(-)     & (0)       & ---       & L0      & 29  & 4    & 3      \\
2MASS J12185957$-$0550282     & 2003 Dec 22 &  300 & 1.03($<$0) & 0.56($<$0) & 0.79(0)    & 4.25(-)     & ($<$0)    & ---       & M8      & 38  & 6    & $<$2   \\
2MASS J14090310$-$3357565     & 2001 Feb 20 & 1200 & 1.37 (1)   & 1.10(2-3)  & 1.00(2)    & 6.63(-)     & (2)       & ---       & L2      & 24  & $<$2 & $<$1   \\
DENIS-P J1441$-$0945          & 2003 Jan 03 &  600 & 1.29(0-1)  & 0.84(1)    & 0.86(1)    & 6.54(-)     & (0)       & ---       & L0.5    & 26  & $<$1 & $<$0.5 \\
DENIS-P J153941.96$-$052042.4 & 2000 Aug 23 & 1200 & 1.48(2)    & ---        & ---        & 10.93(5)    & (3-4)     & (4)       & L4:     & 14: & $<$10& $<$10  \\
2MASS J16184503$-$1321297     & 2000 Aug 23 & 1200 & 1.44(1-2)  & ---        & ---        & 5.89(-)     & (0)       & ($\sim$0) & L0:     & 30: & $<$5 & $<$5   \\
Gl 618.1B                     & 2000 Aug 23 & 2400 & 1.86(4)    & ---        & ---        & 4.97(-)     & (2-3)     & (2-3)     & L2.5    & {\it 30} & $\le$2 & $<$2 \\
2MASS J20575409$-$0252302     & 2000 Aug 23 & 1200 & 1.33(1)    & 0.94(1-2)  & 0.88(1)    & 6.61(-)     & (2)       & ---       & L1.5    & 15  & 11   & 5      \\
2MASS J21041491$-$1037369     & 2000 Aug 23 & 1200 & 1.46(1-2)  & 1.21(3-4)  & 0.98(2)    & 6.39(-)     & (2-3)     & ---       & L2.5    & 18  & $<$1 & $<$1   \\
2MASS J21073169$-$0307337     & 2000 Aug 23 & 1200 & 1.25(0-1)  & 0.59($<$0) & 0.91(1-2)  & 5.24(-)     & ($<$0)    & ---       & d/sdM9  & 32: & 2    & $<$1   \\
2MASS J21304464$-$0845205     & 2000 Aug 23 &  768 & 1.60(2-3)  & ---        & ---        & 6.54(-)     & (1-2)     & (1)       & L1.5    & 25  & $<$4 & $<$4   \\
2MASS J21580457$-$1550098     & 2000 Aug 23 & 1200 & 1.69(3)    & ---        & ---        & 8.52(-)     & (4)       & (4)       & L4:     & 23  & $<$5 & $\le$11 \\
2MASS J23440624$-$0733282     & 2003 Jan 03 &  420 & 2.01(4-5)  & 1.55(4-5)  & 1.38(4-5)  & 11.42(5)    & (4-5)     & ---       & L4.5    & 19  & $<$1 & $<$0.5 \\
\enddata
\tablenotetext{a}{SDSSp J0423$-$0414 was observed on four separate dates: 2001 Nov 13 (900s integration), 2002 Jan 01 (1200s), 2002 Jan 02 
(1200s), and 2002 Jan 03 (1800s). All analysis was performed on a coadded spectrum combining all 5100s of integration.}
\tablenotetext{b}{2MASS J0144$-$0716 shows variable H$\alpha$ emission when spectra from different epochs are compared. See \cite{liebert2003} 
for further discussion.}
\tablenotetext{c}{2MASS J0318$-$3421 shows a possible {\it absorption} trough of $\sim$10 \AA\ EW at the location of H$\alpha$, but this is
likely a data artifact since it is aphysically broad.}
\tablenotetext{d}{Entries in italics are distances as measured through trignonometric parallax. All other entries are distances derived from
spectrophotometric parallaxes. For SDSSp J0423$-$0414, the trigonometric parallax is
from \cite{vrba2004}; for Gl 618.1B the distance is from its association with the primary, Gl 618.1A, whose trigonometric parallax was measured
by Hipparcos (\citealt{perryman1997}).}
\tablenotetext{e}{The optical spectrum of 2MASS J0014$-$4844 is slightly peculiar, perhaps indicative of lower metallicity.}
\end{deluxetable}

\clearpage

\begin{deluxetable}{lccclllllllcc}
\tabletypesize{\scriptsize}
\rotate
\tablewidth{9in}
\tablecaption{Supporting Observations of L Dwarfs\label{new_spec_ml}}
\tablehead{
\colhead{Object} & 
\colhead{Disc.} &
\colhead{Obs.\ Date} &
\colhead{Int.} &
\colhead{CrH-a} &
\colhead{Rb-b/} &
\colhead{Cs-a/} &
\colhead{Color-d} &
\colhead{KI} &
\colhead{Oxide} &
\colhead{Final opt.} &
\colhead{H$\alpha$} &
\colhead{Li I} \\
\colhead{name} & 
\colhead{ref.} &
\colhead{(UT)} &
\colhead{(s)} &
\colhead{} &
\colhead{TiO-b} &
\colhead{VO-b} &
\colhead{} &
\colhead{fit} &
\colhead{fit} &
\colhead{type} &
\colhead{EW (\AA)} &
\colhead{EW (\AA)} \\
\colhead{(1)} &
\colhead{(2)} &
\colhead{(3)} &
\colhead{(4)} &
\colhead{(5)} &
\colhead{(6)} &
\colhead{(7)} &
\colhead{(8)} &
\colhead{(9)} &
\colhead{(10)} &
\colhead{(11)} &
\colhead{(12)} &
\colhead{(13)} 
}
\startdata 
2MASS J05185995$-$2828372     & 1 & 2003 Dec 24                & 1200 & 1.63(7)   & ---        & ---       & 26.58(7-8) & ---       & (7)   & L7      & $<$7   & $<$5 \\
SDSSp J083008.12+482847.4     & 2 & multidate\tablenotemark{a} & 2400 & 1.14(8+)  & ---        & ---       & 35.07(8+)  & ---       & (8+)  & L8      & $<$6   & $<$2 \\
SDSSp J085758.45+570851.4     & 2 & 2002 Jan 02                & 1200 & 1.35(8)   & ---        & ---       & 26.83(7-8) & ---       & (8)   & L8      & $<$8   & 14: \\
Gl 337CD\tablenotemark{b}     & 3 & 2000 Dec 26                & 2400 & 1.24(8)   & 2.54(8)    & 1.80(8+)  & 44.67(8+)  & ---       & (8)   & L8      & $<$2   & $<$2 \\
2MASSI J1315309$-$264951      & 4 & 2003 Jan 02                & 1200 & 2.01(5-6) & 1.83(6)    & 1.47(5)   & 15.42(6)   & ($\sim$5) & (5-6) & L5.5    & 160    & $<$2 \\
2MASSW J2244316+204343        & 5 & 2000 Dec 26                & 2400 & 1.44(7-8) & 1.85(6)    & 1.57(6-7) & 15.23(6)   & ---       & (6-7) & L6.5    & $<$5   & $\le$5 \\
\enddata
\tablecomments{Discovery references are (1) \citealt{cruz2004}, 
(2) \citealt{geballe2002}, (3) \citealt{wilson2001}, (4) \citealt{gizis2002}, (5) \citealt{dahn2002}.}
\tablenotetext{a}{SDSS J0830+4828 was observed on two separate dates: 2002 Jan 01 (1200s integration) and 2002 Jan 02 (1200s). 
All analysis was performed on a coadded spectrum combining all 2400s of integration.}
\tablenotetext{b}{Gl 337CD is also known as 2MASS J0912146+145940 and was identified as a close double by \cite{burgasser2005a}.}
\end{deluxetable}

\clearpage
\thispagestyle{empty}

\begin{deluxetable}{lcccllllllcc}
\tabletypesize{\scriptsize}
\rotate
\tablewidth{9in}
\tablecaption{Supporting Observations of T Dwarfs\label{new_spec_t}}
\tablehead{
\colhead{Object} & 
\colhead{Disc} &
\colhead{Obs.\ Date} &
\colhead{Int.} &
\colhead{Cs I (A)} &
\colhead{CrH(A)/} &
\colhead{FeH(B)} &
\colhead{Color-e} &
\colhead{By-eye} &
\colhead{Final opt.} &
\colhead{H$\alpha$} &
\colhead{Li I} \\
\colhead{name} & 
\colhead{ref.} &
\colhead{(UT)} &
\colhead{(s)} &
\colhead{} &
\colhead{H$_2$O} &
\colhead{} &
\colhead{} &
\colhead{type} &
\colhead{type} &
\colhead{EW (\AA)} &
\colhead{EW (\AA)} \\
\colhead{(1)} &
\colhead{(2)} &
\colhead{(3)} &
\colhead{(4)} &
\colhead{(5)} &
\colhead{(6)} &
\colhead{(7)} &
\colhead{(8)} &
\colhead{(9)} &
\colhead{(10)} &
\colhead{(11)} &
\colhead{(12)} 
}
\startdata 
2MASS J04151954$-$0935066     & 1 & 2000 Dec 26                &  3600 & 1.18       & 0.24       & 0.89          & 4.59(-)    & (8 std) & T8 std & ---\tablenotemark{c}  & ---\tablenotemark{c} \\
SDSSp J083717.22$-$000018.3   & 2 & multidate\tablenotemark{a} & 14400 & 1.84($<$2) & 0.89($<$2) & 0.89(-)       & 3.35($<$2) & (0)     & T0     & $<$6  & $<$6 \\
SDSSp J102109.6$-$030419      & 2 & 2000 Dec 26                &  2400 & 1.88(2-5)  & 0.67(2-5)  & 1.37(5)       & 3.85(2)    & (2(-5)) & T3.5   & $<$13 & $<$5 \\
2MASS J12095613$-$1004008     & 3 & multidate\tablenotemark{b} &  9600 & 1.86(2-5)  & 0.63(5)    & 1.24($\sim$5) & 4.22(-)    & (2-5)   & T3.5   & $<$8  & $<$4 \\
SDSSp J125453.9$-$012247      & 2 & 2000 Dec 26                &  2400 & 1.96       & 0.77       & 1.09          & 4.20       & (2 std) & T2 std & 28    & $<$5 \\
%
%2MASS J05591914$-$1404488\tablenotemark{aa}       & 2000 Mar 05                & 3600  & 1.77       & 0.57       & 1.38          & 4.10(-)    & (5 std) & T5 std &       &      \\
%2MASS J15031961+2525196                           & 2003 Jan 02                &  1200 & 1.68(6)    & 0.50(6)    & 1.30(5)       & 4.75(-)    & (6?)    &        & $<$4  & $<$4 \\ 
%SDSSp J162414.37+002915.6\tablenotemark{bb}       & 1999 Jul 16                &       &            &            &               &            & (6 std) & T6 std &       &      \\
%
\enddata
\tablecomments{Discovery references are (1) \citealt{burgasser2002}, (2) \citealt{leggett2000}, (3) \citealt{burgasser2004}.}
\tablenotetext{a}{SDSS J0837$-$0000 was observed on three separate dates: 2000 Dec 26 (7200s integration), 2000 Dec 27 (3600s),
and 2000 Dec 28 (3600s). All analysis was performed on a coadded spectrum combining all 14400s of integration.}
\tablenotetext{b}{2MASS J1209$-$1004 was observed on three separate dates: 2003 Dec 22 (2400s integration), 2003 Dec 23 (3600s),
and 2003 Dec 24 (3600s). All analysis was performed on a coadded spectrum combining all 9600s of integration.}
\tablenotetext{c}{2MASS J0415$-$0935 has a flux measurement at or very close to zero at these wavelengths, so meaningful equivalent width measures are not possible.}
%
%\tablenotetext{aa}{Spectrum previously published in Burgasser et al.\ (2003).}
%\tablenotetext{bb}{Spectrum previously published in Burgasser et al.\ (2000).}
%
\end{deluxetable}

\clearpage

\begin{deluxetable}{llclclccl}
\tabletypesize{\scriptsize}
\rotate
\tablewidth{9in}
\tablecaption{Ultra-cool Field\tablenotemark{e} Dwarfs ($\ge$M7) with Spectroscopic Signatures of Low Gravity (Youth)\tablenotemark{a}\label{young_dwarfs}}
\tablehead{
\colhead{Object Name} &
\colhead{Optical} &
\colhead{Opt.} &
\colhead{Near-IR} &
\colhead{Near-IR} &
\colhead{2MASS} &
\colhead{Disc.} &
\colhead{Youth} &
\colhead{Diagnostics} \\
\colhead{} &
\colhead{Sp.\ Type} &
\colhead{Ref.} &
\colhead{Sp.\ Type} &
\colhead{Ref.} &
\colhead{J-K$_s$} &
\colhead{Ref.} &
\colhead{Ref.} &
\colhead{} \\
\colhead{(1)} &
\colhead{(2)} &
\colhead{(3)} &
\colhead{(4)} &
\colhead{(5)} &
\colhead{(6)} &
\colhead{(7)} &
\colhead{(8)} &
\colhead{(9)}
}
\startdata
2MASS J00332386$-$1521309 & L2 pec & 1 & ---           & --- & 1.88$\pm$0.07 & 18& 1 & (see Figure~\ref{lowg_L2}) \\
2MASS J01415823$-$4633574 & L0 pec & 2 & L0 pec        & 2   & 1.74$\pm$0.05 & 2 & 2 & (see Figure~\ref{lowg_L0}) \\
2MASSI J0241115$-$032658  & L1 pec & 8 & ---           & --- & 1.76$\pm$0.08 & 8 & 8 & (see Figure~\ref{lowg_L0}) \\
2MASSI J0253597+320637    & M7 pec?& 14& ---           & --- & 1.07$\pm$0.03 & 14& 14& marginally weak CaH, possibly weak \ion{Na}{1} \\
DENIS J035726.9$-$441730  & L0 pec & 1 & ---           & --- & 1.46$\pm$0.04 & 3 & 1 & (see Figure~\ref{lowg_L0}) \\
DENIS J0436278$-$411446   & M8 pec?& 8 & ---           & --- & 1.04$\pm$0.04 & 17& 8 & slightly weak \ion{Na}{1}, slightly strong VO? \\
SDSS J044337.61+000205.1  & M9 pec & 1 & ---           & --- & 1.29$\pm$0.03 & 4 & 1,8 & (see Figure~\ref{lowg_lateM}) \\
2MASSI J0608528$-$275358  & M9 pec & 14 & ---          & --- & 1.22$\pm$0.04 & 14& 14& (see Figure~\ref{lowg_lateM}) \\
2MASS J10220489+0200477   & L0 pec?& 1 & ---           & --- & 1.20$\pm$0.04 & 5 & 1 & (see Figure~\ref{lowg_L0}) \\
2MASS J10224821+5825453   & L1 pec & 1 & ---           & --- & 1.34$\pm$0.04 & 5 & 1 & (see Figure~\ref{lowg_L1}) \\
SDSS J102552.43+321234.0  & ---   & ---& L7.5$\pm$2.5  & 13  & ---\tablenotemark{c}  & 13& 1 & weak near-IR \ion{K}{1}, weak H$_2$O \\
SSSPM J1102$-$3431\tablenotemark{b} & M8.5 pec& 16& --- & --- & 1.15$\pm$0.03 & 16& 16& weak \ion{Na}{1}, strong VO, weak CaH \\
2MASSW J1139511$-$315921  & M8 pec & 15& M9 pec        & 19  & 1.18$\pm$0.03 & 15& 15& weak CaH, weak \ion{Na}{1} doublet, strong VO \\
2MASSW J1207334$-$393254  & M8 pec & 15& M8 pec        & 19  & 1.05$\pm$0.04 & 15& 15& weak CaH, weak \ion{Na}{1} doublet, strong VO \\
DENIS J124514.1$-$442907  & M9.5 pec& 19& M9 pec       & 19  & 1.15$\pm$0.05 & 19& 19& weak CaH, weak \ion{Na}{1} doublet, strong VO, etc. \\
2MASSI J1615425+495321   & L4 pec\tablenotemark{d}  & 8 & ---           & --- & 2.48$\pm$0.15 & 8 & 8 & (see Figure~\ref{lowg_L4}) \\
2MASSW J2208136+292121    & L2 pec & 6 & ---           & --- & 1.65$\pm$0.11 & 6 & 1 & (see Figure~\ref{lowg_L2}) \\
2MASS J22134491$-$2136079 & L0 pec & 1 & ---           & --- & 1.62$\pm$0.05 & 8 & 1 & (see Figure~\ref{lowg_L0}) \\
2MASSW J2244316+204343    & L6.5   & 1 & ???           & 7   & 2.45$\pm$0.16 & 9 & 7 & (see Figure~\ref{lowg_L6.5}) \\
SDSSp J224953.45+004404.2 & L3     & 4 & L5$\pm$1.5    & 10  & 2.23$\pm$0.14 & 11& 12& peaky H-band, weak 1.1 $\mu$m H$_2$O, weak near-IR \ion{K}{1} \\
\enddata
\tablecomments{Key to references:
(1) this paper, (2) \citealt{kirkpatrick2006}, (3) \citealt{bouy2003}, (4) \citealt{hawley2002}, (5) \citealt{reid2008},
(6) \citealt{kirkpatrick2000}, (7) \citealt{looper2008}, (8) \citealt{cruz2007}, (9) \citealt{dahn2002}, (10) \citealt{knapp2004}, 
(11) \citealt{geballe2002}, (12) \citealt{nakajima2004}, (13) \citealt{chiu2006}, (14) \citealt{cruz2003}, 
(15) \citealt{gizis2002}, (16) \citealt{scholz2005}, (17) \citealt{phan-bao2003}, 
%(18) \citealt{schneider1991}, 
%(19) \citealt{martin1999b}, (20) \citealt{tinney1998}, (21) \citealt{kirkpatrick1999}, (22) \citealt{luyten1975},
(18) \citealt{gizis2003}, (19) \citealt{looper2007}.}
\tablenotetext{a}{Other possible low-g dwarfs have been noted due to color anomalies only, e.g., SDSS 0107+0041 in \cite{knapp2004}.
For T dwarfs \cite{knapp2004} and \cite{burgasser2006} have found that for fixed T$_{eff}$ the H-K color is diagnostic of gravity 
because it measures the importance of gravity- (pressure-) sensitive CIA H$_2$. Based on the combined work of these two groups, the
lowest gravities probed by the currently known set of T dwarfs is $4.5 < log(g) < 5.0$.}
\tablenotetext{b}{SSSPM J1102$-$3431: Also known as 2MASS J11020983$-$3430355.}
\tablenotetext{c}{SDSS J1025+3212: Not detected at 2MASS J-band.}
\tablenotetext{d}{2MASS J1615+4953: Spectrum has poor signal-to-noise and may prove to be dusty rather than low gravity.}
\tablenotetext{e}{Includes any field objects not known to be companions to other stars.}

\end{deluxetable}

\clearpage

\begin{deluxetable}{lllllllll}
\tabletypesize{\scriptsize}
\rotate
\tablewidth{9in}
\tablecaption{Log of Spectroscopic Observations for Objects Shown in Figure~\ref{lateM_gravity_sequence} - Figure~\ref{lowg_L6.5}\label{gravity_obs}}
\tablehead{
\colhead{2MASS Name} &
\colhead{Other Name} &
\colhead{Obs.\ Date (UT)} &
\colhead{Teles.} &
\colhead{Inst.} &
\colhead{Int.(s)} &
\colhead{Spec.\ Ref.} &
\colhead{Sp.\ Type} &
\colhead{Tell.\ Corr.?\tablenotemark{e}} \\
\colhead{(1)} &
\colhead{(2)} &
\colhead{(3)} &
\colhead{(4)} &
\colhead{(5)} &
\colhead{(6)} &
\colhead{(7)} &
\colhead{(8)} &
\colhead{(9)}
}
\startdata
2MASS J00332386$-$1521309 &                        &  2003 Dec 24 &  Keck    & LRIS   &  1200     &  8  & L2 pec        & no \\
2MASS J01415823$-$4633574 &                        &  2003 Dec 23 &  Keck    & LRIS   &  2400     &  4  & L0 pec        & no \\
2MASS J02052940$-$1159296 & DENIS-P J0205.4$-$1159 &  1997 Nov 09 &  Keck    & LRIS   &  1200     &  1  & L7 (std)      & no \\
2MASS J02411151$-$0326587 &                        &  2005 Oct 10 &  Gemini-S& GMOS   &  1187     &  5  & L0 pec        & no \\
2MASS J03435353+2431115 & Roque 4                  &  2003 Jan 03 &  Keck    & LRIS   &  2400     &  8  & late-M Pleiad & yes \\
2MASS J03454316+2540233 &                          &  1997 Nov 09 &  Keck    & LRIS   &  1200     &  1  & L0 (std)      & no \\
2MASS J03471791+2422317 & Teide 1                  &  2003 Jan 02 &  Keck    & LRIS   &  1800     &  8  & late-M Pleiad & yes \\
----- J0348306 +224450  & Roque 25                 &  2003 Jan 02 &  Keck    & LRIS   &  6000     &  8  & L0 Pleiad     & yes \\
2MASS J03572695$-$4417305 & DENIS-P J035729.6$-$441730 &  2003 Dec 24 &  Keck    & LRIS   &  1200     &  8  & L0 pec        & no \\
2MASS J04300724+2608207 & KPNO-Tau6                &  2003 Jan 02 &  Keck    & LRIS   &  1200     &  8  & M8.5          & yes \\
2MASS J04433761+0002051 & SDSS J044337.61+000205.1 &  2003 Dec 24 &  Keck    & LRIS   &  600      &  8  & M9 pec        & no \\
2MASS J06085283$-$2753583 &                        &  2003 Dec 23 &  Keck    & LRIS   &  1200     &  6  & M8.5 pec      & yes \\
2MASS J07065882+0852370 & V CMi                    &  2003 Dec 23 &  Keck    & LRIS   &  1        &  8  & late-M giant  & yes \\
2MASS J08503593+1057156 &                          &  multidate\tablenotemark{a} &  Keck    & LRIS   &  10800    &  1  & L6 (std)      & no \\
2MASS J10042066+5022596 & G 196-3B                 &  multidate\tablenotemark{b} &  Keck    & LRIS   &  3600/6000     &  3  & L2 pec        & both\tablenotemark{b} \\
2MASS J10220489+0200477 &                          &  multidate\tablenotemark{c} &  CTIO-4m & RC-Spec&  1200     &  7  & L0 pec        & no \\
2MASS J10224821+5825453 &                          &  2004 Feb 10 &  KPNO-4m & RC-Spec&  600      &  7  & L1 pec        & no \\
2MASS J11122567+3548131 & Gl 417B                  &  multidate\tablenotemark{d} &  Keck    & LRIS   &  6000     &  2  & L4.5 pec      & no \\
2MASS J11463449+2230527 &                          &  1997 Dec 07 &  Keck    & LRIS   &  1200     &  1  & L3 (std)      & no \\
2MASS J11550087+2307058 &                          &  1998 Jan 22 &  Keck    & LRIS   &  2400     &  1  & L4 (std)      & no \\
2MASS J12073346$-$3932539 &                        &  2003 Jan 02 &  Keck    & LRIS   &   600     &  8  & M8            & yes \\
2MASS J12281523$-$1547342 & DENIS-P J1228.2$-$1547 &  1997 Dec 09 &  Keck    & LRIS   &  1200     &  1  & L5 (std)      & no \\
2MASS J12391934+2029519 &                          &  1997 Dec 08 &  Keck    & LRIS   &  1200     &  1  & M9            & no \\
2MASS J13054019$-$2541059 & Kelu-1AB               &  1998 Jan 22 &  Keck    & LRIS   &  1200     &  1  & L2 (std)      & no \\
2MASS J14325988$-$1056035 & IRAS 14303$-$1042      &  2003 Jan 03 &  Keck    & LRIS   &  10(or 15)&  8  & late-M giant  & yes \\
2MASS J14392836+1929149 &                          &  1997 Dec 08 &  Keck    & LRIS   &  1200     &  1  & L1 (std)      & no \\
2MASS J16154255+4953211 &                          &  2004 Sep 12 &  Gemini-N& GMOS   &  960      &  5  & L4 pec        & no \\
%2MASS J18410861+3117279 &                          &  2007 Aug 21 &  Subaru  & FOCAS  &  2400     &  8  & L4 pec        & no \\
2MASS J19165762+0509021 & vB 10                    &  2004 Jul 15 &  Keck    & LRIS   &  300      &  8  & M8 (std)      & yes \\
2MASS J22081363+2921215 &                          &  2007 Aug 21 &  Subaru\tablenotemark{f}  & FOCAS  &  1200     &  8  & L2 pec        & no \\
2MASS J22134491$-$2136079 &                        &  2007 Aug 21 &  Subaru\tablenotemark{f}  & FOCAS  &  1200     &  8  & L0 pec        & no \\
2MASS J22443167+2043433 &                          &  2000 Dec 26 &  Keck    & LRIS   &  2400     &  8  & L6.5          & no \\
\enddata
\tablecomments{Key to reference for spectral acquisition:
(1) \citealt{kirkpatrick1999}, (2) \citealt{kirkpatrick2000}, (3) \citealt{kirkpatrick2001}, (4) \citealt{kirkpatrick2006},
(5) \citealt{cruz2007}, (6) \citealt{cruz2003}, (7) \citealt{reid2008} and DwarfArchives.org, (8) this paper.}
\tablenotetext{a}{2MASS J0850+1057 was observed on five different dates; the plotted spectrum is a sun of all five. Exposure times
were 2400s on 1997 Nov 09 (UT), 1200s on 1997 Dec 07 (UT), 1200s on 1997 Dec 08 (UT), 1200s on 1997 Dec 09 (UT), and 4800s on 1998 Jan 24 (UT).}
\tablenotetext{b}{G 196-3B was observed using a blue (3900-8700\AA) on one night and our standard red
setup (6300-10100 \AA) on two other nights. Observations for the blue setup were done in superior conditions, so the plotted spectrum
shortward of 8500 \AA\ is from this set, and data longward of 8500 \AA\ come from
the poorer data taken with the red setup. Data taken with the blue setup were corrected for telluric absorption; data with the red
setup were not telluric corrected. The exposure time for the blue setup was 3600s on 2001 Feb 19 
(UT). Exposure times for the red setup were 2400s on 1999 Mar 04 (UT) and 3600s on 1999 Mar 05 (UT).}
\tablenotetext{c}{2MASS J1022+0200 was observed for 600s on two different dates, 2003 Apr 04 (UT) and 2006 Jan 15 (UT). The plotted spectrum
is the sum of these two observations.}
\tablenotetext{d}{Gl 417B was observed on three different dates; the plotted spectrum is a sum of all three. Exposure times were
1200s on 1998 Dec 14 (UT), 1200s on 1999 Mar 04 (UT), and 3600s on 1999 Mar 05 (UT).}
\tablenotetext{e}{Indicates whether or not the spectrum was telluric corrected using a 
  G dwarf spectrum acquired near in time and near on sky to the program object.}
\tablenotetext{f}{Observation and reduction procedures for Subaru-FOCAS data are discussed in \cite{looper2008}.}
\end{deluxetable}

\clearpage

\begin{deluxetable}{clcclcclcclc}
\tabletypesize{\tiny}
\rotate
\tablewidth{9in}
\tablecaption{Lithium Equivalent Width Measurements for L0, L2, L4, and L6.5 Dwarfs\label{lithium_age_grid}}
\tablehead{
\colhead{} &
\multicolumn{2}{c}{Normal Dwarfs (without Lithium)} &
\colhead{} &
\multicolumn{2}{c}{Normal Dwarfs (with Lithium)} &
\colhead{} &
\multicolumn{2}{c}{Peculiar Dwarfs (Low-g)} &
\colhead{} &
\multicolumn{2}{c}{Very Peculiar Dwarfs (Very Low-g)} \\
\cline{2-3} 
\cline{5-6}
\cline{8-9}
\cline{11-12} \\
\colhead{Sp.} &
\colhead{Name} &
\colhead{EW} &
\colhead{} &
\colhead{Name} &
\colhead{EW} &
\colhead{} &
\colhead{Name} &
\colhead{EW} &
\colhead{} &
\colhead{Name} &
\colhead{EW} \\
\colhead{Ty.} &
\colhead{} &
\colhead{(\AA)} &
\colhead{} &
\colhead{} &
\colhead{(\AA)} &
\colhead{} &
\colhead{} &
\colhead{(\AA)} &
\colhead{} &
\colhead{} &
\colhead{(\AA)} \\
\colhead{(1)} &
\colhead{(2)} &
\colhead{(3)} &
\colhead{} &
\colhead{(4)} &
\colhead{(5)} &
\colhead{} &
\colhead{(6)} &
\colhead{(7)} &
\colhead{} &
\colhead{(8)} &
\colhead{(9)}
}
\startdata
L0 &  2MASSW J0058425$-$065123& $<$1   &   &2MASS J11544223$-$3400390& 2.9$\pm$0.7&  &DENIS-P J035729.6$-$441730& $\le$2&  &2MASS J01415823$-$4633574\tablenotemark{e}& $\le$1 \\
   &  2MASP J0345432+254023   & $<$0.5 &   &                         &            &  &Roque 25                  & $<$3&    &2MASS J02411151$-$0326587& $<$2 \\
   &  DENIS-P J0909$-$0658    & $<$0.5 &   &                         &            &  &2MASS J10220489+0200477   & $<$10&   &2MASS J22134491$-$2136079& $<$1 \\
   &  2MASSW J1449378+235537  & $<$10  & \\
   &  2MASS J16184503$-$1321297 & $<$5 & \\
   & \\
L2 &  2MASSW J0015447+351603  & $<$0.5 &   &Kelu-1\tablenotemark{b}  & 1.7$\pm$0.5& &2MASS J00332386$-$1521309  & $<$2 &   \\
   &  2MASSW J0030438+313932  & $<$1.0 &   &2MASSI J1726000+153819   & 6$\pm$2    & &G 196-3B\tablenotemark{c}  & 6.0$\pm$0.7 \\ 
   &  2MASSI J0224367+253704  & $<$1   &   &2MASSW J2206540$-$421721 & 8$\pm$2    & &2MASSW J2208136+292121     & 4$\pm$1  \\
   &  2MASSI J0753321+291711  & $<$1   & \\
   &  2MASSW J0829066+145622  & $<$0.5 & \\
   &  2MASSW J0928397$-$160312& $<$1   & \\
   &  2MASSW J0944027+313132  & $<$1   & \\
   &  2MASSI J1332286+263508  & $<$2   & \\
   &  2MASS J14090310$-$3357565& $<$1  & \\
   & \\
L4 &  2MASSW J1155009+230706  & $<$0.5 &   &2MASSW J0129122+351758 & 3.3$\pm$1.2 & &Gl 417B\tablenotemark{d} & 9.4$\pm$0.5 &  \\
   &  GD 165B\tablenotemark{a} & $<$0.7 &   &2MASSW J1246467+402715 & 10.5$\pm$1.5  & &2MASS 16154255+4953211 & $\le$8\\
   &  DENIS-P J153941.96$-$052042.4& $<$10 \\
   &  2MASS J21580457$-$1550098 & $\le$11 \\
   &  \\
L6.5& 2MASSW J0801405+462850 & $<$2 &      & 2MASSW J0829570+265510 & 18$\pm$5 &  &2MASS J22443167+2043433\tablenotemark{f} & 5$\pm$3 \\
   &  2MASSW J0920122+351742 & $<$0.5&     & 2MASS J21481628+4003593\tablenotemark{g} & 12.6$\pm$0.2 \\
   &  2MASSI J1711457+223204 & $\le$7&    \\
\enddata
\tablenotetext{a}{GD 165B: Age estimate using the white dwarf primary is 1.2-5.5 Gyr (\citealt{kirkpatrick1999b}).}
\tablenotetext{b}{Kelu-1: Age estimate from model fits is 300-800 Myr (\citealt{liu2005}).}
\tablenotetext{c}{G 196-3B: Age estimates from the presence of the lithium in G 196-3B itself as well as activity diagnostics in the primary is 20-300 Myr (\citealt{rebolo1998}); an age estimate using indicators from the early-M primary alone is 60-300 Myr (\citealt{kirkpatrick2001}).}
\tablenotetext{d}{Gl 417B: Age estimate from the G dwarf primary is 80-300 Myr (\citealt{kirkpatrick2001}). Gl 417B is actually typed as L4.5 but is included here as an L4 
so that this bin of peculiar, low-gravity L4 dwarfs could be populated.}
\tablenotetext{e}{2MASS J0141$-$4633: Age estimate from fitting of theoretical models to spectra is 1-50 Myr (\citealt{kirkpatrick2006}).}
\tablenotetext{f}{2MASS J2244+2043: This object has a normal L6.5 optical spectrum, but its near-infrared spectrum is extremely peculiar. In this case, the peculiarity
may be attributable to low gravity (\citealt{mclean2003}).}
\tablenotetext{f}{2MASS J2148+4003: This object has a normal L6.5 optical spectrum, but its near-infrared spectrum is extremely peculiar. In this case, the peculiarity
is best explained by excessive atmospheric dust (\citealt{looper2008}).}

\end{deluxetable}

\clearpage

\clearpage 

\clearpage 

\clearpage

\clearpage 

\clearpage 

\clearpage 

\clearpage 

\clearpage 

\clearpage 

\clearpage

\clearpage

\clearpage 

\clearpage 

\clearpage 

\clearpage 

\clearpage 

\clearpage

\clearpage

\clearpage

\clearpage

\clearpage 

\end{document}